\documentclass[preprint,aps,prc,showpacs,nofootinbib,secnumarabic,floatfix]{revtex4-1}

\usepackage{graphicx}
\usepackage{bm}
\usepackage{color}
\usepackage{gensymb}
\usepackage{amsmath}
\usepackage{slashed}
\usepackage{appendix}
\usepackage{hyperref}

\newcommand{\ltsim}{\protect\raisebox{-0.5ex}{$\:\stackrel{\textstyle <}
	{\sim}\:$}}
\newcommand{\gtsim}{\protect\raisebox{-0.5ex}{$\:\stackrel{\textstyle >}
	{\sim}\:$}}

\newcommand{\bvec}[1]{\ensuremath{\boldsymbol{#1}}}

\begin{document}

\title{Color coherence effects in the reaction $d(p,2p)n$}  

\author{A.B. Larionov}
\email{e-mail: larionov@theor.jinr.ru}
\affiliation{Bogoliubov Laboratory of Theoretical Physics, Joint Institute for Nuclear Research, 141980 Dubna, Russia}

\begin{abstract}
The hard proton knock-out by the proton from the deuteron at relativistic energies is considered with a focus on   
the color transparency (CT) effect which influences the initial and final state interactions.
The calculations are performed in the framework of the generalized eikonal approximation supplemented by the quantum diffusion model of CT.
The main results of the previous calculations \cite{Frankfurt:1996uz} at the beam momentum below 20 GeV/c are confirmed, 
including the dependence of the nuclear transparency on the transverse momentum of the spectator neutron, $p_{st}$,
and on the relative azimuthal angle $\phi$ between proton and neutron:
absorption at $p_{st} < 0.2$ GeV/c and enhancement at $p_{st} > 0.3$ GeV/c
due to rescattering on the neutron, the change of $\phi$-dependence between these two regions, and the enhancement of CT effects
with $p_{\rm lab}$. The study is then generalized to higher beam momenta, up to 75 GeV/c.
It is shown that such behavior of the transparency is mainly preserved up to $p_{\rm lab}\simeq 50$ GeV/c, but changes significantly
at higher beam momenta due to the interference of valence quark configurations of small and large sizes.
As a result, the transparency at small $p_{st}$ exhibits oscillations as a function of the beam momentum (the nuclear filtering effect).
The tensor analyzing power due to the longitudinal polarization of the deuteron is calculated.
The event rate at NICA-SPD is estimated.
\end{abstract}

\maketitle

\section{Introduction}
\label{intro}

The search for the CT phenomenon started about four decades ago after prediction \cite{Brodsky,Mueller} that in the hadron-induced semi-exclusive reactions
$h+A \to h +p + (A-1)^*$ with large momentum transfer ($Q^2 = -t \simeq -u \gg 1$ GeV$^2$) in the elastic $hp$ scattering
the nucleus becomes transparent for the incoming and outgoing particles.
As a result, the cross section will be $A$ times the cross section on the nucleon. 
CT is a consequence of the reduced, $\sim 1/Q$, transverse size of the quark configurations participating in the hard process
and their color neutrality, see reviews \cite{Frankfurt:1994hf,Jain:1995dd,Dutta:2012ii}.
So far, the most successful attempts to find CT
were undertaken for reactions with mesons in the initial or final states, such as the coherent diffractive dissociation of $\pi^-$ into two jets \cite{Aitala:2000hc},
incoherent  electroproduction of $\pi^+$ \cite{Clasie:2007aa} and $\rho^0$ \cite{ElFassi:2012nr}.
This is apparently explained by the smaller number of quarks in the meson. Thus, for the same $Q^2$, the pair $q\bar q$ has more chances to end up in a small-size,
so-called point-like configuration (PLC), as compared to the $qqq$ triple.

The key quantity in the studies of CT is the nuclear transparency
\begin{equation}
  T = \frac{\sigma}{\sigma_{\rm IA}}~,    \label{T}
\end{equation}
where $\sigma$ is the production cross section on the nucleus in the specific kinematics and
$\sigma_{\rm IA}$ is the same cross section but evaluated in the impulse approximation (IA).
Neglecting nuclear Fermi motion, $\sigma_{\rm IA}=A \sigma_p$ where $\sigma_p$ is the production cross sections on the
free proton.
The recent measurements of the nuclear transparency for the $^{12}\mbox{C}(e,e^\prime p)$ reaction \cite{HallC:2020ijh} exclude CT for $Q^2$ up to 14.2 GeV$^2$.
Calculations based on the Glauber model, including short-range $NN$ correlations \cite{Das:2021oig}, describe these data without taking into account the effects of CT.
On the other hand, in quasielastic proton knock-out reactions, the onset of CT might be shifted to larger $Q^2$ values  \cite{Brodsky:2022bum}.

In the AGS experiments (see Ref. \cite{Aclander:2004zm} and refs. therein), the nuclear transparency in the proton knock-out reaction
$A(p,pp)$ at $\Theta_{c.m.}=90\degree$ on heavy targets (Li, C, Al, Cu, Pb) in the beam momentum range from 5.9 to 14.4 GeV/c
(that corresponds to $Q^2=4.8-12.7$ GeV$^2$) clearly deviates from the Glauber model predictions.
The transparency starts rising from Glauber level at 5.9 GeV/c
reaching the maximum at 9.5 GeV/c and then returns back to the Glauber level or below at 12 GeV/c and higher beam momentum.
The rise is clearly consistent with CT \cite{Lee:1992rd}.
However, the decrease of $T$ at larger beam momenta can not be explained within
a pure CT mechanism and requires to introduce the interference of small- and large-size configurations
\cite{Brodsky:1987xw,Ralston:1988rb,Jennings:1991rw,Jennings:1993hw,VanOvermeire:2006tk}. 

The deuteron, being the lightest (and simplest) nucleus may be a natural starting point to look for nuclear medium effects of any kind.
The deuteron target has been suggested for the studies of CT in several wide-angle processes:
$d(e,e^\prime p)n$ \cite{Anisovich:1993eh,Frankfurt:1994kt}, $d(p,2p)n$ \cite{Frankfurt:1996uz}, and $d(\bar p, \pi^- \pi^0)p$ \cite{Larionov:2019xdn}.
By varying the transverse momentum of the spectator nucleon, it is possible to regulate the strength of initial- (ISI) and final (FSI) state interactions.
This gives more control over CT effects as compared to the case of heavier nuclear targets where the nuclear remnant is usually not detected.
On the other hand, the ISI/FSI in the deuteron are small effects in a quasi-free kinematics. To amplify them, one has to look into the kinematical ranges
where the production cross sections might be extremely small and have to be at least estimated for planning new experiments.

The first stage of the experimental program of the NICA Spin Physics Detector (SPD) project includes studies of $pp$, $pd$ and $dd$ collisions
at $\sqrt{s_{NN}}=3.4-10$ GeV with possible polarization of one or both colliding particles \cite{Abramov:2021vtu}.
In the full operation mode of NICA-SPD, $pd$ collisions with $\sqrt{s_{NN}}$ up to 19 GeV will be possible \cite{Arbuzov:2020cqg}.
This will allow to study both the generalized parton distributions of the proton and neutron and the CT effects.
In fact, the generalized parton distributions and CT are closely related, since CT is a necessary condition for the applicability of
factorization \cite{Strikman:2000qn}.
This is because in the absence of CT the multiple gluon exchanges before and after hard process can not be suppressed.
Thus, the CT behavior for the $d(p,2p)n$ reaction is regulated by the perturbative and nonperturbative
QCD mechanisms of the wide-angle elastic $pp$ scattering.

The present work extends the studies of Ref. \cite{Frankfurt:1996uz} for the reaction $d(p,2p)n$ at large center-of-mass (c.m.) angle
to the energy range covered by NICA. The focus is on the CT effects including the interference of small- and large-size configurations.
The calculations are performed within the generalized eikonal approximation (GEA) complemented by the quantum diffusion model (QDM)
to account for CT, see Refs. \cite{Frankfurt:1996uz,Frankfurt:1996xx,Sargsian:2001ax,Larionov:2019xdn}.

The structure of the paper is as follows. The underlying theoretical approach is explained in sec. \ref{model}
starting from Feynman diagrams. The parameterizations of elementary amplitudes
are described, including the separation of small- and large-size quark configurations to the hard elastic $pp$ scattering.
Then, on the basis of QDM, the effects of CT are included in the soft elastic $pn$ amplitudes.
The used observables, i.e. four-differential cross section, transparency, and tensor analyzing power, are explained in sec. \ref{observ}.
Sec. \ref{results} contains the results of numerical calculations for several values of beam momentum in the range from 6 to 75 GeV/c.
In sec. \ref{nonpol}, the four-differential cross section and transparency
are calculated as functions of the transverse momentum of the spectator neutron
and the relative azimuthal angle between the scattered proton and the neutron
in the kinematics when the neutron momentum is almost transverse to the beam direction
($\alpha_s=1$, see Eq.(\ref{alpha_s})).
In sec. \ref{pol}, the tensor analyzing power for the longitudinally polarized deuteron is calculated.
In sec. \ref{rates}, the event rate at NICA-SPD is estimated based on the calculated cross sections.
The summary and conclusions are given in sec. \ref{summary}.
Appendix \ref{d^4sigma} includes the calculational details of the four-differential cross section.

\section{Basic formulas}
\label{model}

The GEA formalism for the $d(p,2p)n$ reaction has been already given in Ref. \cite{Frankfurt:1996uz}.
Nevertheless, for completeness, we sketch the derivations and provide explicit
expressions for the partial amplitudes used in the present calculations.
Some moderate differences with the formalism of Ref. \cite{Frankfurt:1996uz} are also described here.   

\begin{figure}
  \begin{center}
    \begin{tabular}{cc}
  \includegraphics[scale = 0.50]{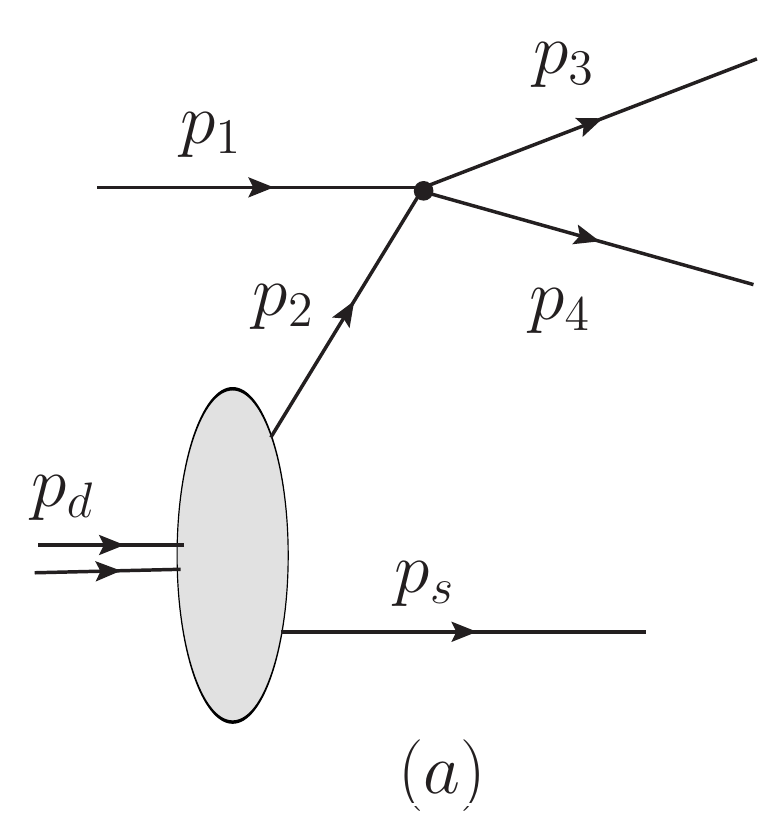} &
  \includegraphics[scale = 0.50]{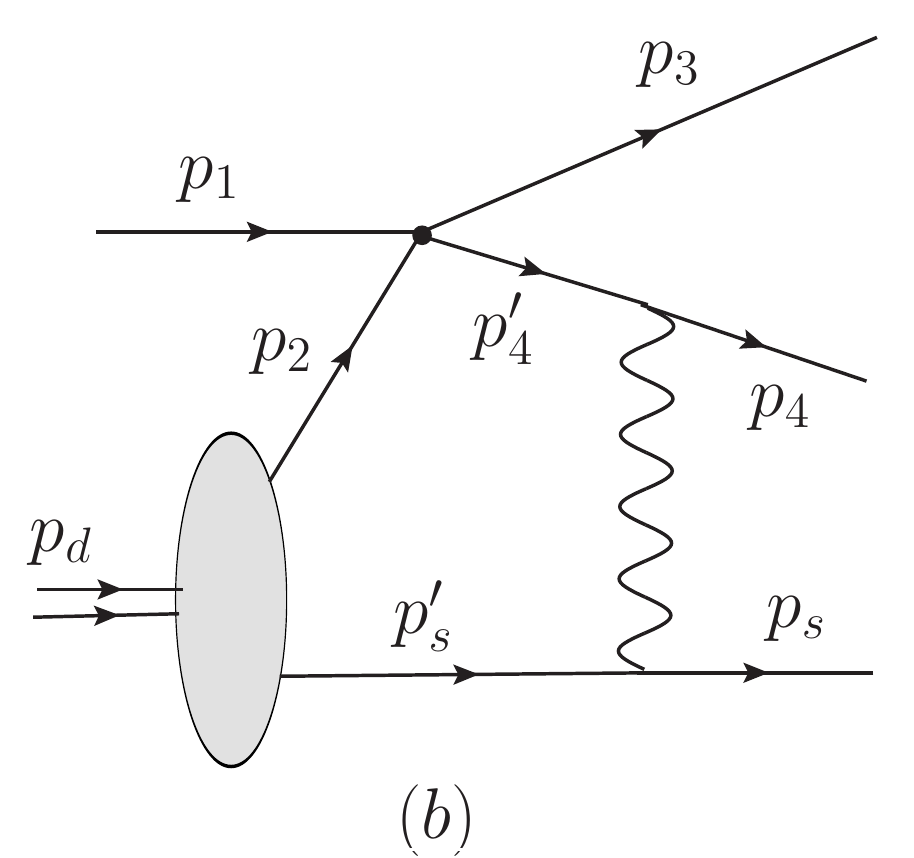}\\
  \includegraphics[scale = 0.50]{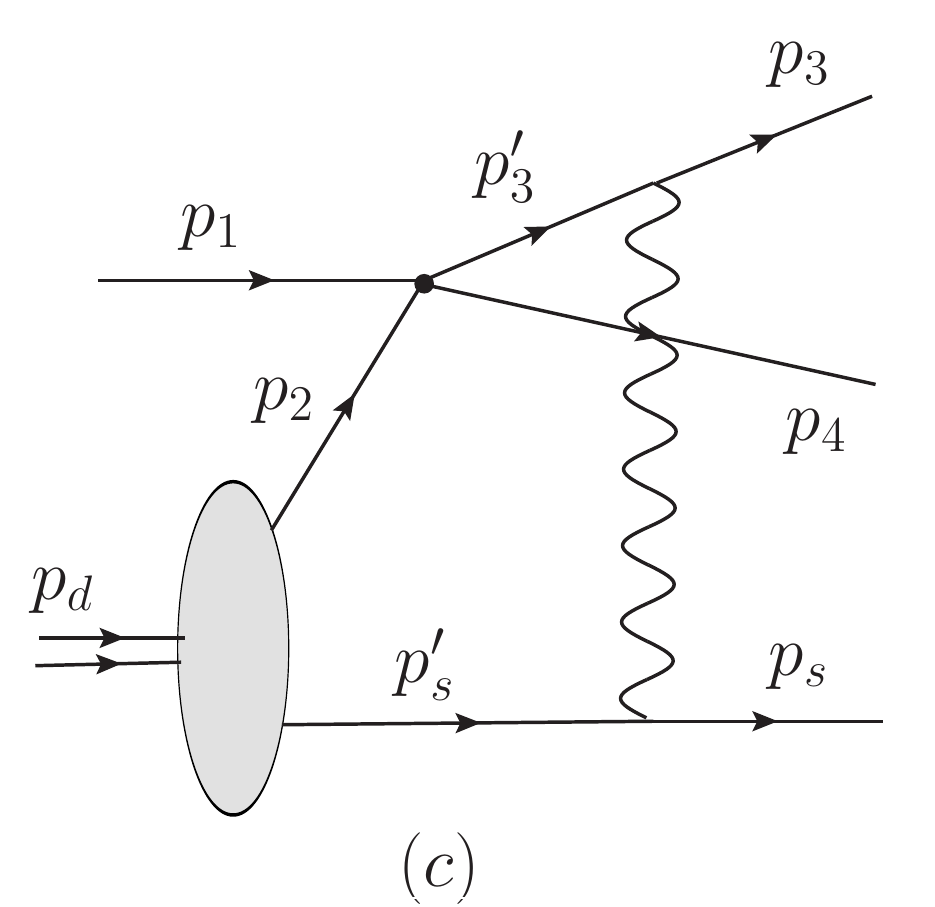} &
  \includegraphics[scale = 0.50]{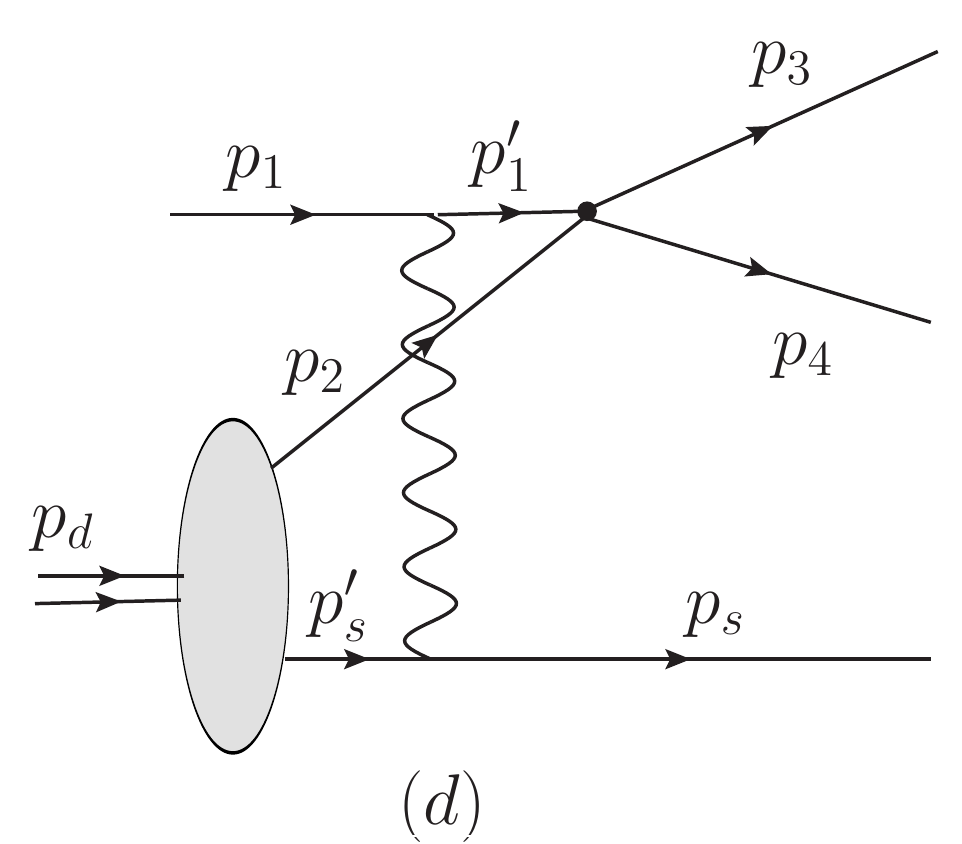}\\
  \includegraphics[scale = 0.50]{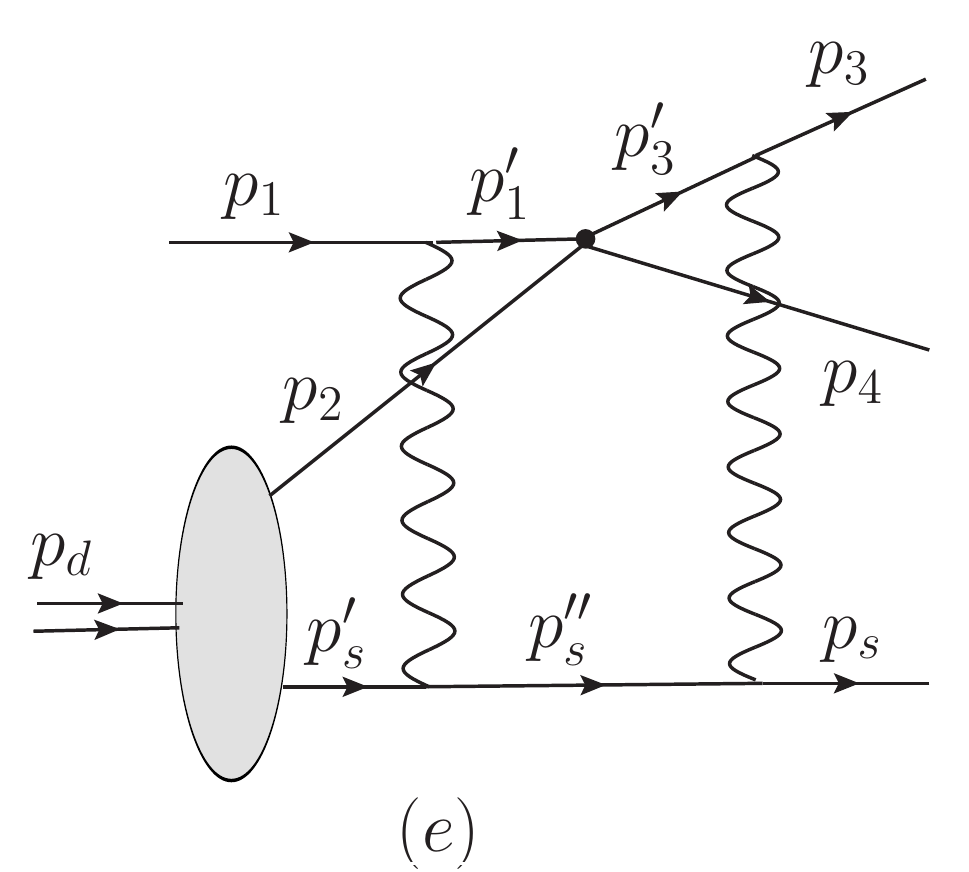} &
  \includegraphics[scale = 0.50]{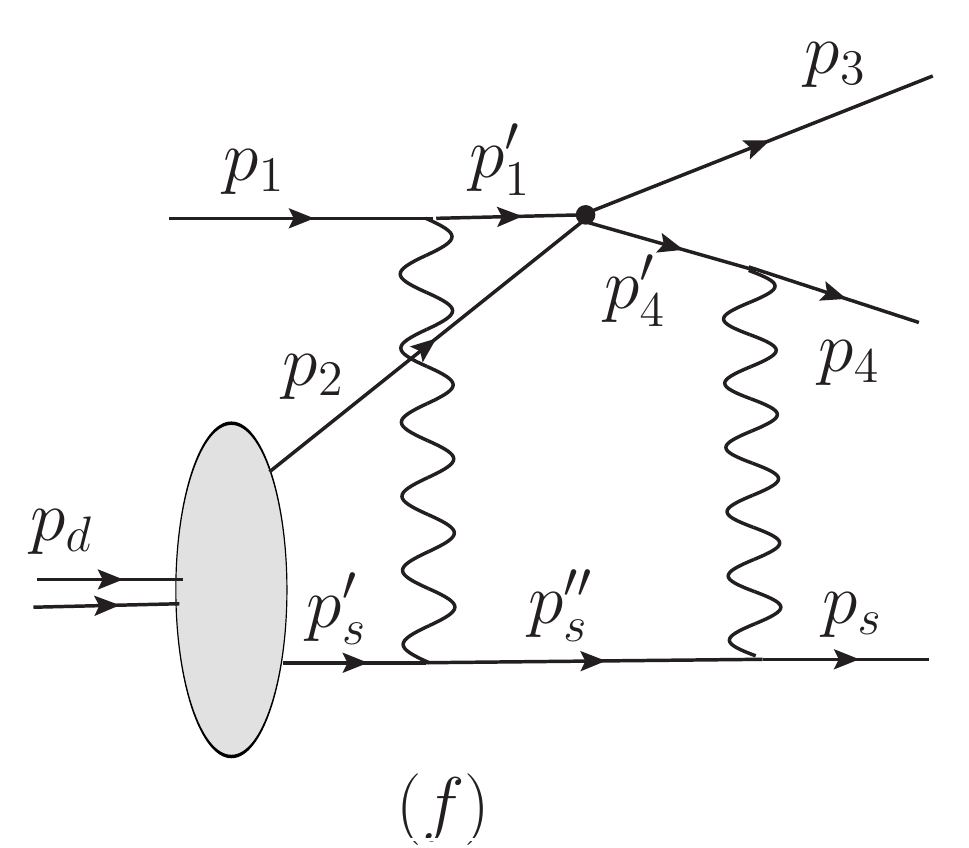}\\
  \includegraphics[scale = 0.50]{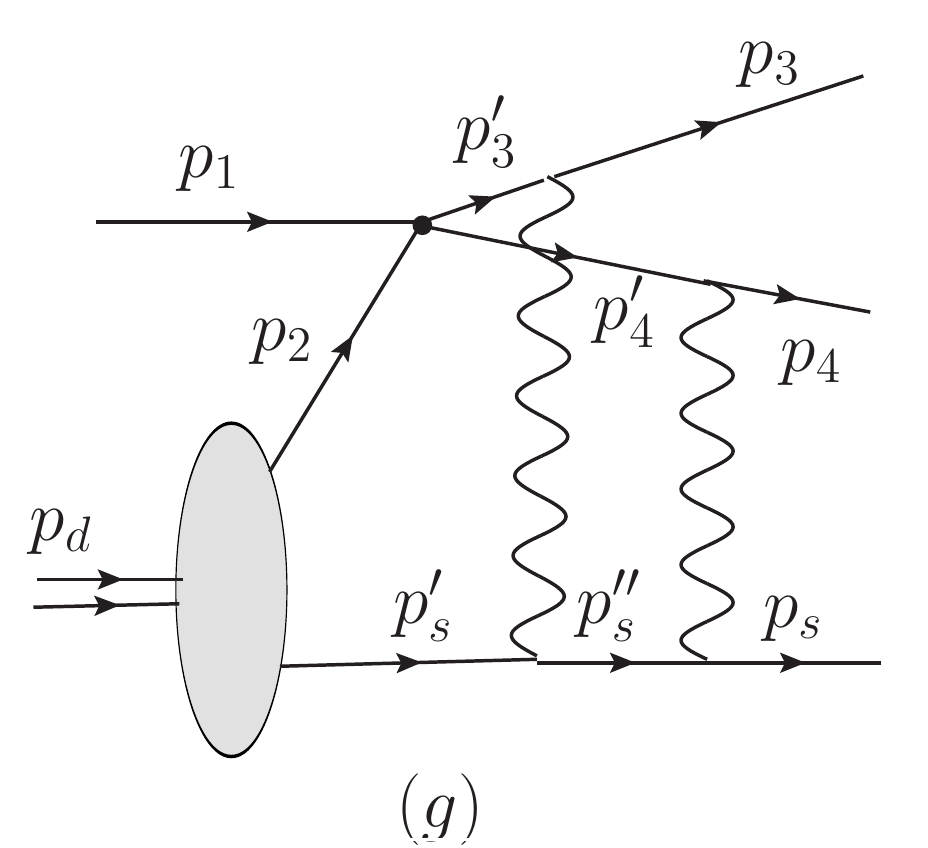} &
  \includegraphics[scale = 0.50]{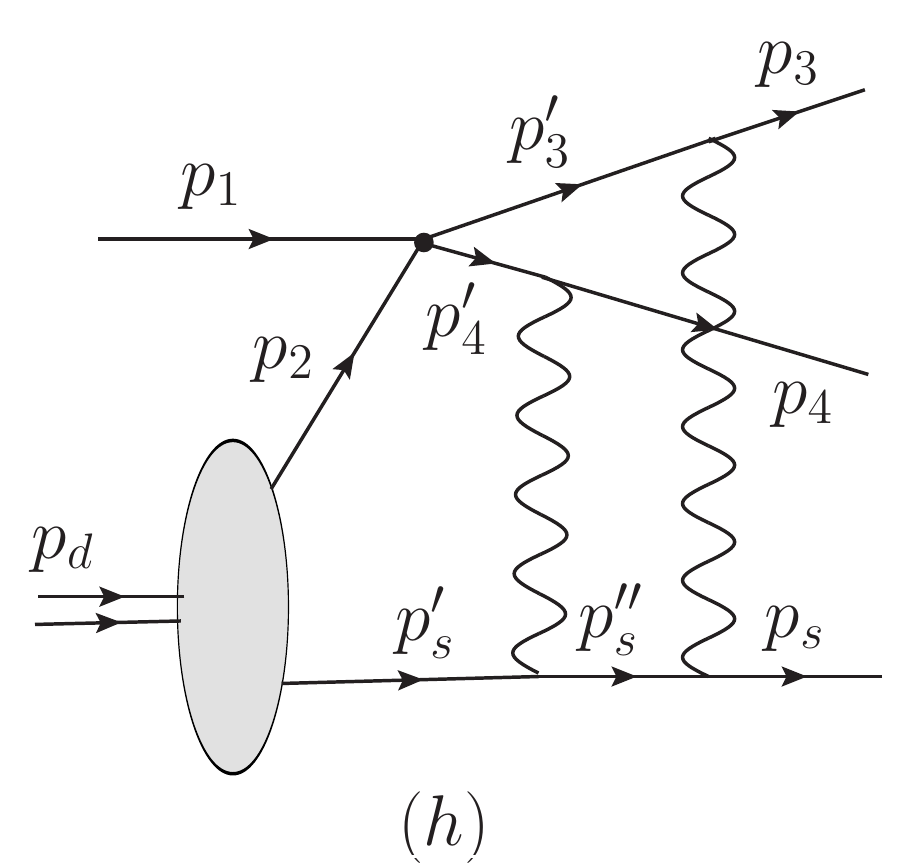}\\
    \end{tabular}
    \end{center}
\caption{\label{fig:diagr} Feynman diagrams for the process $p d  \to p p n$ with a slow spectator neutron.
  The wavy lines denote soft elastic scattering amplitudes.
  The four-momenta of the deuteron, neutron, beam proton, struck proton, and outgoing protons are denoted as
  $p_d$, $p_s$, $p_1$, $p_2$, $p_3$, $p_4$, respectively.
  The primed and double-primed quantities denote the four-momenta of
  intermediate particles in the amplitudes (b),(c),(d),(e),(f),(g),(h) with rescatterings.}
\end{figure}

The processes included in our calculations are shown in Fig.~\ref{fig:diagr}.
We take into account the IA amplitude (a), the amplitudes with single rescattering
(b),(c),(d), and the amplitudes with double rescattering (e),(f),(g),(h). 
In the notations of Ref. \cite{BLP}, the invariant matrix element of the IA amplitude  is expressed as
\begin{equation}
    M^{(a)} = M_{\rm hard}(p_3,p_4,p_1) 
              \frac{i\Gamma_{d \to pn}(p_d,p_s)}{p_2^2-m^2+i\epsilon}~,               \label{M^(a)}
\end{equation}
where $M_{\rm hard}(p_3,p_4,p_1)$ is the invariant matrix element of the hard elastic $pp$ scattering amplitude,
$\Gamma_{d \to pn}(p_d,p_s)$ is the deuteron decay vertex function, and $m$ is the nucleon mass.
Summation over intermediate spin indices is always assumed implicitly.
If the spectator neutron is on the mass shell, i.e. $p_s^{0}=E_s \equiv (m^2+\bvec{p}_s^2)^{1/2}$, then one can show by considering
the Fermi motion in the deuteron non-relativistically and projecting the deuteron wave function (DWF)
onto plane wave states (cf. Ref. \cite{Larionov:2018lpk}) that the following relation holds in the deuteron rest frame:
\begin{equation} 
 \frac{i\Gamma_{d \to pn}(p_d,p_s)}{p_2^2-m^2+i\epsilon} 
 = \left(\frac{2E_sm_d}{p_2^0}\right)^{1/2} (2\pi)^{3/2} \phi^{\lambda_d}(\bvec{p}_2)~,~~~\bvec{p}_2=-\bvec{p}_s~,     \label{Gamma_d}
\end{equation}
where $\phi^{\lambda_d}(\bvec{p}_2)$ is the DWF, $\lambda_d=0,\pm1$ is the $z$-projection of the deuteron spin,
$m_d$ is the deuteron mass
\footnote{The deuteron binding energy is unimportant on the energy scale of the studied processes.
Thus, we will set $m_d=2m$ in final expressions.},
and $p_2^0 = m_d - E_s$ is the energy of the off-shell struck proton.

The DWF contains the S- and D-state components and can be represented as follows \cite{Buck:1979ff,Gilman:2001yh}:
\begin{equation}
     \phi^{\lambda_d}(\bvec{p})=\frac{1}{\sqrt{4\pi}}\left[u(p) + \frac{w(p)}{\sqrt{8}} S(\bvec{p})\right] \chi^{\lambda_d}~,      \label{phi_decomp}
\end{equation}
where
\begin{equation}
   S(\bvec{p})=\frac{3(\bvec{\sigma}_p\bvec{p})(\bvec{\sigma}_n\bvec{p})}{p^2} 
               -\bvec{\sigma}_p\bvec{\sigma}_n    \label{Sspin}
\end{equation}
is the spin tensor operator and $\chi^{\lambda_d}$ is the eigenfunction of the spin~=~1 state with spin projection $\lambda_d$.
$\bvec{\sigma}_p$ and $\bvec{\sigma}_n$ are the Pauli matrices acting, respectively, on the proton and neutron spin indices
of $\chi^{\lambda_d}$. The DWF is normalized as
\begin{equation}
   \int d^3p |\phi^{\lambda_d}(\bvec{p})|^2 = 1~.    \label{DWFnorm}
\end{equation}
In calculations, the Paris potential model \cite{Lacombe:1981eg} will be used for the DWF. 
(Note that the D-state wave function is defined with negative sign, i.e. $w(p) < 0$.)
Spin quantum numbers and indices will be omitted below if this does not lead to misunderstandings.

Since the DWF quickly drops with increasing momentum, the momentum dependence of the prefactor in Eq.(\ref{Gamma_d}) can be neglected
which results in the following expression for the IA amplitude:
\begin{equation}
   M^{(a)} = 2m^{1/2} (2\pi)^{3/2} \phi(-\bvec{p}_s) M_{\rm hard}(s,t)~,    \label{M^(a)_fin}
\end{equation}
where the hard scattering amplitude is expressed in terms of the Mandelstam variables defined as
\begin{equation}
  s=(p_3+p_4)^2~,~~~t=(p_1-p_3)^2~,~~~u=(p_1-p_4)^2~.      \label{stu}
\end{equation}
Eq.(\ref{M^(a)_fin}) and all expressions below where the DWF appears
are applicable in the deuteron rest frame. The $z$-axis directed along the beam momentum $\bvec{p}_1$ is chosen
as the spin quantization axis.

The amplitude (b) with rescattering of the outgoing proton 4 can be written as
\begin{equation}
  M^{(b)} = \int \frac{d^4p_s^\prime}{(2\pi)^4}
  \frac{i\Gamma_{d \to pn}(p_d,p_s^\prime)}{p_2^2-m^2+i\epsilon} M_{\rm hard}(p_3,p_4^\prime,p_1)
  \frac{i}{p_4^{\prime 2}-m^2+i\epsilon} \frac{i}{p_s^{\prime 2}-m^2+i\epsilon} iM_{\rm el}(p_4,p_s,p_4^\prime)~,  \label{M^(b)}
\end{equation}
where $M_{\rm el}(p_4,p_s,p_4^\prime)$ is the invariant matrix element of the $pn$ elastic scattering amplitude.
The integration over $dp_s^{\prime 0}$ can be performed by closing the integration contour in the lower part
of the $p_s^{\prime 0}$ complex plane where the positive energy pole, $E_s^\prime=(m^2+\bvec{p}_s^{\prime 2})^{1/2}$,
of the intermediate neutron propagator provides the main contribution.
Thus, by putting the intermediate neutron on the mass shell and then using Eq.(\ref{Gamma_d})
the following formula can be obtained:
\begin{equation}
  M^{(b)} =  m^{-1/2} \int \frac{d^3p_s^\prime}{(2\pi)^{3}} (2\pi)^{3/2} \phi(-\bvec{p}_s^\prime) M_{\rm hard}(p_3,p_4^\prime,p_1)
  \frac{i}{p_4^{\prime 2}-m^2+i\epsilon} iM_{\rm el}(p_4,p_s,p_4^\prime)~.  \label{M^(b)_3d}
\end{equation}
The inverse propagator of the fast intermediate proton allows for further simplification:
\begin{equation}
  p_4^{\prime 2}-m^2+i\epsilon = (p_4+p_s-p_s^\prime)^2 -m^2 + i\epsilon = 2p_4(p_s-p_s^\prime) + (p_s-p_s^\prime)^2 + i\epsilon
  = 2|\bvec{p}_4| (p_s^{\prime \tilde z} - p_s^{\tilde z} + \Delta_4 + i\epsilon)~,       \label{4_inv_prop}
\end{equation}
where the $\tilde z$-axis is directed along $\bvec{p}_4$ and
\begin{equation}
  \Delta_4 \equiv \frac{E_4(E_s-E_s^\prime)}{|\bvec{p}_4|} + \frac{(p_s-p_s^\prime)^2}{2|\bvec{p}_4|}
          \simeq \frac{(E_4-m)(E_s-m)}{|\bvec{p}_4|}~.               \label{Delta_4}
\end{equation}
In the last approximate equality we neglected the Fermi motion in the deuteron that is essentially the GEA
where the propagator of a fast particle depends on the momentum transfer along the particle
momentum.
Note that in Ref. \cite{Frankfurt:1996uz}, the inverse propagator of
the fast intermediate proton was linearized with respect to the momentum  transfer along
the beam direction and neglecting the term $(p_s-p_s^\prime)^2$.

The transition to the coordinate representation is reached by applying the identity
\begin{equation}
   \frac{i}{p+i\epsilon}
     =\int dz^0 \Theta(z^0) \mbox{e}^{i p z^0}~,   \label{Dcoord}
\end{equation}
where $\Theta(x)$ is the Heaviside step function, and the relation between the DWFs
in the momentum and coordinate representation 
\begin{equation}
   (2\pi)^{3/2} \phi(\bvec{p}_2) = \int d^3 r \mbox{e}^{-i\bvec{p}_2\bvec{r}} 
                   \phi(\bvec{r})~, ~~~\bvec{r}=\bvec{r}_2-\bvec{r}_s~.     \label{phi(p_2)}
\end{equation}
After some algebra, this leads to the amplitude (b) in the coordinate representation:
\begin{equation}
  M^{(b)} = \frac{i}{2 |\bvec{p}_4| m^{1/2}} \int d^3r \Theta(-\tilde{z}) \phi(\bvec{r}) 
  \mbox{e}^{i\bvec{p}_s\bvec{r}-i\Delta_{4}\tilde{z}}
  \int \frac{d^2k_t}{(2\pi)^2} \mbox{e}^{-i\bvec{k}_t\bvec{\tilde b}}
  M_{\rm hard}(p_3,p_4^\prime,p_1) M_{\rm el}(p_4,p_s,p_4^\prime)~,   \label{M^(b)_coord}
\end{equation}
where $\tilde z = \bvec{r} \bvec{p}_4 / |\bvec{p}_4|$ and
$\bvec{\tilde b} = \bvec{r} - (\bvec{r} \bvec{p}_4) \bvec{p}_4 / |\bvec{p}_4|^2$
are the relative positions of the proton and neutron along and perpendicular to $\bvec{p}_4$,
respectively.
$\bvec{k}_t = \bvec{k} - (\bvec{k} \bvec{p}_4) \bvec{p}_4 / |\bvec{p}_4|^2$
is the transverse to $\bvec{p}_4$ component of the momentum transfer 
$\bvec{k} = \bvec{p}_s-\bvec{p}_s^\prime$ to the spectator neutron. 
The hard scattering amplitude can be factorized out of the momentum integral,
since its change on the scale of the transferred momentum in the soft rescattering is small.
This allows to perform the integration over angle between $\bvec{k}_t$ and $\bvec{\tilde b}$ and leads to the
following expression: 
\begin{equation}
  M^{(b)} = \frac{i M_{\rm hard}(s,t)}{4\pi |\bvec{p}_4| m^{1/2}} \int d^3r \Theta(-\tilde{z}) \phi(\bvec{r})  
  \mbox{e}^{i\bvec{p}_s\bvec{r}-i\Delta_{4}\tilde{z}}
  \int\limits_0^{+\infty} dk_t  k_t M_{\rm el}(|\bvec{p}_4|,k_t)  J_0(k_t \tilde{b})~,  \label{M^(b)_coord_fin}
\end{equation}
where we take into account that the soft elastic $pn$ amplitude depends on the proton momentum
and on the transverse momentum transfer. The quantity
\begin{equation}
  J_0(x)=\frac{1}{2\pi}\int\limits_0^{2\pi} d\phi\, \mbox{e}^{-ix\cos \phi}    \label{J_0}
\end{equation}
is the Bessel function of the first kind.

The amplitude (c) with rescattering of the outgoing proton 3 is obtained by simply replacing
$4 \to 3$ in Eq.(\ref{M^(b)_coord_fin}). The $\tilde z$ axis is directed along $\bvec{p}_3$
in this case.

The amplitude (d) with rescattering of the incoming proton is
\begin{equation}
  M^{(d)} = \int \frac{d^4p_s^\prime}{(2\pi)^4}
  \frac{i\Gamma_{d \to pn}(p_d,p_s^\prime)}{p_2^2-m^2+i\epsilon} M_{\rm hard}(p_3,p_4,p_1^\prime)
  \frac{i}{p_1^{\prime 2}-m^2+i\epsilon} \frac{i}{p_s^{\prime 2}-m^2+i\epsilon} iM_{\rm el}(p_1^\prime,p_s,p_1)~.  \label{M^(d)}
\end{equation}
The integration over $dp_s^{\prime 0}$ can be again performed by closing the contour in the lower part of the $p_s^{\prime 0}$
complex plane and keeping only the contribution of the positive energy pole of the intermediate neutron propagator.
\footnote{The positive energy pole of the propagator of fast intermediate proton provides the contribution $\propto 1/E_1^\prime$,
where $E_1^\prime=[m^2+(\bvec{p}_1+\bvec{p}_s^\prime-\bvec{p}_s)^2]^{1/2}$. This contribution disappears in the high-energy limit
and, thus, can be neglected.}
This leads to the expression
\begin{equation}
   M^{(d)} = m^{-1/2} \int \frac{d^3p_s^\prime}{(2\pi)^{3}} (2\pi)^{3/2} \phi(-\bvec{p}_s^\prime) iM_{\rm el}(p_1^\prime,p_s,p_1) 
            \frac{i}{p_1^{\prime 2}-m^2+i\epsilon} M_{\rm hard}(p_3,p_4,p_1^\prime)~.   \label{M^(d)_3d}
\end{equation}
The inverse propagator can now be simplified to the eikonal form:
\begin{equation}
  p_1^{\prime 2}-m^2+i\epsilon = (p_1+p_s^\prime-p_s)^2-m^2+i\epsilon
  = 2p_1(p_s^\prime-p_s) + (p_s^\prime-p_s)^2 +i\epsilon
  = 2 |\bvec{p}_1| (p_s^z - p_s^{\prime z} - \Delta_1)~,         \label{1_inv_prop}
\end{equation}
where
\begin{equation}
  \Delta_1 = \frac{E_1(E_s-E_s^\prime)}{|\bvec{p}_1|} - \frac{(p_s^\prime-p_s)^2}{2|\bvec{p}_1|}
           \simeq \frac{(E_1+m)(E_s-m)}{|\bvec{p}_1|}~.               \label{Delta_1}
\end{equation}
Here, in the last step we again neglected the Fermi motion in the deuteron. By using Eqs.(\ref{Dcoord}),(\ref{phi(p_2)})
and (\ref{1_inv_prop}) and assuming that the hard scattering amplitude can be factorized out of the momentum integral
we can express the amplitude (\ref{M^(d)_3d}) in the coordinate representation:
\begin{equation}
   M^{(d)} = \frac{i M_{\rm hard}(s,t)}{4\pi |\bvec{p}_1| m^{1/2}} \int d^3r \Theta(z) \phi(\bvec{r})  
  \mbox{e}^{i\bvec{p}_s\bvec{r}-i\Delta_{1}z}
  \int\limits_0^{+\infty} dk_t  k_t M_{\rm el}(|\bvec{p}_1|,k_t)  J_0(k_t b)~.  \label{M^(d)_coord_fin}
\end{equation} 

Let us now calculate the amplitudes with double rescattering. As shown in Ref. \cite{Frankfurt:1996uz},
the amplitudes (e) and (f) with rescattering of incoming and outgoing particles are equal to zero in the pole approximation.
The amplitudes (g) and (h) with rescattering of both outgoing protons, however, do not disappear and have to be taken into account.
For the amplitude (g), by putting both intermediate neutron propagators on the mass shell, we have:
\begin{eqnarray}
  M^{(g)} &=&  \frac{1}{2m^{3/2}} \int \frac{d^3p_s^\prime}{(2\pi)^{3}} (2\pi)^{3/2} \phi(-\bvec{p}_s^\prime)
    \int \frac{d^3p_s^{\prime\prime}}{(2\pi)^{3}} M_{\rm hard}(p_3^\prime,p_4^\prime,p_1)
      \frac{i}{p_3^{\prime 2}-m^2+i\epsilon} \frac{i}{p_4^{\prime 2}-m^2+i\epsilon} \nonumber \\
      && \times iM_{\rm el}(p_3,p_s^{\prime\prime},p_3^\prime) iM_{\rm el}(p_4,p_s,p_4^\prime)~.  \label{M^(g)_3d}
\end{eqnarray}
The transformation of inverse propagators to the eikonal form can now be performed:
\begin{eqnarray}
  p_3^{\prime 2}-m^2+i\epsilon &=& (p_3+p_s^{\prime\prime}-p_s^\prime)^2 -m^2 +i\epsilon
  = 2p_3(p_s^{\prime\prime}-p_s^\prime) + (p_s^{\prime\prime}-p_s^\prime)^2 +i\epsilon \nonumber \\
  &=& 2p_3^z(p_s^{\prime z} - p_s^{\prime\prime z}  + \Delta_3 +i\epsilon)~,  \label{3_inv_prop_d}\\
  p_4^{\prime 2}-m^2+i\epsilon &=& (p_4+p_s-p_s^{\prime\prime})^2 -m^2+i\epsilon
  = 2p_4(p_s-p_s^{\prime\prime}) + (p_s-p_s^{\prime\prime})^2 +i\epsilon \nonumber \\
  &=& 2p_4^z(p_s^{\prime\prime z} - p_s^z  + \Delta_4 +i\epsilon)~, \label{4_inv_prop_d} 
\end{eqnarray}
where 
\begin{eqnarray}
  \Delta_3 &=& \frac{E_3(E_s^{\prime\prime}-E_s^\prime)}{p_3^z} + \frac{(p_s^{\prime\prime}-p_s^\prime)^2}{2p_3^z} -\frac{\bvec{p}_{3t}\bvec{k}_t^\prime}{p_3^z}~,  \label{Delta_3_d}\\
  \Delta_4 &=& \frac{E_4(E_s - E_s^{\prime\prime})}{p_4^z} +  \frac{(p_s-p_s^{\prime\prime})^2}{2p_4^z} -\frac{\bvec{p}_{4t}\bvec{k}_t^{\prime\prime}}{p_4^z}~.   \label{Delta_4_d}  
\end{eqnarray}
and $\bvec{k}_t^\prime = \bvec{p}_{st}^{\prime\prime} -\bvec{p}_{st}^\prime$, $\bvec{k}_t^{\prime\prime} = \bvec{p}_{st} - \bvec{p}_{st}^{\prime\prime}$
are the transverse momentum transfers to the spectator neutron in the two successive soft rescatterings.
The transverse momentum transfers enter Eqs.(\ref{Delta_3_d}),(\ref{Delta_4_d}) since we choose now
the original (non-rotated) $z$-axis along the beam momentum.
Neglecting the Fermi motion allows us to approximate:
\begin{equation}
  \Delta_3 \simeq \frac{(E_3-m)(E_s^{\prime\prime}-m)}{p_3^z} -\frac{\bvec{p}_{3t}\bvec{k}_t^\prime}{p_3^z}
                        \equiv \Delta_3^0 -\frac{\bvec{p}_{3t}\bvec{k}_t^\prime}{p_3^z}~.  \label{Delta_3_appr}
\end{equation}
Since the longitudinal momentum transfer in soft elastic rescattering is much less than the transverse momentum transfer,
we can set $\bvec{p}_s^{\prime\prime} \simeq \bvec{p}_s - \bvec{k}_t^{\prime\prime}$. This leads to the approximate expression:
\begin{equation}
  \Delta_4 \simeq \frac{(E_4+E_s) (E_s - E_s^{\prime\prime})}{p_4^z} -\frac{(\bvec{p}_{4t}+\bvec{p}_{st})\bvec{k}_t^{\prime\prime}}{p_4^z}
   \equiv \tilde\Delta_4^0 -\frac{(\bvec{p}_{4t}+\bvec{p}_{st})\bvec{k}_t^{\prime\prime}}{p_4^z}~.   \label{Delta_4_appr}  
\end{equation}
In the same approximation, the energy of the intermediate neutron $E_s^{\prime\prime} \simeq \sqrt{m^2+(\bvec{p}_s - \bvec{k}_t^{\prime\prime})^2}$.
This allows us to integrate over $dp_s^{\prime\prime z}$ in Eq.(\ref{M^(g)_3d}) closing the integration contour in the lower part
of the complex plane $p_s^{\prime\prime z}$ which gives
\begin{eqnarray}
  M^{(g)} &=&  \frac{1}{8p_4^zp_3^z m^{3/2}} \int \frac{d^3p_s^\prime}{(2\pi)^{3}} (2\pi)^{3/2} \phi(-\bvec{p}_s^\prime)
    \int \frac{d^2p_{st}^{\prime\prime}}{(2\pi)^{2}} 
    \frac{i}{p_s^{\prime z} - p_s^z  -\frac{\bvec{p}_{3t}\bvec{k}_t^\prime}{p_3^z} -\frac{(\bvec{p}_{4t}+\bvec{p}_{st})\bvec{k}_t^{\prime\prime}}{p_4^z}
             + \Delta_3^0 + \tilde\Delta_4^0 + i\epsilon}\nonumber \\
      && \times M_{\rm hard}(p_3^\prime,p_4^\prime,p_1) iM_{\rm el}(p_3,p_s^{\prime\prime},p_3^\prime) iM_{\rm el}(p_4,p_s,p_4^\prime)~.  \label{M^(g)_3d_simp}
\end{eqnarray}
A similar expression for the diagram (g) has been obtained in Ref. \cite{Frankfurt:1996uz}. In the kinematics of hard scattering, we have $E_3 \simeq p_3^z \gg m$
and $E_4 \simeq p_4^z \gg E_s$ and thus $E_s^{\prime\prime}$ practically does not influence the sum $\Delta_3^0 + \tilde\Delta_4^0$.
\footnote{For numerical reasons, it is convenient to set $E_s^{\prime\prime}$ to be independent on the momentum transfer.
By default, the value $E_s^{\prime\prime} = (E_s+m)/2$ was used but we checked that increasing $E_s^{\prime\prime}$
by 50\% does not lead to visible changes in the transparencies.} 
Finally, by using Eqs.(\ref{Dcoord}),(\ref{phi(p_2)}) we express Eq.(\ref{M^(g)_3d_simp}) in the coordinate space:
\begin{eqnarray}
  M^{(g)} &=&  -\frac{1}{8p_4^zp_3^z m^{3/2}} \int d^3r \Theta(-z) \phi(\bvec{r}) \mbox{e}^{i\bvec{p}_s\bvec{r}-i(\tilde\Delta_4^0+\Delta_3^0)z} 
              \int \frac{d^2k_{t}^{\prime}}{(2\pi)^{2}} \mbox{e}^{-i\bvec{k}_t^{\prime} \bvec{b}^\prime} M_{\rm el}(p_3,p_s^{\prime\prime},p_3^\prime) \nonumber \\
  && \times \int \frac{d^2k_{t}^{\prime\prime}}{(2\pi)^{2}}  \mbox{e}^{-i\bvec{k}_t^{\prime\prime} \bvec{b}^{\prime\prime}} M_{\rm el}(p_4,p_s,p_4^\prime)
  M_{\rm hard}(p_3^\prime,p_4^\prime,p_1)~,  \label{M^(g)_3d_coord} 
\end{eqnarray}
where $\bvec{b}^\prime = \bvec{b} - \bvec{p}_{3t} z/p_3^z$, $\bvec{b}^{\prime\prime} = \bvec{b} - (\bvec{p}_{4t} + \bvec{p}_{st}) z/p_4^z$
are the modified impact parameter vectors of the outgoing protons due to finite transverse momenta. 
In Eq.(\ref{M^(g)_3d_coord}), the internal four-momenta $p_3^\prime, p_4^\prime$ are fixed by the on-shell conditions $(p_3^\prime)^2=(p_4^\prime)^2=m^2$
and by three-momentum transfers $(\Delta_3,\bvec{k}_t^{\prime})$, $(\Delta_4,\bvec{k}_{t}^{\prime\prime})$.
In principle, this fully determines the elementary amplitudes. However, for practical calculations, similar to the case of single-rescattering amplitudes,
we further simplify Eq.(\ref{M^(g)_3d_coord})
by factorizing the hard amplitude out of the momentum integrals and by assuming that the soft rescattering amplitudes depend only on the longitudinal momenta
of outgoing protons and on the transverse momentum transfers which gives:
\begin{eqnarray}
  M^{(g)} &=&  -\frac{M_{\rm hard}(s,t)}{32\pi^2p_4^zp_3^z m^{3/2}} \int d^3r \Theta(-z) \phi(\bvec{r}) \mbox{e}^{i\bvec{p}_s\bvec{r} -i(\tilde\Delta_4^0+\Delta_3^0)z } \nonumber \\
         && \times \int\limits_0^{+\infty} dk_t^{\prime} k_t^{\prime} M_{\rm el}(p_3^z,k_t^\prime) J_0(k_t^\prime b^\prime)
                   \int\limits_0^{+\infty} dk_t^{\prime\prime} k_t^{\prime\prime} M_{\rm el}(p_4^z,k_t^{\prime\prime}) J_0(k_t^{\prime\prime} b^{\prime\prime})~.  \label{M^(g)_3d_coord_fin} 
\end{eqnarray}
Finally, the amplitude (h) is obtained by exchanging $3 \leftrightarrow 4$ in Eq.(\ref{M^(g)_3d_coord_fin}). There is an asymmetry of Eq.(\ref{M^(g)_3d_coord_fin})
with respect to this exchange which disappears in the limit of high energy. In contrast, $M^{(h)}=M^{(g)}$ was used in calculations of Ref. \cite{Frankfurt:1996uz}.

\subsection{Elementary amplitudes}

The differential cross section of hard elastic $pp$ scattering was parameterized in Ref. \cite{Frankfurt:1994nw} as follows:
\begin{equation}
  \frac{d\sigma_{pp}}{dt} = \frac{d\sigma_{pp}^{\rm QC}}{dt} |1+R(s)|^2  F(s,\Theta_{c.m.})~,      \label{dsigma_pp/dt_par}
\end{equation}
where
\begin{equation}
  \frac{d\sigma_{pp}^{\rm QC}}{dt} = 45 \frac{\mu{\rm b}}{{\rm GeV}^2} \left(\frac{10\,{\rm GeV}^2}{s}\right)^{10}
                                  \left(\frac{4m^2-s}{2t}\right)^{4\gamma}                     \label{dsigma_pp^QC/dt_par}
\end{equation}
with $\gamma=1.6$ is the formula motivated by the quark counting (QC) prediction \cite{Matveev:1973ra,Brodsky:1973kr}. 
\begin{equation}
  R(s) = M_{\rm L}/M_{\rm QC} = \frac{\rho_1\sqrt{s}}{2} \mbox{e}^{\pm i(\phi(s)+\delta_1)}         \label{R}
\end{equation}
is the ratio of the Landshoff and QC contributions to the full scattering amplitude \cite{Ralston:1988rb}
with $\rho_1=0.08$ GeV$^{-1}$, $\delta_1=-2$ and
\begin{equation}
  \phi(s) = \frac{\pi}{0.06} \log\left[\log\left(\frac{s}{\Lambda_{\rm QCD}^2}\right)\right]~,     \label{phi}
\end{equation}
where $\Lambda_{\rm QCD}=0.1$ GeV. The sign of the phase in Eq.(\ref{R}) can not be determined
from $pp$ scattering and thus both signs are equally applicable. In our default calculations, the ``+''
sign was used. In some selected cases we used both signs and realized that the sensitivity to this uncertainty is very weak.
The function $F(s,\Theta_{c.m.})$ was introduced in Ref. \cite{Frankfurt:1994nw} to fit the elastic $pp$ scattering data 
at $60\degree \leq \Theta_{c.m.} \leq 90\degree$ for $s < 15$ GeV$^2$. Since we are primarily interested in larger values of
$s$ where $F(s,\Theta_{c.m.})$ approaches unity, we set $F(s,\Theta_{c.m.})=1$ in calculations.

\begin{figure}
  \includegraphics[scale = 0.60]{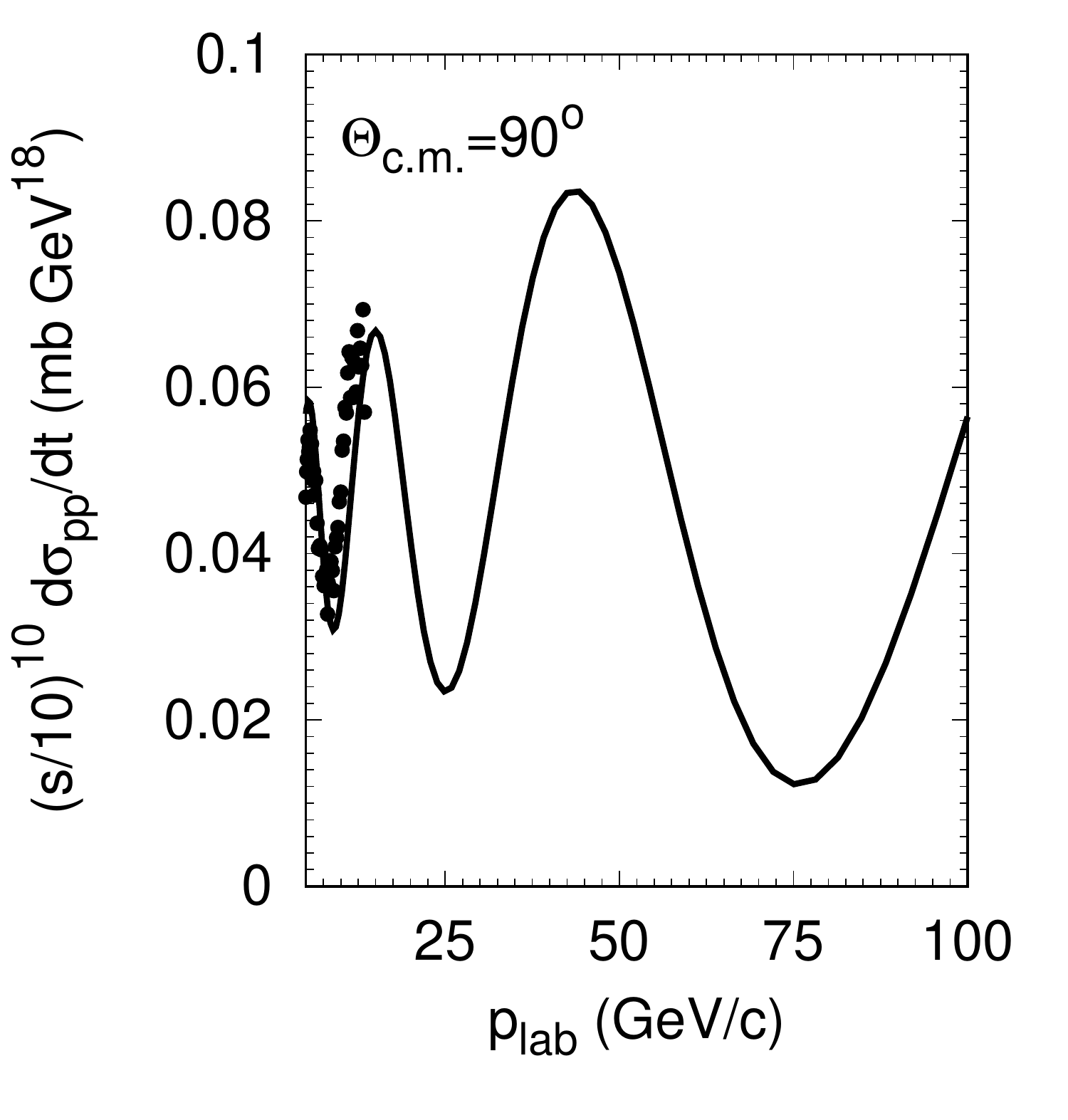}
  \caption{\label{fig:R} The scaled $pp$ elastic scattering cross section, Eq.(\ref{dsigma_pp/dt_par}),
    at $\Theta_{c.m.}=90\degree$ vs beam momentum. Experimental data are from Ref. \cite{Akerlof:1967zz}.}
\end{figure}

Fig.~\ref{fig:R} shows the $p_{\rm lab}$ dependence of $pp$ elastic cross section at $t=(4m^2-s)/2$,
i.e. at $\Theta_{c.m.}=90\degree$. To better see the deviations from the QC prediction, the cross section is multiplied
by a factor of $(s/10)^{10}$.
The parameterization (\ref{dsigma_pp^QC/dt_par}) predicts oscillations of the cross section with the beam momentum
and describes the available data \cite{Akerlof:1967zz} at $p_{\rm lab}=5-13$ GeV/c reasonably well.
In the maxima (minima) the ratio $R$ is pure real and positive (negative). 
The maxima are located at $p_{\rm lab} = 1.7, 5.4, 14.8,$ and 42.6 GeV/c
while minima -- at $p_{\rm lab} = 3.2, 8.9, 24.8,$ and 75.1 GeV/c. 

Relation to the hard $pp \to pp$ scattering amplitude is given by a standard formula (cf. Ref. \cite{BLP}):
\begin{equation}
  \frac{d\sigma_{pp}}{dt} = \frac{\overline{|M_{\rm hard}|^2}}{64\pi I_{pp}^2}~,    \label{dsigma_pp/dt}
\end{equation}
where overline means averaging over spin projections of the incoming particles and
summation over spin projections of the outgoing ones,
and $I_{pp} = [(p_1p_2)^2 - m^4]^{1/2} = [(s/2-2m^2)s/2]^{1/2}$ is the flux factor.
The relation (\ref{dsigma_pp/dt}) allows us to estimate the matrix element $M_{\rm hard}$:
\begin{equation}
  M_{\rm hard} = M_{\rm QC} + M_{\rm L} = \left(16\pi(s-4m^2)s \frac{d\sigma_{pp}^{\rm QC}}{dt}\right)^{1/2} [1+R(s)] \delta_{\lambda_1\lambda_3} \delta_{\lambda_2\lambda_4},     \label{M_hard}
\end{equation}
where, for brevity, we omit the unknown common phase factor and approximate the spin part by a diagonal matrix in the spin projections $\lambda_i, i=1,\ldots,4$.
The last assumption is rather rough, taking into account spin polarization studies in $\Theta_{c.m.}=90\degree$ elastic $pp$ scattering \cite{Crabb:1978km}, but still reasonable
for the unpolarized proton beam discussed in the present work.

The soft elastic $pn$ scattering amplitude can be parameterized in the standard high-energy form used in Glauber  theory
\cite{Glauber:1970jm}:
\begin{equation}
    M_{\rm el}(p_{\rm lab},k_t)= 2 p_{\rm lab} m \sigma_{p n}^{\rm tot}
     (i+\rho_{\rm p n}) \mbox{e}^{-B_{\rm p n}k_t^2/2}~,           \label{M_el}
\end{equation}
where $\sigma_{p n}^{\rm tot}$ is the total $pn$ cross section,
$\rho_{\rm p n}=\mbox{Re}M_{\rm el}(p_{\rm lab},0)/\mbox{Im}M_{\rm el}(p_{\rm lab},0)$
is the ratio of the real part of the forward $pn$ scattering amplitude to the imaginary one,
and $B_{\rm p n}$ is the slope of the dependence on the transverse momentum transfer.
Eq.(\ref{M_el}) takes into account only the central part of the elastic $pn$ amplitude neglecting smaller spin-orbit and spin-flip parts
(see Refs. \cite{Alkhazov:1978et,Platonova:2010wjt,Uzikov:2020zho} for a more general form of the elastic $NN$ amplitude).
Energy dependent $\sigma_{p n}^{\rm tot}$ and $\rho_{\rm p n}$ are taken as Regge-Gribov fits from PDG \cite{Patrignani:2016xqp}
which describe well the experimental data at $p_{\rm lab} > 2$ GeV/c.
At the beam momentum $p_{\rm lab} < 9$ GeV/c, the slope $B_{\rm p n}$ is described by the parameterization from Ref. \cite{Cugnon:1996kh}
while at higher beam momentum the PYTHIA parameterization for elastic  $pp$  scattering is applied (see Appendix A in Ref. \cite{Falter:2004uc}):
\begin{equation}
  B_{\rm p n} =  5.0 + 4s^{0.0808}~,             \label{B_pn}
\end{equation}
where $s$ is in GeV$^2$ and $B_{\rm p n}$ is in GeV$^{-2}$.

\subsection{Color transparency effects}

The hard scattering amplitude, Eq.(\ref{M_hard}), includes the QC and Landshoff contributions. The last one does
not lead to the formation of PLCs as it is described by disconnected graphs corresponding to the independent scattering
of constituents \cite{Landshoff:1974ew} (see also discussion in Ref. \cite{Carlson:1992jw}).
Therefore, the Landshoff mechanism alone should lead to conventional ISI and FSI of full-sized hadrons.
In contrast, the QC contribution is governed by the connected graphs where all propagators
are of the order of $1/s$ that gives an estimate of the relative distance squared between constituents.
This results in the formation of PLCs that experience reduced ISI and FSI due to CT.
For the first time, different descriptions of nuclear absorption for the QC and Landshoff contributions were introduced
in Ref. \cite{Ralston:1988rb}, which led to a successful description of the anomalous dependence of the transparency
of nuclei on the beam momentum in $A(p,pp)$ reactions on heavy targets.
This idea was later developed in Refs. \cite{Jennings:1991rw,Jennings:1993hw,VanOvermeire:2006tk}
in more complex models for $A(p,pp)$ reactions.

Following Ref. \cite{VanOvermeire:2006tk}, we split each of the amplitudes with rescattering, i.e. (b),(c),(d),(g),(h) in Fig.~\ref{fig:diagr},
into two terms:
The first term is proportional to $M_{\rm QC}$, so the $pn$ rescattering amplitude in it should be affected by CT.
The second one is proportional to $M_{\rm L}$ and the $pn$ amplitude in it remains unchanged.
Thus, unless specifically mentioned, in the present calculations CT effects are included {\it only in the QC part}
of the rescattering amplitudes.

The effects of CT are implemented within the QDM of Ref. \cite{Farrar:1988me}.
This model accounts for the decrease in the transverse size of quark configurations at the hard interaction point
and the gradual increase with distance from this point.
Thus, the total $pn$ cross section in the expression for the
elastic $pn$ amplitude (\ref{M_el}) is replaced by the position-dependent effective cross section:
\begin{equation}
   \sigma_{pn}^{\rm eff}(l)
  = \sigma_{pn}^{\rm tot}\left(\left[ \frac{l}{l_c}
    + \frac{\langle n^2 k_t^2\rangle}{Q^2} \left(1-\frac{l}{l_c}\right) \right]
    \Theta(l_c-l) +\Theta(l-l_c)\right)~,                 \label{sigma_pn_eff}
\end{equation}
where $l = |(\bvec{r}_2-\bvec{r}_s) \cdot \bvec{p}|/p$ is the distance between a proton and a neutron in a deuteron
along the momentum of a fast proton
\footnote{For the amplitude (d) with rescattering of the incoming proton this is equivalent to $l=|z|=|z_2-z_s|$.
According to Ref. \cite{Frankfurt:1996uz}, we set $l=|z|$ also for the amplitudes (g),(h) with double rescattering.},
\begin{equation}
  l_c = \frac{2p}{\Delta M^2}~,   \label{l_c}
\end{equation}
is the coherence length with $\Delta M^2 \simeq 1$ GeV$^2$ being the standard setting
as determined from nuclear transparency
studies for $A(e,e^\prime \pi)$ reactions \cite{Larson:2006ge} (see, however, discussion in this section below),
$Q^2=\min(-t,-u)$ is the hard scale, $\sqrt{\langle k_t^2\rangle}=0.35$ GeV/c is the average transverse momentum of a quark
in a proton, $n=3$ is the number of valence quarks in the proton.
Eq.(\ref{sigma_pn_eff}) describes the reduction of the interaction cross section of the fast proton
-- that propagates from/to the hard interaction point -- with the spectator neutron within the interval
$l < l_c$, while for $l \geq l_c$ the proton and neutron interact with the usual total cross section.

According to the recent JLab data on the nuclear transparency in the $^{12}$C$(e,e^\prime p)$ reaction \cite{HallC:2020ijh}
the proton coherence length might be much shorter as compared to the formula (\ref{l_c}) with $\Delta M^2 = 1$ GeV$^2$.
The analysis of Ref.~\cite{Li:2022uvf} within the relativistic multiple scattering Glauber approximation (RMSGA) 
demonstrates that the data \cite{HallC:2020ijh} is compatible with $\Delta M^2 = 2-3$ GeV$^2$.
On the other hand, the description of the AGS data on the nuclear transparency in $A(p,2p)$ reactions by the RMSGA calculations
\cite{VanOvermeire:2006tk} is reached for $\Delta M^2 = 0.7-1.1$ GeV$^2$
with $\Delta M^2 = 0.7$ GeV$^2$ being a preferred value. Thus, the range of acceptable values
of $\Delta M^2$ remains currently quite broad.
In order to show the sensitivity of our results to the choice of coherence length, the calculations are performed
with two values of $\Delta M^2=0.7$ and 3 GeV$^2$ which estimate the boundaries of the uncertainty range
of the proton coherence length.

CT affects not only the total interaction cross section but also the slope of the momentum-transfer dependence
of the $pn$ scattering amplitude.
According to Ref.\cite{Frankfurt:1994kt}, this can be taken into account by rescaling the proton form factor.
To this end, the amplitude (\ref{M_el}) is multiplied by the ratio
$G(t \cdot \frac{\sigma_{pn}^{\rm eff}(l)}{\sigma_{pn}^{\rm tot}})/G(t)$ where
\begin{equation}
  G(t)=\frac{1}{(1-t/0.71~\mbox{GeV}^2)^2}~,     \label{SachsFF}
\end{equation}
is the Sachs electric form factor of a proton, $t=-k_t^2$.

\section{Observables}
\label{observ}

We will consider the non-polarized as well as the polarized along the beam axis deuteron.
In both cases the modulus squared of the reaction amplitude is rotation-invariant about the beam axis.
Therefore, the reaction cross section
is fully determined by four independent variables which can be chosen according to Ref. \cite{Frankfurt:1996uz} as follows:
\begin{equation}
  \alpha_s = \frac{2(E_s-p_s^z)}{m_d}                 \label{alpha_s}
\end{equation}
-- the light cone variable defined such that, in the infinite momentum frame (IMF) where the deuteron moves fast backward,
$\alpha_s/2$ is the fraction of the deuteron momentum carried by the spectator neutron;
$t$ - the squared four-momentum transfer in the hard scattering, Eq.(\ref{stu});
$\phi=\phi_3-\phi_s$ -- the azimuthal angle between transverse momenta of the scattered proton
\footnote{Here and below, "scattered proton" can be understood as an outgoing proton for which $t$ is measured.} 
and the neutron;
$p_{st}$ -- the transverse momentum of the neutron.
It is possible to show after a somewhat lengthy but straightforward derivation
(see Appendix~\ref{d^4sigma}) that the four-differential cross section has the following form:
\begin{equation}
  \alpha_s \frac{d^4\sigma}{d\alpha_s\, dt\, d\phi\, p_{s t} dp_{s t}}
  = \frac{\overline{|M|^2} \, p_{3t}}%
         {16(2\pi)^4 p_{\rm lab} m_d \kappa_t \kappa_t^\prime}~,                   \label{dsig/dalpha}
\end{equation}
where the matrix element is given by the sum of the partial matrix elements, i.e
\begin{equation}
  M = M^{(a)} + M^{(b)} + M^{(c)} + M^{(d)} + M^{(g)} + M^{(h)}~.    \label{M}
\end{equation}
The overline means averaging over spin projections of the incoming proton and of the deuteron (unless it is fixed)
and the sum over spin projections of outgoing particles.
\begin{equation}
  \kappa_t = 2\left|\frac{2p_{3t}}{\beta} + p_{s t} \cos\phi \right|    \label{kappa_t}
\end{equation}
and 
\begin{equation}
  \kappa_t^\prime = \frac{2(E_3+p_3^z)(p_{\rm lab}E_3 - \lambda E_1 - p_s^z E_1)}%
        {\lambda + E_3+p_3^z}         \label{kappa_t^prime}
\end{equation}
are the phase space factors.
\begin{equation}
  \beta    = \frac{2(E_3+p_3^z)}{E_1+m_d-E_s+p_{\rm lab}-p_s^z}            \label{beta}
\end{equation}
is the light cone variable defined such that $\beta/2$ is the fraction of the momentum of the two colliding protons
carried by the scattered proton in the IMF where the beam proton moves fast forward,
and
\begin{equation}
  \lambda = \frac{E_3 p_4^z - E_4 p_3^z}{E_3 + E_4 + \frac{E_3 p_{s t} \cos\phi}{p_{3 t}}}~.   \label{lambda}
\end{equation}

The effects of ISI and FSI are well visible in the nuclear transparency $T$.
The definition of $T$, Eq.(\ref{T}), uses the differential cross section in the IA that is expressed in the
present kinematics as follows:
\begin{equation}
  \alpha_s \frac{d^4\sigma_{\rm IA}}{d\alpha_s\, dt\, d\phi\, p_{s t} dp_{s t}}
  = \frac{\overline{|M^{(a)}|^2} \, p_{3t}}%
  {16(2\pi)^4 p_{\rm lab} m_d \kappa_t \kappa_t^\prime}                   \label{dsig/dalpha_IA}
\end{equation}
with
\begin{equation}
  \overline{|M^{(a)}|^2} = 4m (2\pi)^3 \, \overline{|M_{\rm hard}|^2} \, |\phi^{\lambda_d}(-\bvec{p}_s)|^2    \label{|M^(a)|^2}
\end{equation}
being the modulus squared of the matrix element in the IA summed over spin projections of outgoing particles and averaged over spin projection
of the incoming proton. The averaging of Eq.(\ref{|M^(a)|^2}) over the deuteron spin projection can be easily done by using the relation
\begin{equation}
  \frac{1}{3} \sum_{\lambda_d=0,\pm1}|\phi^{\lambda_d}(\bvec{p})|^2 = \frac{u^2(p) +w^2(p)}{4\pi}~.        \label{DWF^2}
\end{equation}

Tensor analyzing power, also known as spin asymmetry, is given by the following expression
(see Refs. \cite{Frankfurt:1994kt,Platonova:2010wjt,Arbuzov:2020cqg,Uzikov:2020zho}):
\begin{equation}
   A_{zz} = \frac{\sigma(+1)+\sigma(-1)-2\sigma(0)}{\sigma(+1)+\sigma(-1)+\sigma(0)}~,      \label{A_zz}
\end{equation}
where $\sigma(\lambda_d)$ is the differential cross section (\ref{dsig/dalpha}) for fixed $\lambda_d$. If the cross section is calculated
in the IA then, for a spin-independent hard amplitude, the tensor analyzing power (\ref{A_zz}) is expressed as follows:
\begin{eqnarray}
  A_{zz}^{IA} &=& \frac{|\phi^{+1}(-\bvec{p}_s))|^2 + |\phi^{-1}(-\bvec{p}_s))|^2 - 2|\phi^{0}(-\bvec{p}_s))|^2}%
  {|\phi^{+1}(-\bvec{p}_s))|^2 + |\phi^{-1}(-\bvec{p}_s))|^2 + |\phi^{0}(-\bvec{p}_s))|^2}   \nonumber \\
            &=& \frac{(3(p_s^z/p_s)^2 -1)(\sqrt{2}u(p_s)w(p_s) - w^2(p_s)/2)}{u^2(p_s) +w^2(p_s)}~.       \label{A_zz^IA}
\end{eqnarray}
This formula shows that the tensor analyzing power probes the D-state component of the DWF.

\section{Numerical results}
\label{results}

\begin{figure}
  \includegraphics[scale = 0.50]{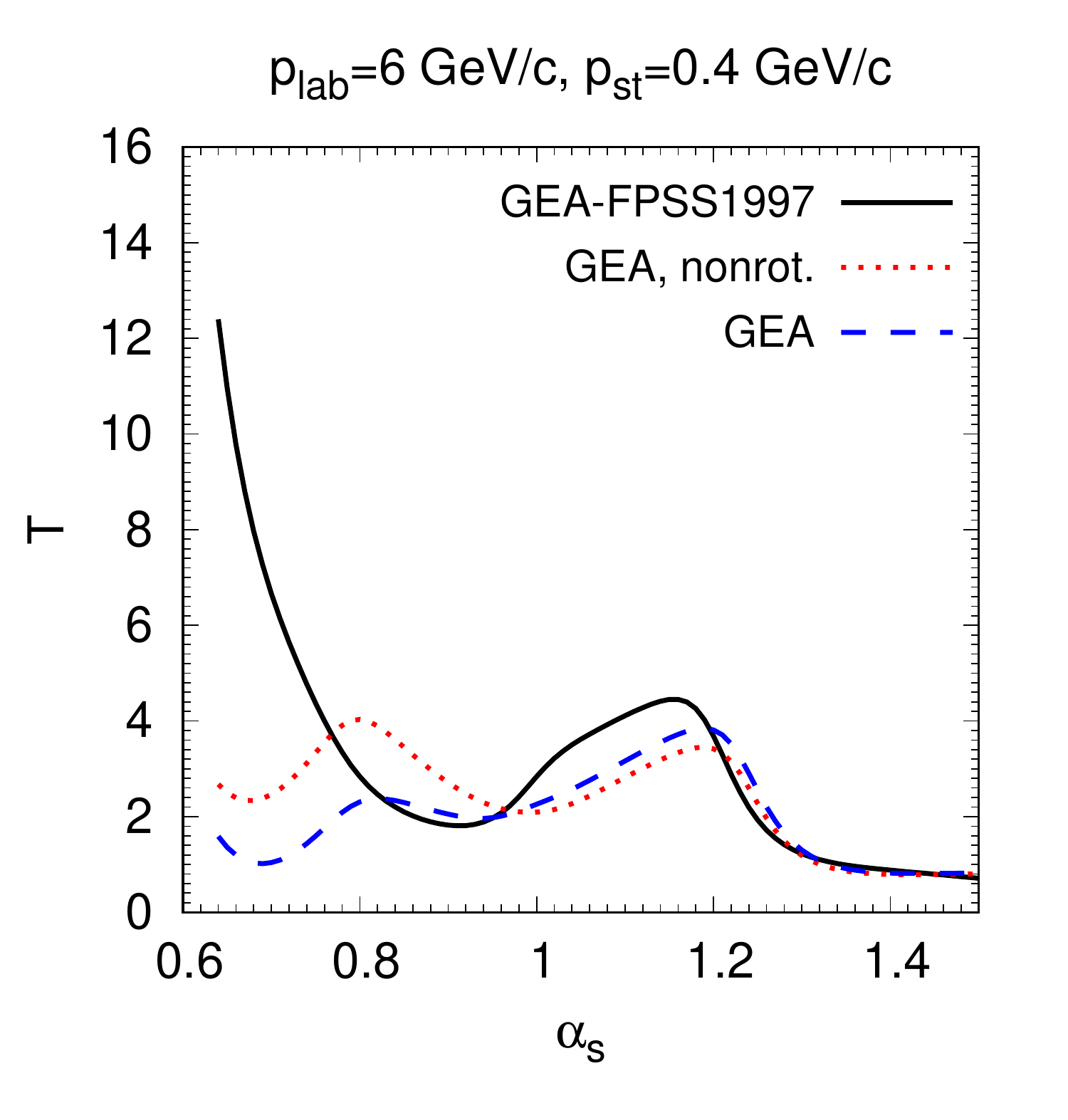}
  \caption{\label{fig:T_6gevc_alphasDep_pst0.4} Transparency, Eq.(\ref{T}), for $p d \to p p n$ at the beam momentum of 6 GeV/c
    as a function of $\alpha_s$, Eq.(\ref{alpha_s}), for the transverse momentum of the spectator neutron
    of 0.4 GeV/c. Chosen is the kinematics with $\phi=180\degree$ and $\Theta_{c.m.}=90\degree$.
    Black solid line -- GEA calculation with $\Delta_1,~\Delta_3$ and $\Delta_4$ from Ref. \cite{Frankfurt:1996uz}.
    Blue dashed line -- default GEA calculation of the present work. 
    Red dotted line -- same as blue dashed line but with single-rescattering amplitudes calculated
    in the original (non-rotated) frame with $z$-axis along the beam momentum.}
\end{figure}

Before to discuss predictions at higher energies, it is instructive to reproduce some of the results
of Ref. \cite{Frankfurt:1996uz}. Fig.~\ref{fig:T_6gevc_alphasDep_pst0.4} shows $T$ as a function
of $\alpha_s$. The black solid line is calculated by using the longitudinal momentum transfers (i.e. $\Delta$'s)
from Ref. \cite{Frankfurt:1996uz} and quite closely reproduces Fig.~12c from that paper, some deviations
are most probably due to slightly different parameters of the $pn$ scattering amplitude in Ref. \cite{Frankfurt:1996uz}.
The difference with our version of the GEA at small $\alpha_s$ is very large. This is explained by the fact that
the outgoing proton under the same azimuthal angle with spectator has a small longitudinal and transverse
momentum (e.g., $p_4^z=1.98$ GeV/c, $p_{4t}=1.56$ GeV/c for $\alpha_s=0.65$).
In these conditions the term $\propto (p_s-p_s^\prime)^2$ in Eq.(\ref{Delta_4})
should be kept, while it was neglected in Ref. \cite{Frankfurt:1996uz}.
However, close to $\alpha_s=1$ all calculations agree with each other. 

Thus, below we will consider the case of transverse kinematics, $\alpha_s=1$, and moderate transverse momenta of the spectator neutron,
$p_{st} < 0.5$ GeV/c that restricts its longitudinal momentum within the range from 0 to 0.13 GeV/c.
\footnote{The longitudinal momentum $p_s^z$ grows quadratically with $p_{st}$
according to the relation $p_s^z = [m^2+p_{st}^2-(\alpha_s m_d/2)^2]/\alpha_s m_d$.}
Moreover, under these conditions, the nonrelativistic description of the deuteron and neglect of the nonnucleon components of the DWF
are reliable approximations \cite{Frankfurt:1996uz,Larionov:2018lpk}.
Calculations were done in the deuteron rest frame for the proton beam momenta $p_{\rm lab}=6, 15, 30, 50, 65$ and 75 GeV/c
which corresponds to the invariant energies $\sqrt{s_{NN}}=3.63, 5.47, 7.62, 9.78, 11.12$ and 11.94 GeV.
Two values of the Mandelstam $t$ in the hard $pp$ scattering amplitude were chosen: $t=(4m^2-s)/2$
and (only for $p_{\rm lab}=30$ GeV/c) $t=0.4(4m^2-s)/2$, which corresponds to $\Theta_{c.m.}=90\degree$ and $\Theta_{c.m.}=53\degree$.
Unless specifically mentioned, $\Theta_{c.m.}=90\degree$ is chosen in the results described below.

\subsection{Non-polarized deuteron}
\label{nonpol}

\begin{figure}
  \includegraphics[scale = 0.55]{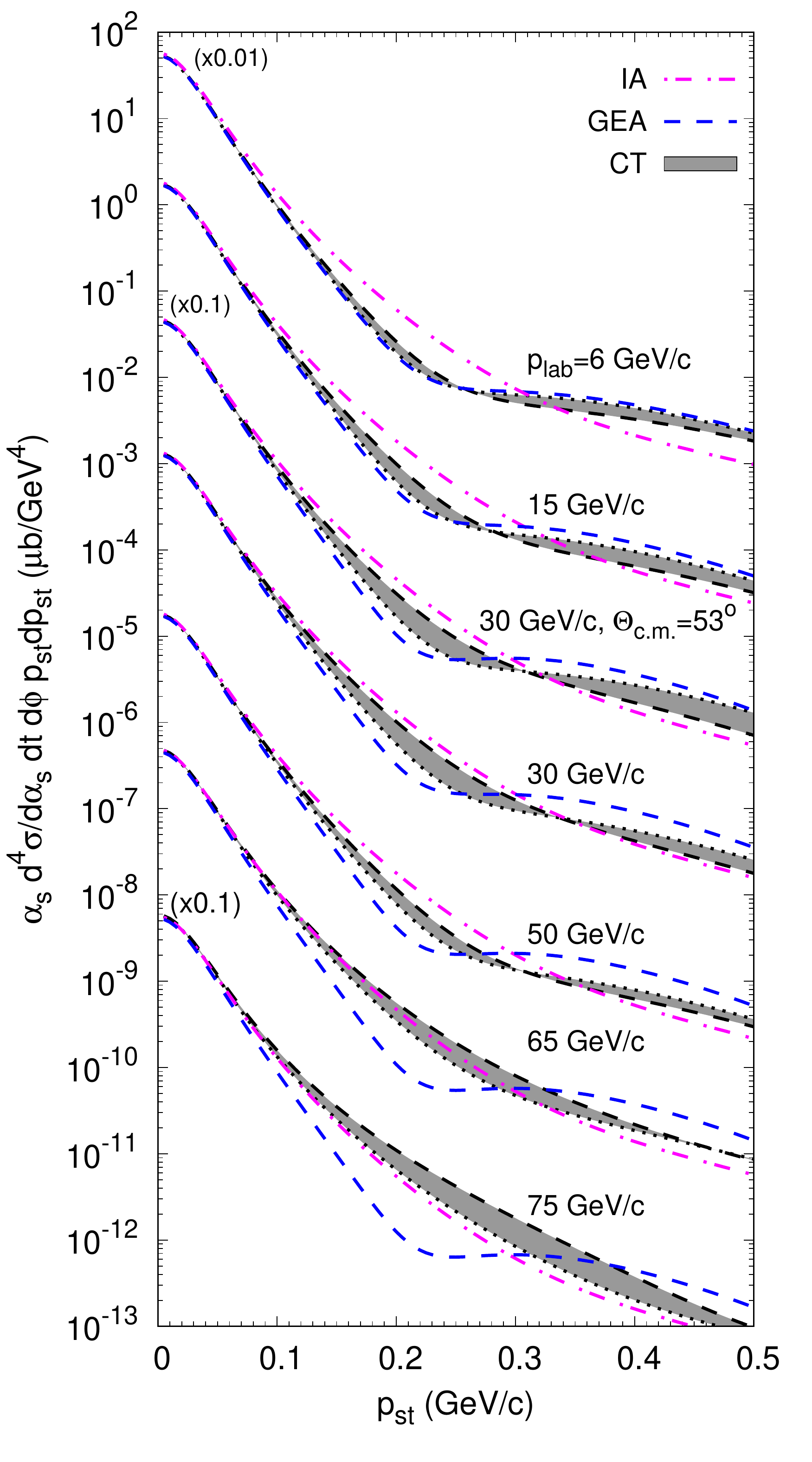}
  \caption{\label{fig:sig_180deg} Differential cross section of $p d \to p p n$ as a function of the transverse
    momentum of spectator neutron for $\alpha_s=1$ and $\phi=180\degree$.
    The magenta dash-dotted
    and blue dashed lines are calculated using the IA and full GEA, respectively.
    The grey band with borders given by the mass denominator of the coherence length,
    $\Delta M^2 = 0.7$ GeV$^2$ (black dashed line) and 3 GeV$^2$ (black dotted line), shows the calculation with CT effects.
    Lines from the uppermost to the lowermost correspond to $p_{\rm lab}=6$ GeV/c, 15 GeV/c, 30 GeV/c with $\Theta_{c.m.}=53\degree$,
    30 GeV/c, 50 GeV/c, 65 GeV/c, and 75 GeV/c.
    Plotted is the cross section times factor shown in parentheses.}
\end{figure}

The dependence of the differential cross section on the transverse momentum of the spectator neutron is shown in Fig.~\ref{fig:sig_180deg}.
For a small transverse momentum $p_{st} \ltsim 0.1$ GeV/c, the rescattering effects are negligibly small, and the cross section
is well described in the IA.
As $p_{st}$ increases up to approximately 0.2 GeV/c, ISI and FSI become more and more absorptive due to the destructive
interference of the IA amplitude and the amplitudes with single rescattering.
At larger transverse momenta, squared amplitudes with rescattering begin to dominate, which leads to the excess of the GEA cross sections
over the IA cross sections.
The overall effect of CT is that the cross section gets closer and closer to the IA limit as the beam energy increases
due to the increase in coherence length. However, the interference of the QC and Landshoff amplitudes distorts this simple behavior
at $p_{\rm lab} > 30$ GeV/c and even leads to an anomaly at $p_{\rm lab} = 75$ GeV/c. We will return to this point a little later.

\begin{figure}
  \begin{center}
  \begin{tabular}{cc}
  \includegraphics[scale = 0.40]{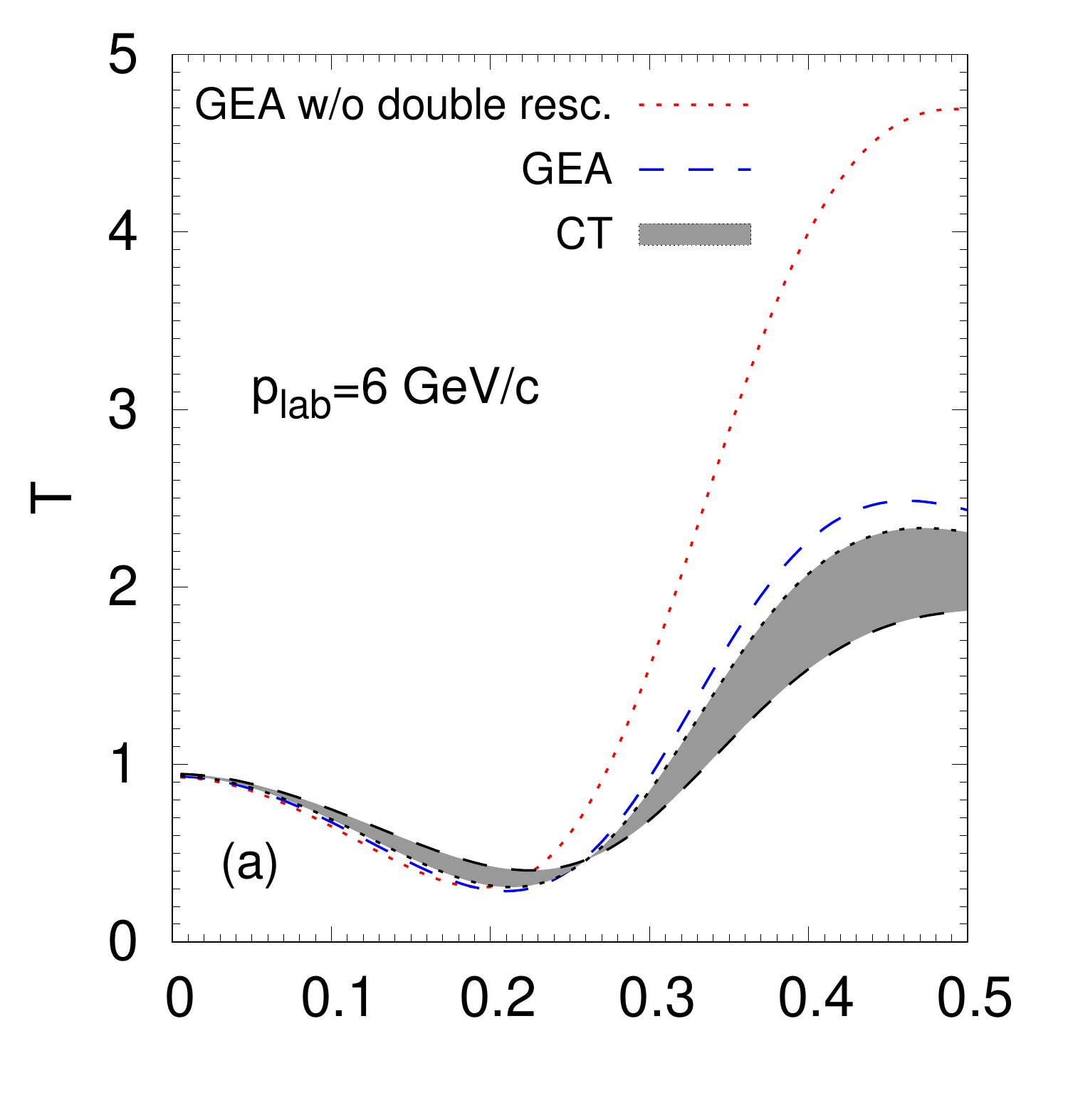} &
  \includegraphics[scale = 0.40]{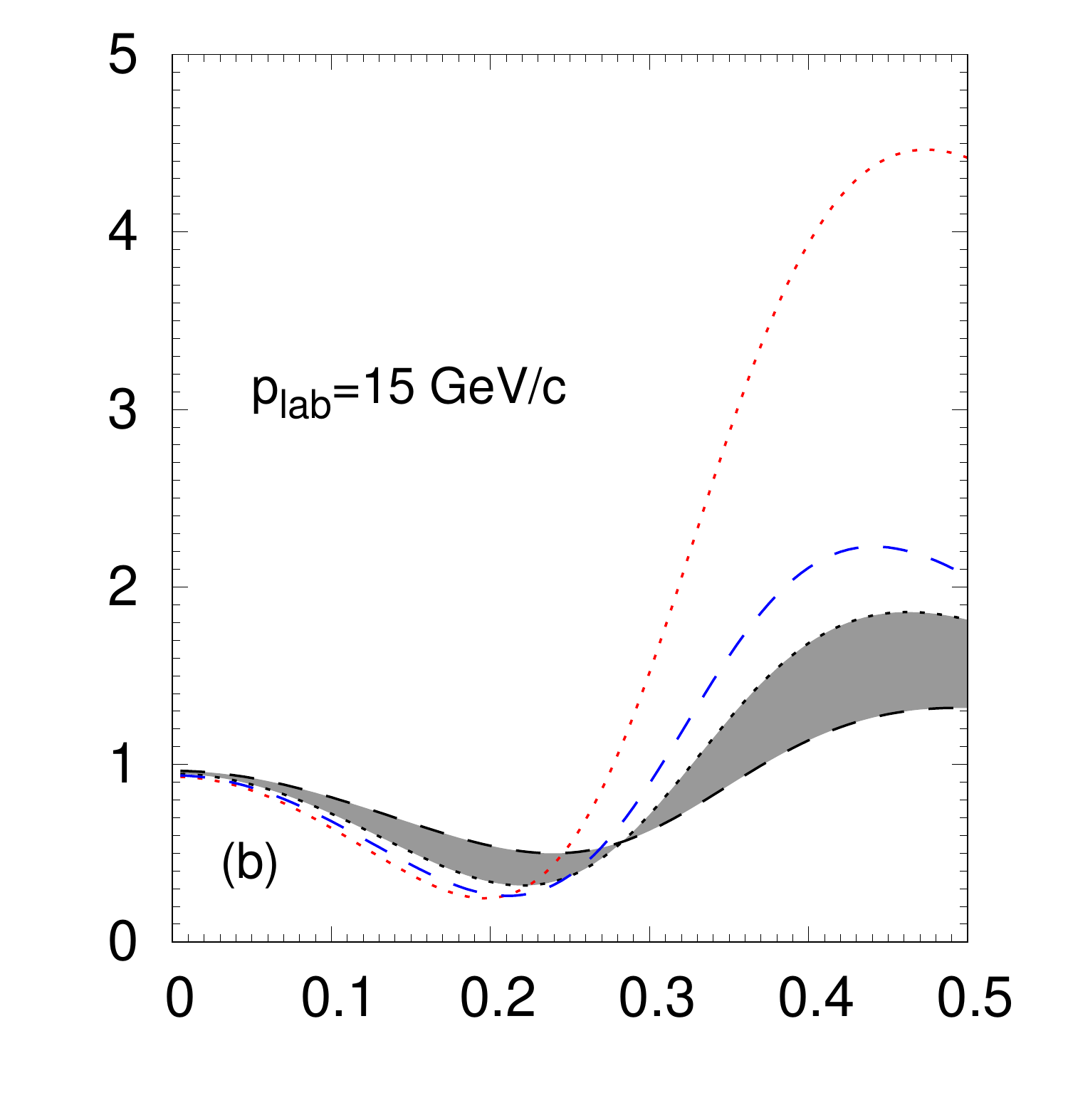} \\ 
  \includegraphics[scale = 0.40]{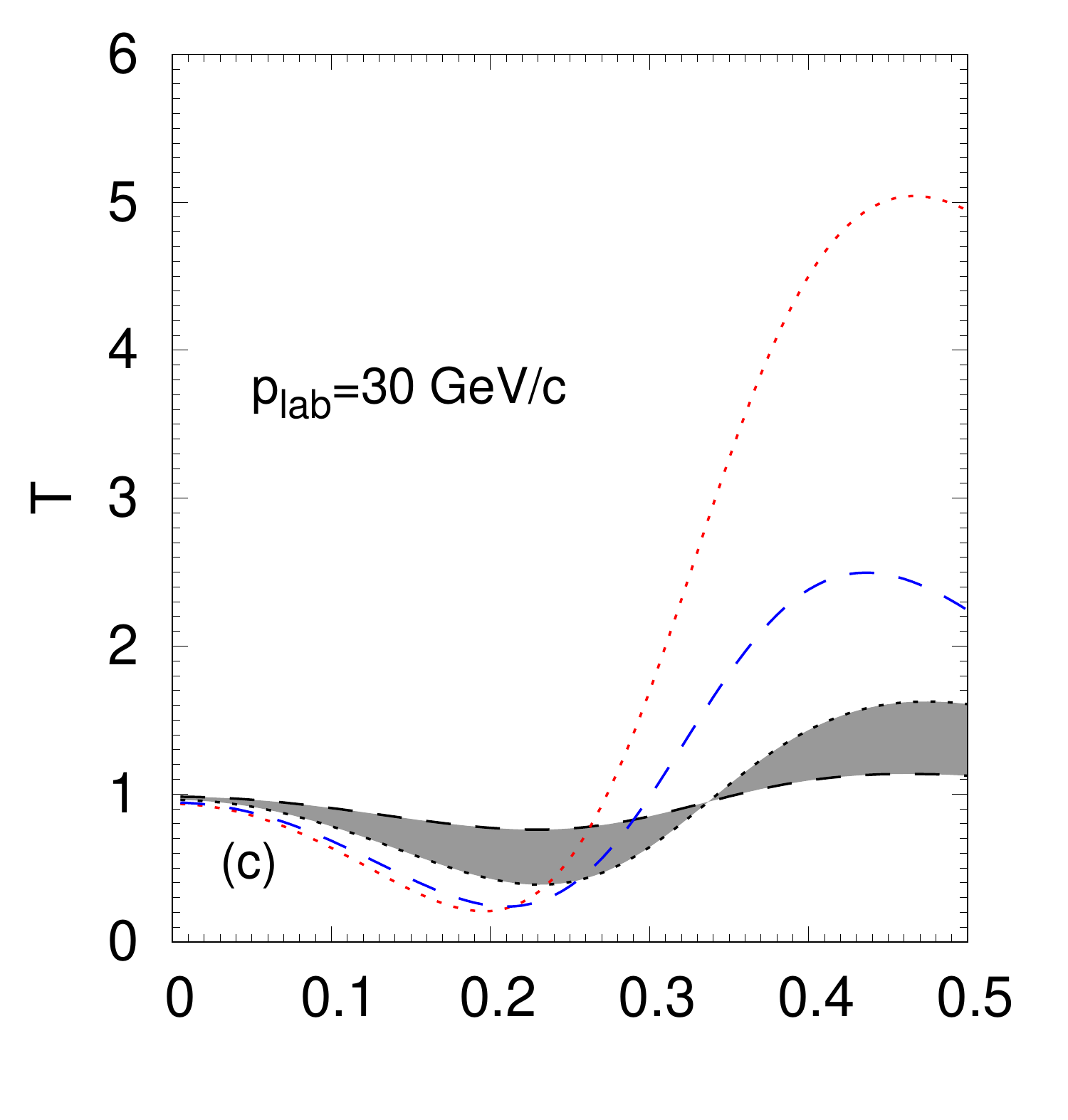} &
  \includegraphics[scale = 0.40]{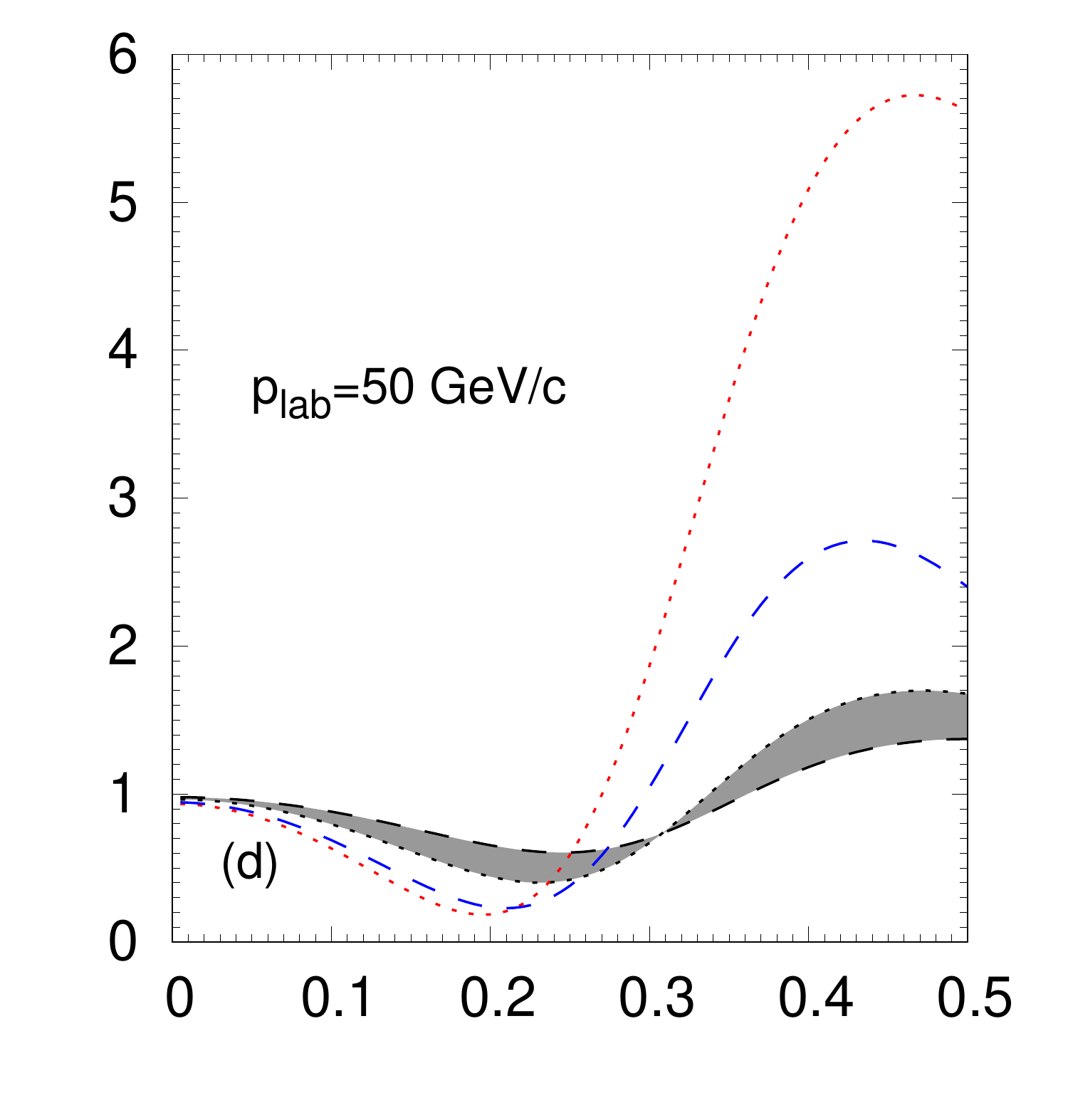} \\
  \includegraphics[scale = 0.40]{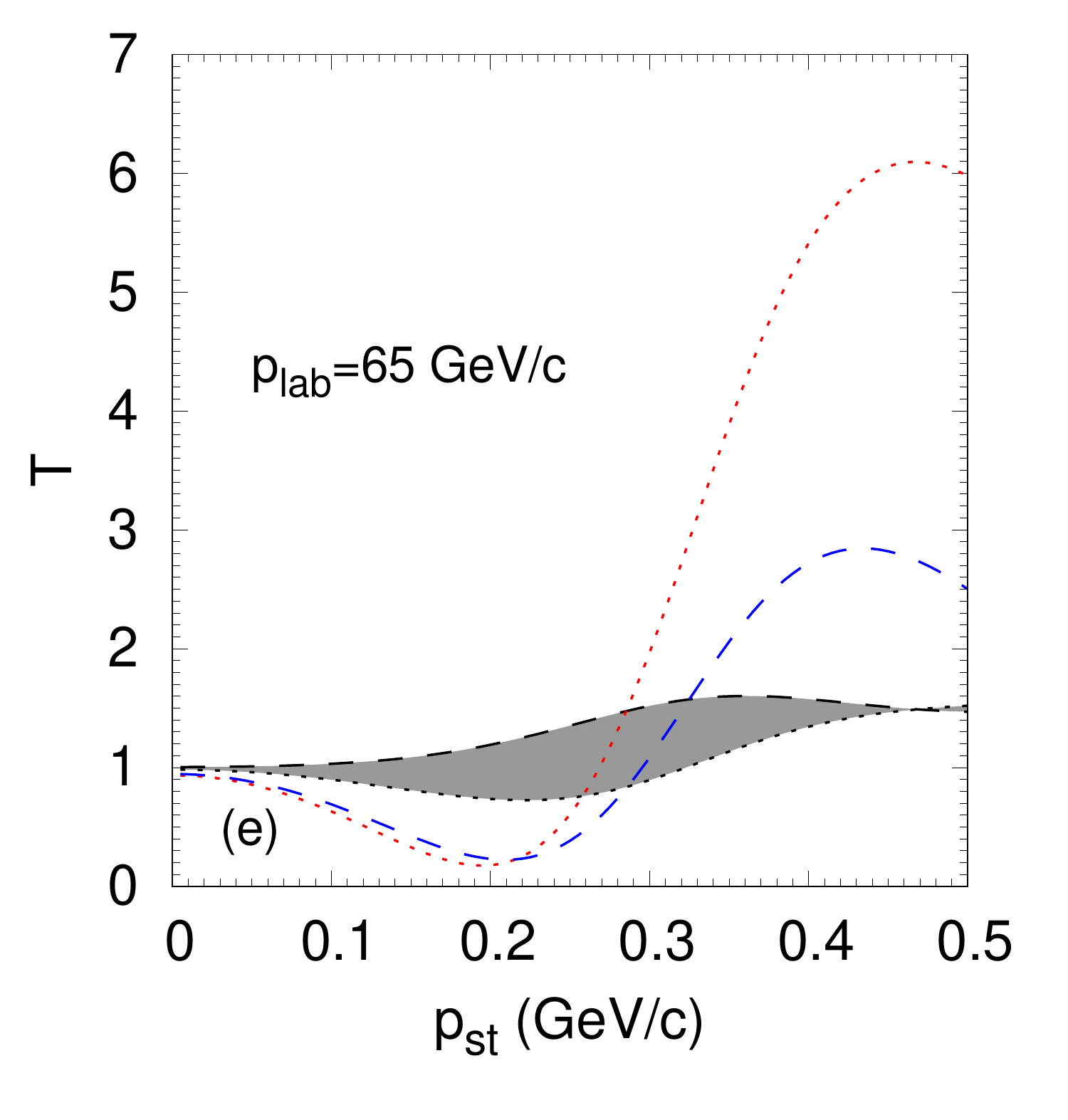} &
  \includegraphics[scale = 0.40]{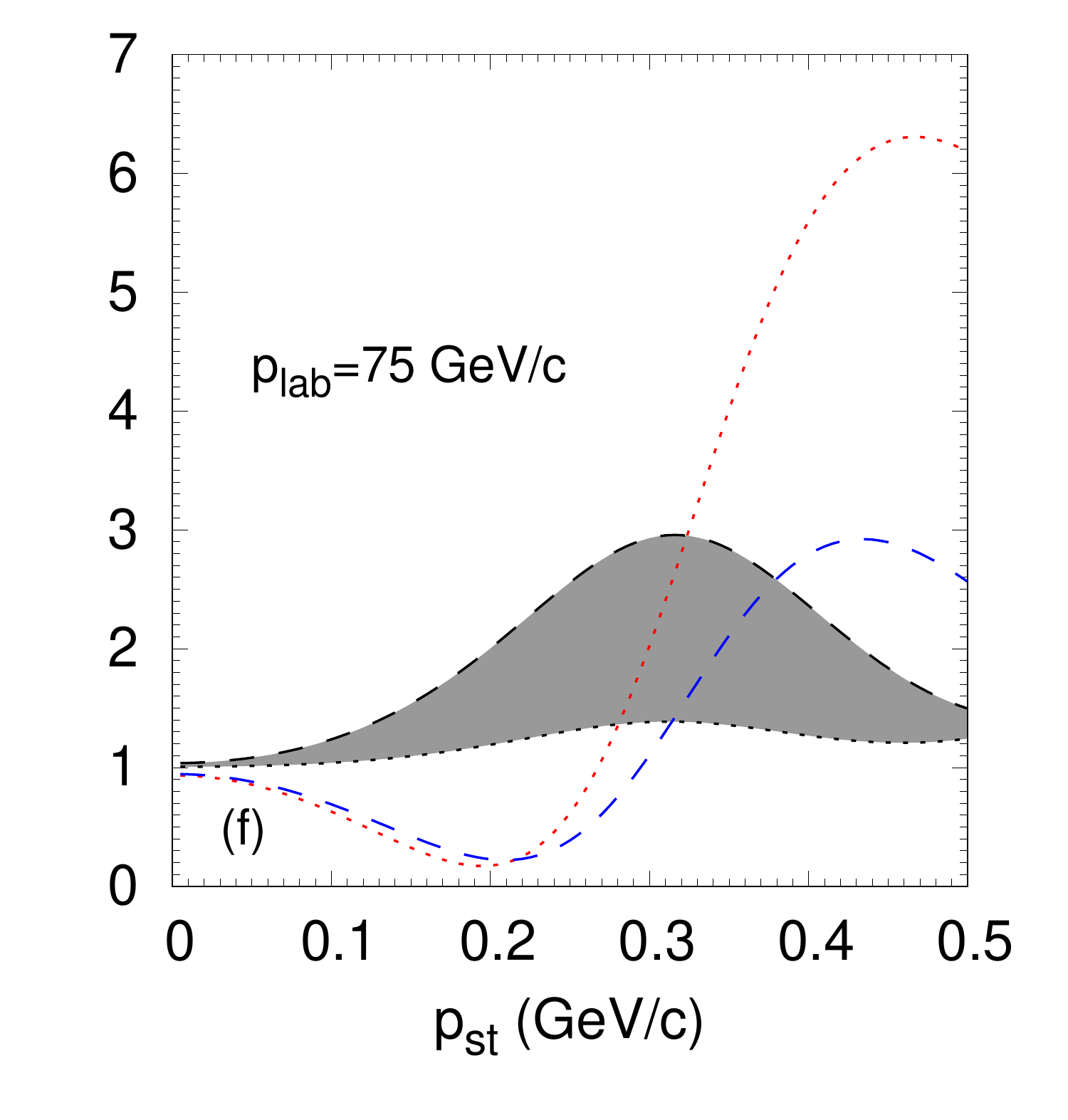} \\
  \end{tabular}
  \end{center}
  \caption{\label{fig:T_180deg} Transparency as a function
    of the spectator transverse momentum. Red dotted line -- GEA calculation including
    only the IA and single rescattering amplitudes. Blue dashed line -- full GEA calculation.
    The grey band with borders given by the mass denominator of the coherence length,
    $\Delta M^2 = 0.7$ GeV$^2$ (black dashed line) and 3 GeV$^2$ (black dotted line)
    -- the calculation with CT effects.
    Different panels display results for different beam momenta as indicated.}
\end{figure}

The effects of ISI and FSI are best seen on the transparency shown in Fig.~\ref{fig:T_180deg}.
Double rescattering amplitudes become important at $p_{st} \gtsim 0.3$ GeV/c due to their destructive interference
with single rescattering amplitudes, which is consistent with previous studies \cite{Frankfurt:1996uz}.
\footnote{In the calculations of Ref. \cite{Frankfurt:1996uz}, $T$ experiences a slight kink
at $p_{st} \simeq 0.4$ GeV/c due to the double rescattering amplitudes.
Most likely, this is due to slightly different parameters of the soft elastic $pn$ scattering amplitude.
}
Indeed, without CT, the hard scattering amplitude $M_{\rm hard}$ (for definiteness, we assume it to be positive)
factorizes out of reaction amplitude.
Neglecting a small real part of soft elastic $pn$ scattering amplitude (\ref{M_el}), expressions 
(\ref{M^(a)_fin}),(\ref{M^(b)_coord_fin}),(\ref{M^(d)_coord_fin}) and (\ref{M^(g)_3d_coord_fin})
allow us to conclude that $M^{(a)}, M^{(g)}, M^{(h)} > 0$ while $M^{(b)}, M^{(c)}, M^{(d)} < 0$.
Therefore, there is a destructive interference between the IA and single rescattering amplitudes
as well as between the double- and single rescattering amplitudes.

The fact that CT modifies only parts of the rescattering amplitudes proportional to the QC amplitude
complicates the $p_{\rm lab}$ dependence of the transparency. For the beam momenta of 6, 15, and 30 GeV/c,
CT influences $T$ in a usual way, i.e. bringing it closer to unity with increasing $p_{\rm lab}$.
However, between $p_{\rm lab}=30$ and 50 GeV/c, $T$ seems to saturate and even slightly moves
away from unity closer to the GEA result.

\begin{figure}
  \includegraphics[scale = 0.50]{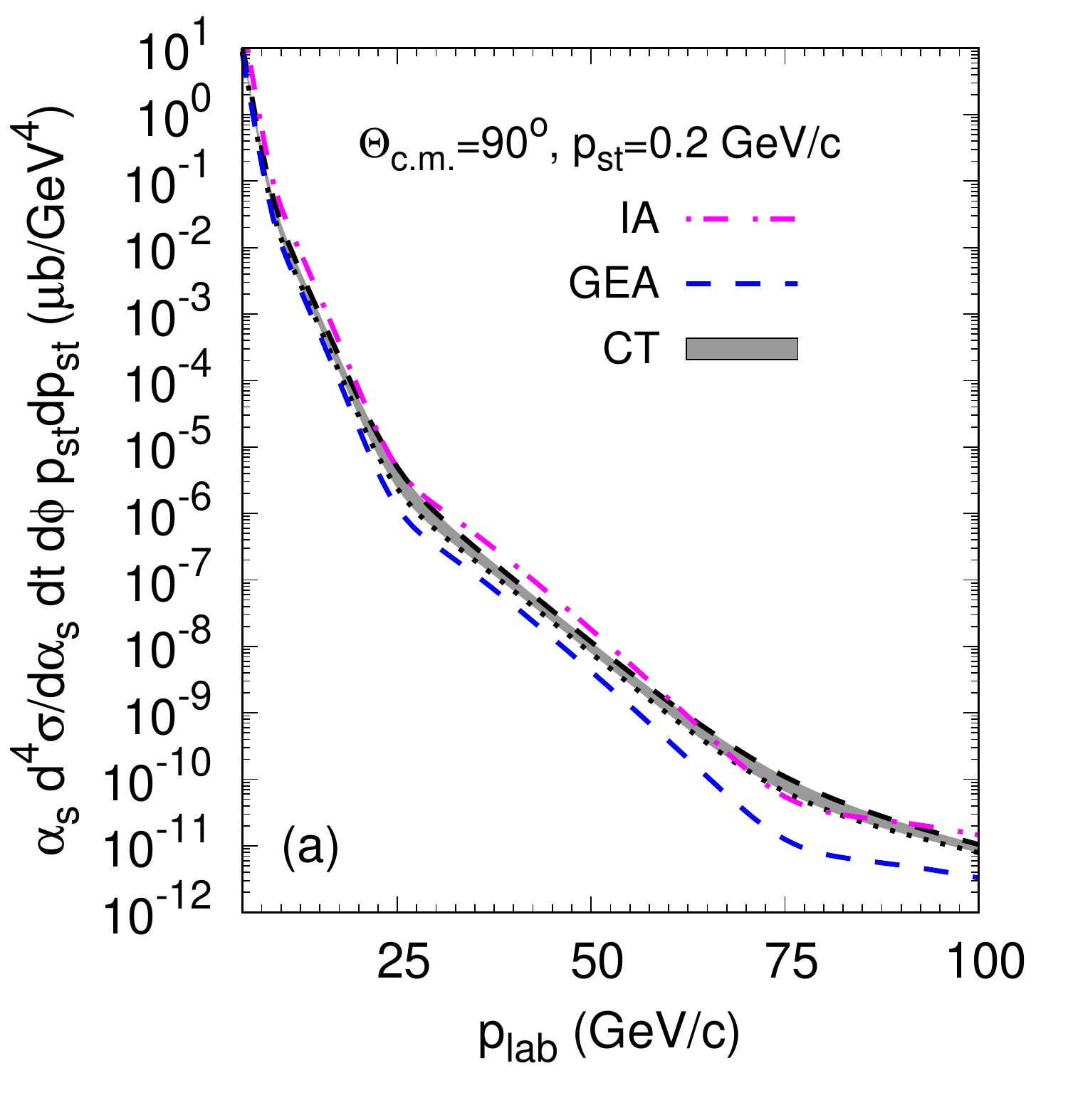}
  \includegraphics[scale = 0.50]{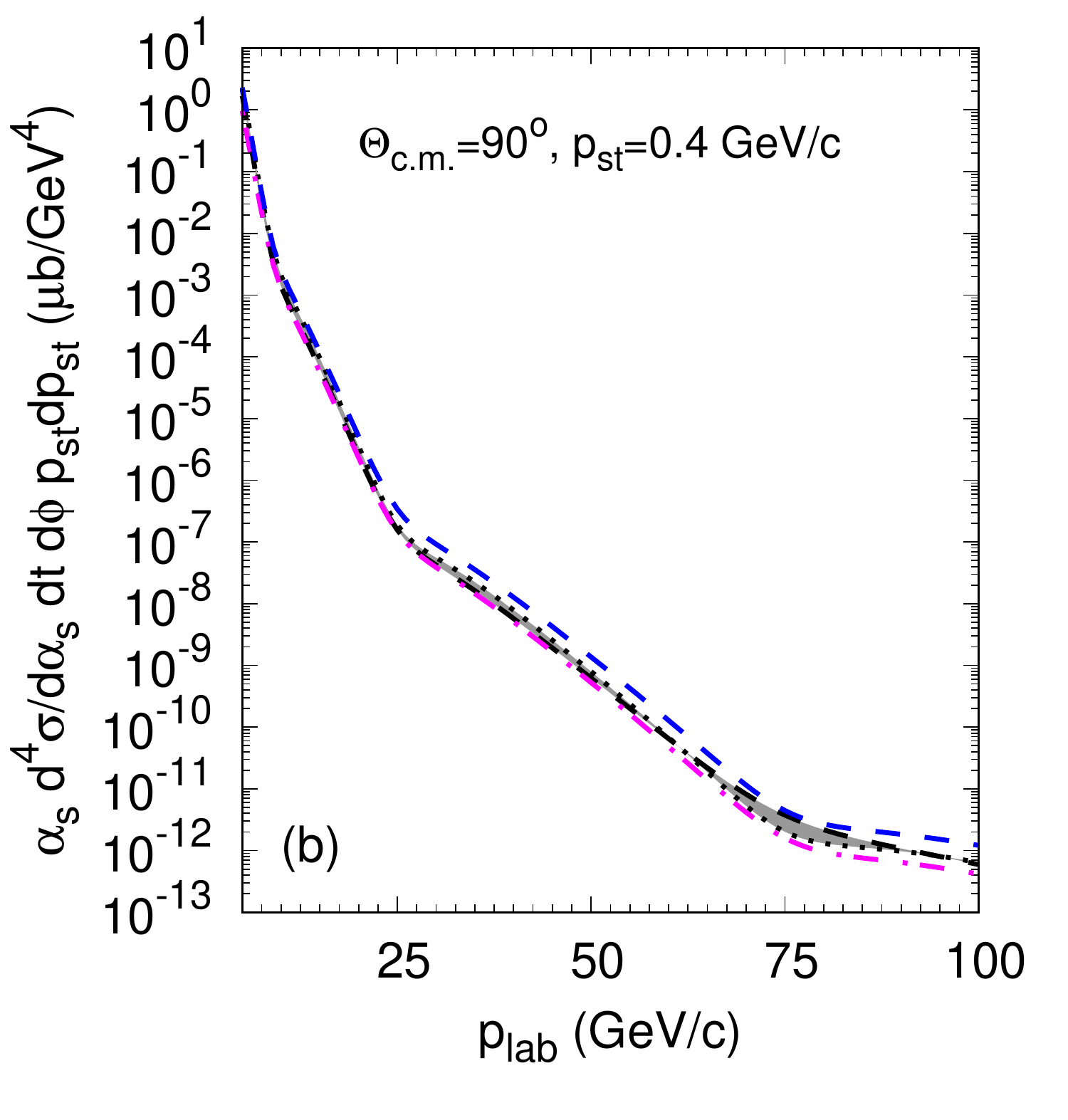}  
  \includegraphics[scale = 0.50]{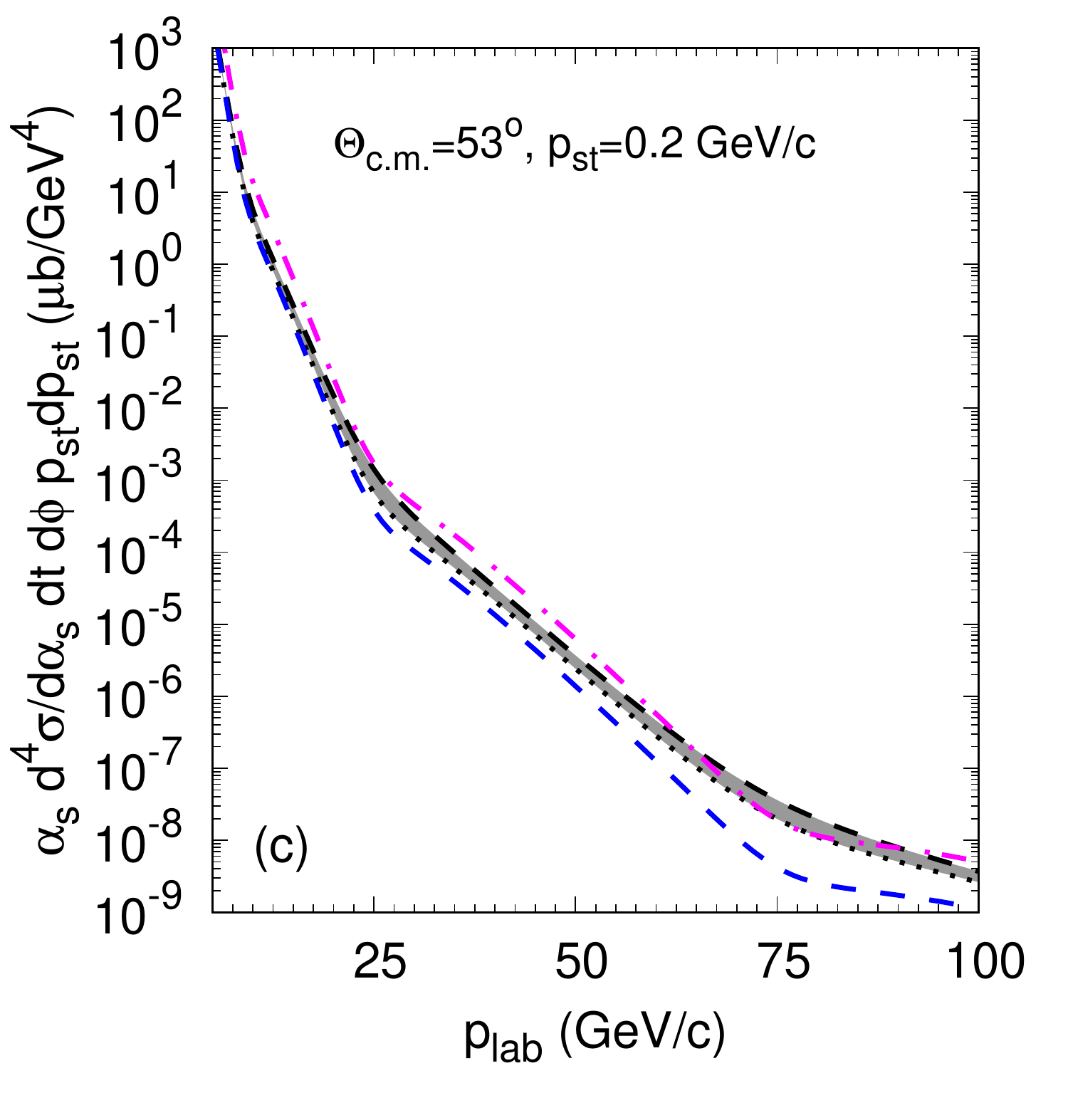}
  \includegraphics[scale = 0.50]{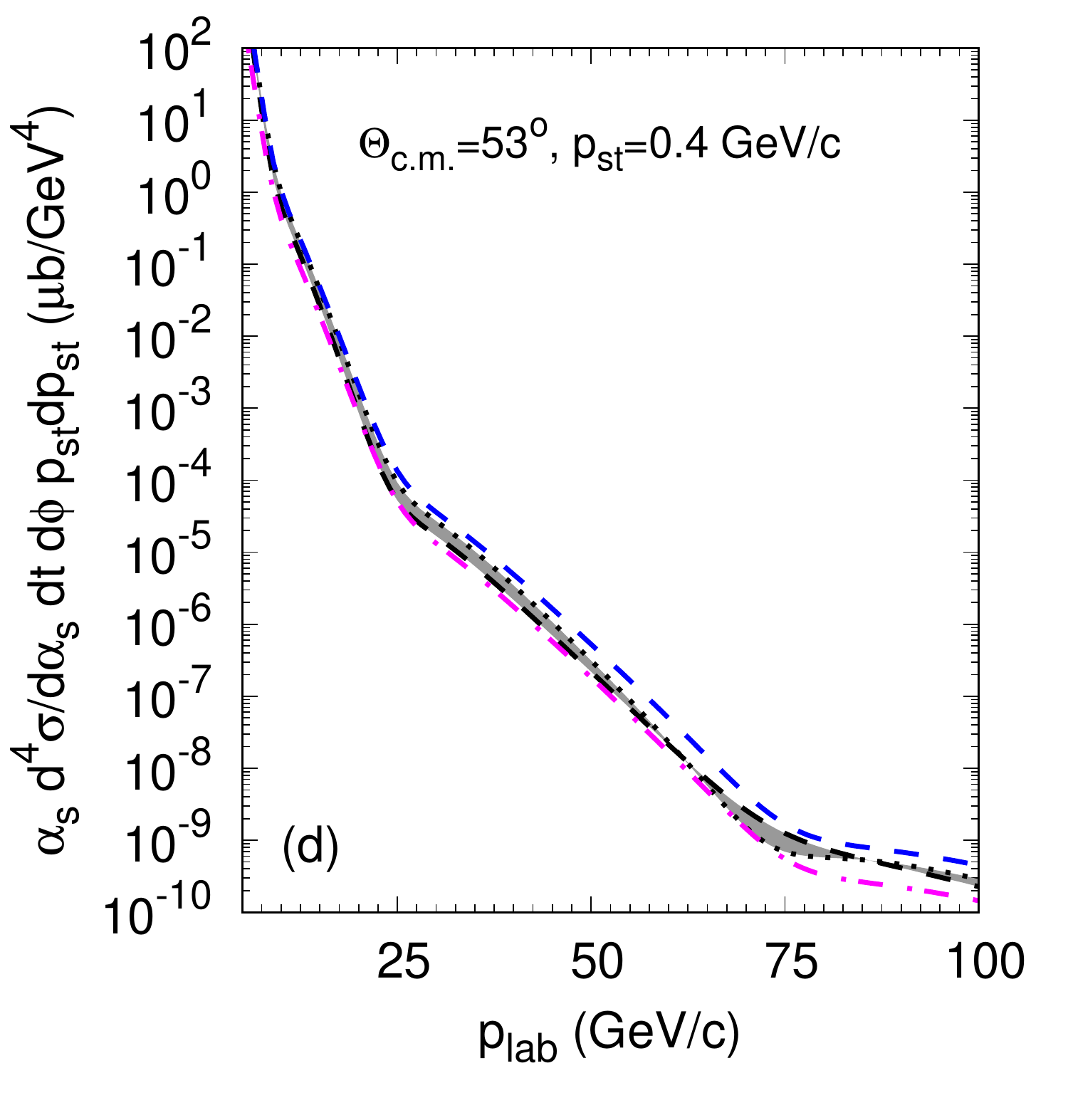}
  \caption{\label{fig:sig_vs_plab} Beam momentum dependence of the differential cross section of $pp \to ppn$, Eq.(\ref{dsig/dalpha}),
    at $\alpha_s=1$, $\phi=180\degree$ for different values of $p_{st}$ and $\Theta_{c.m.}$ as indicated on the panels.
    Notations are the same as in Fig.~\ref{fig:sig_180deg}.}
\end{figure}

This is explained by nuclear filtering out the large-size Landshoff component \cite{Ralston:1988rb}.
Note, however, that for the deuteron target this can not be reduced
to the exponential absorption factors like in the case of heavier nuclear targets \cite{Ralston:1988rb,VanOvermeire:2006tk}
which requires some additional consideration.
Indeed, let us suppose that the QC and Landshoff amplitudes
are out-of-phase and that the QC part of the single rescattering amplitude is suppressed by CT.
Since the IA amplitude is dominated by the QC part,
the dominating Landshoff part of the single rescattering
amplitude is in-phase with the IA amplitude and, thus, at small transverse momenta,
their interference term compensates the diminished modulus
squared of the IA amplitude. Similar consideration for the in-phase QC and Landshoff amplitudes
leads to the negative interference term of the Landshoff rescattering and IA amplitude which again acts as a compensation
of the increased  modulus squared of the IA amplitude. Thus, at small transverse momenta, the oscillations of the cross section
calculated in the IA due to the variation of the relative phase of the QC and Landshoff components are smoothened by the interference
of the single rescattering and the IA amplitude (see Fig.~\ref{fig:sig_vs_plab}a,c)
which corresponds to filtering out the Landshoff component.
\footnote{At large transverse momenta the same argument does not apply since the cross section is dominated
by the rescattering amplitude modulus squared.}

From Fig.~\ref{fig:R} one sees now that at 30 (50) GeV/c the QC and Landshoff amplitudes interfere destructively (constructively)
and, therefore, the denominator of the transparency, Eq.(\ref{T}), is relatively small (large).   
This explains larger $T$ for 30 GeV/c than for 50 GeV/c at small $p_{st}$'s in calculations with $\Delta M^2=0.7$ GeV$^2$
when CT is most pronounced.
In calculations with $\Delta M^2=3$ GeV$^2$ CT is not yet fully pronounced at 30 GeV/c which blures the nuclear filtering mechanism somewhat.

At 75 GeV/c, the QC and Landshoff amplitudes interfere destructively again. Since at this momentum the coherence length
is about $10\div40$ fm, CT practically removes QC contribution to the rescattering amplitudes which now contain
almost only Landshoff components.
Hence, the IA amplitude interferes with single rescattering amplitudes
constructively. This explains the ``anti-absorptive'' behavior of $T$ at small $p_{st}$'s.

\begin{figure}
  \includegraphics[scale = 0.50]{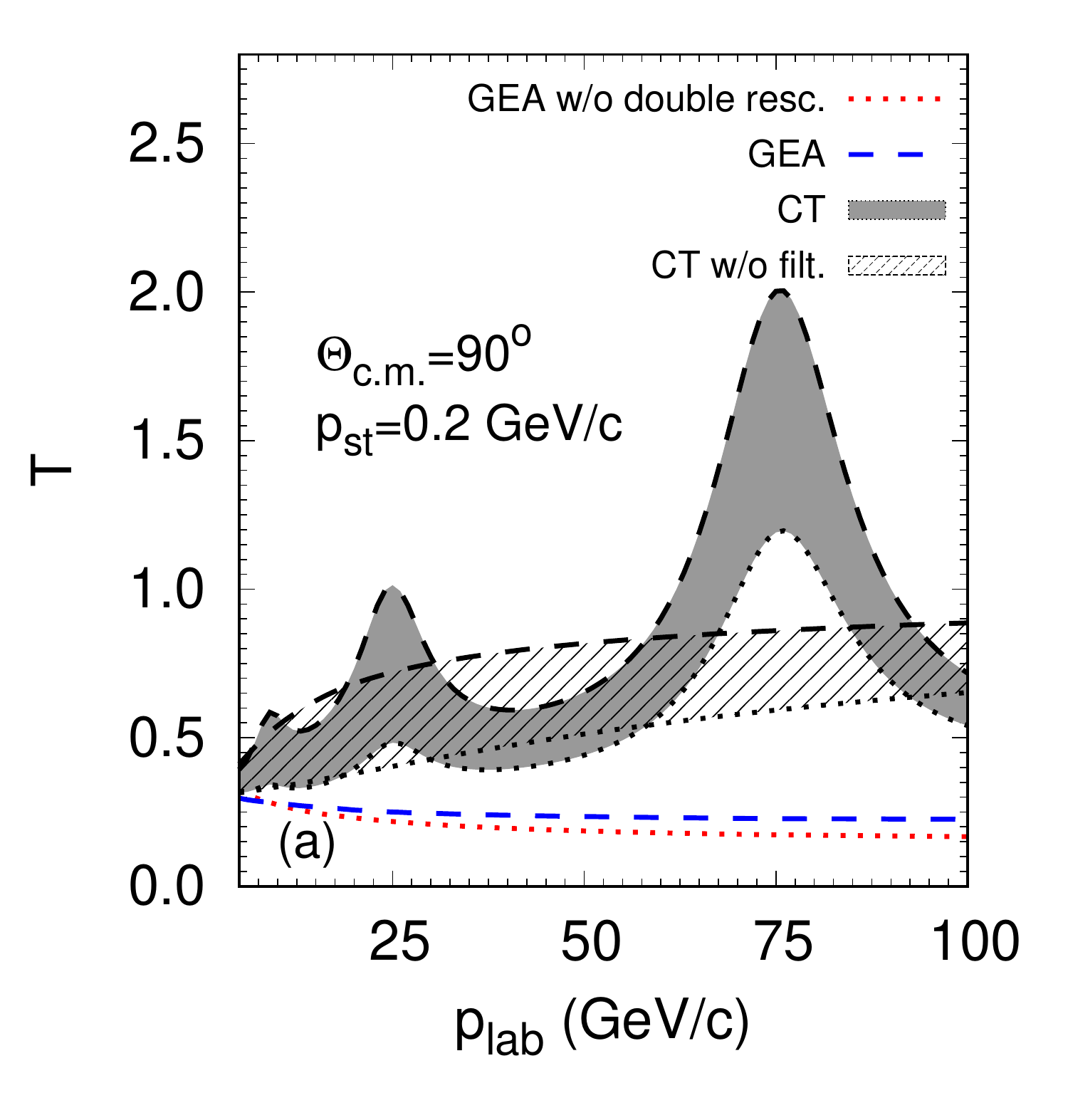}
  \includegraphics[scale = 0.50]{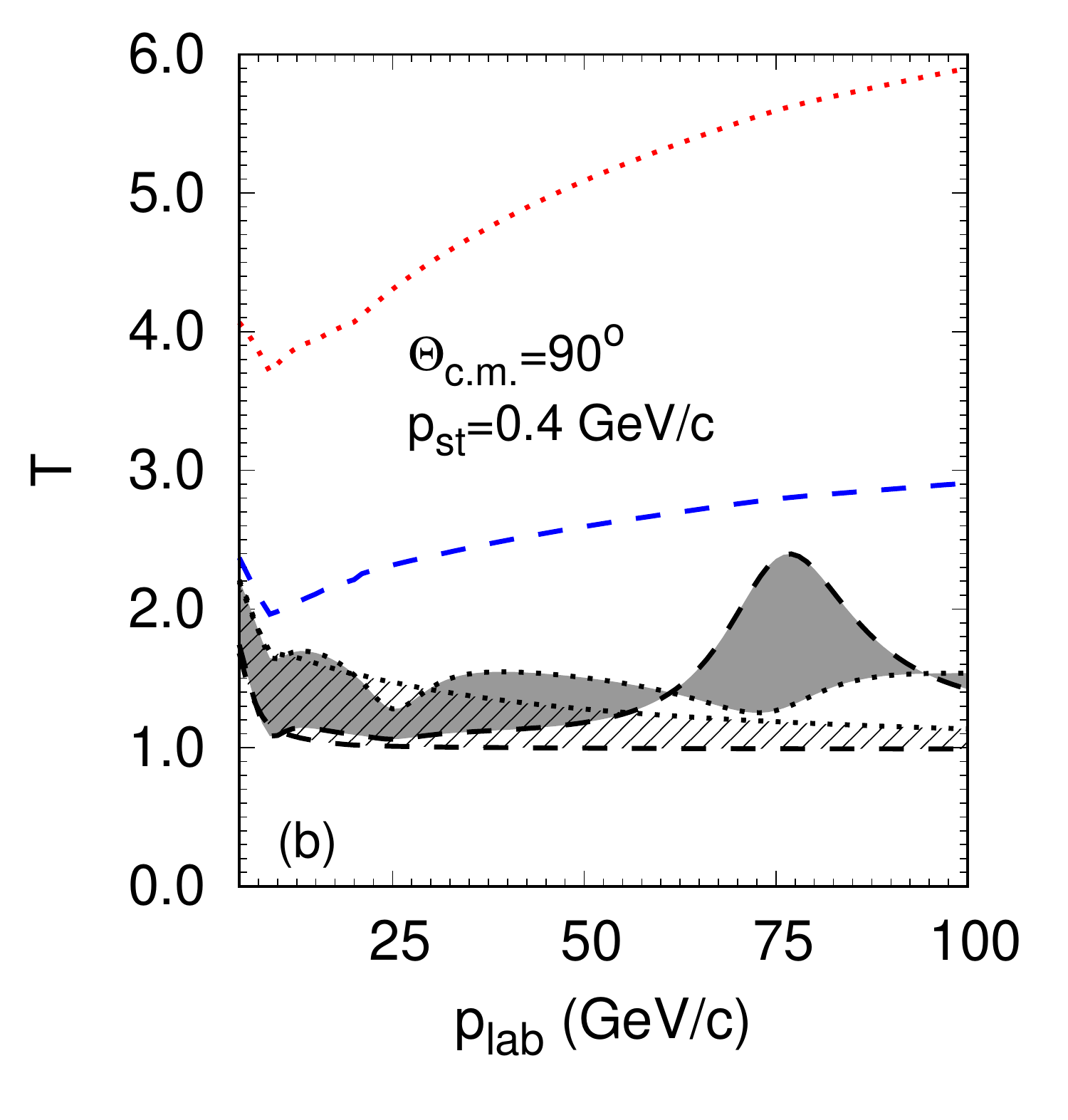}  
  \includegraphics[scale = 0.50]{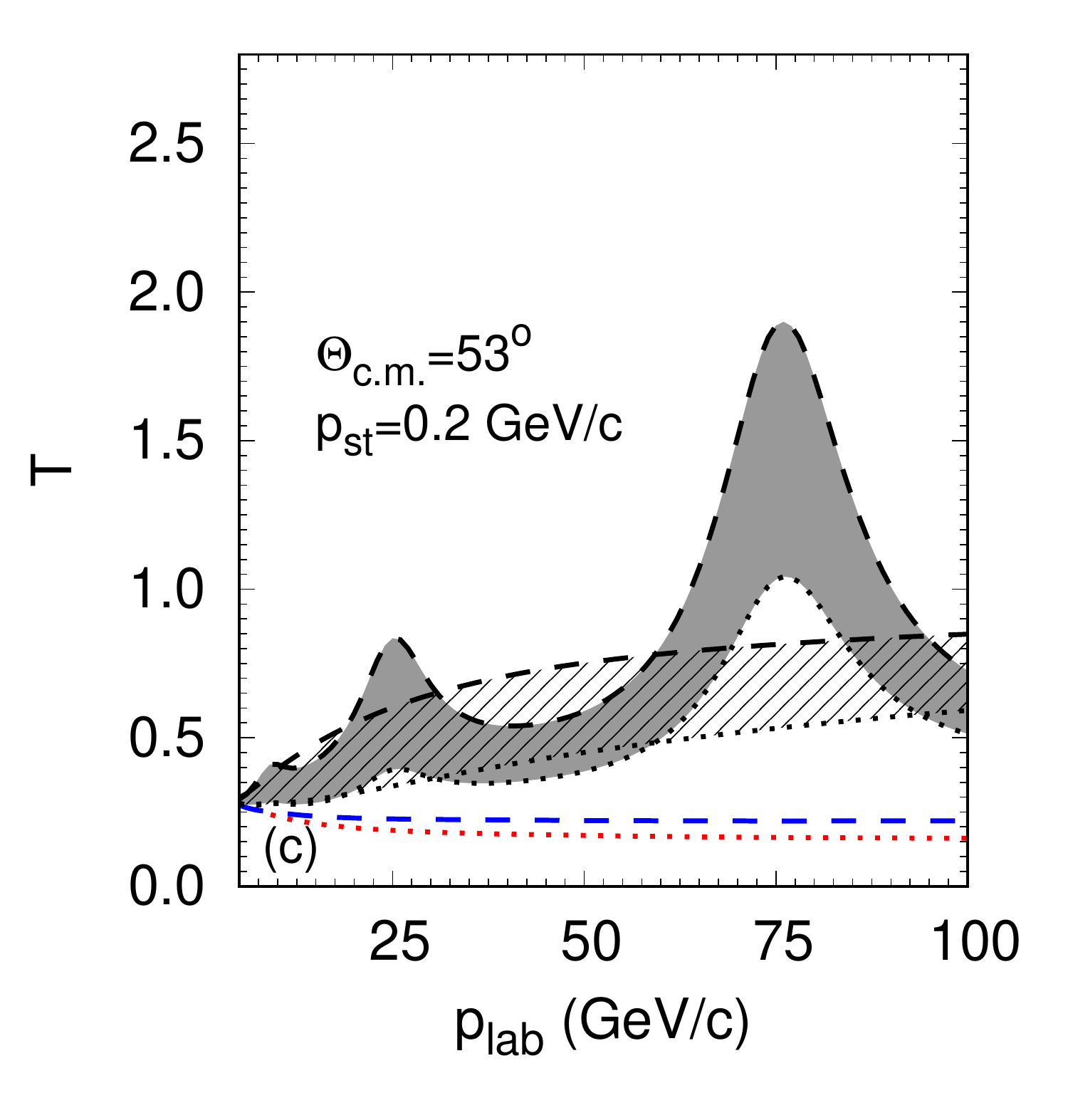}
  \includegraphics[scale = 0.50]{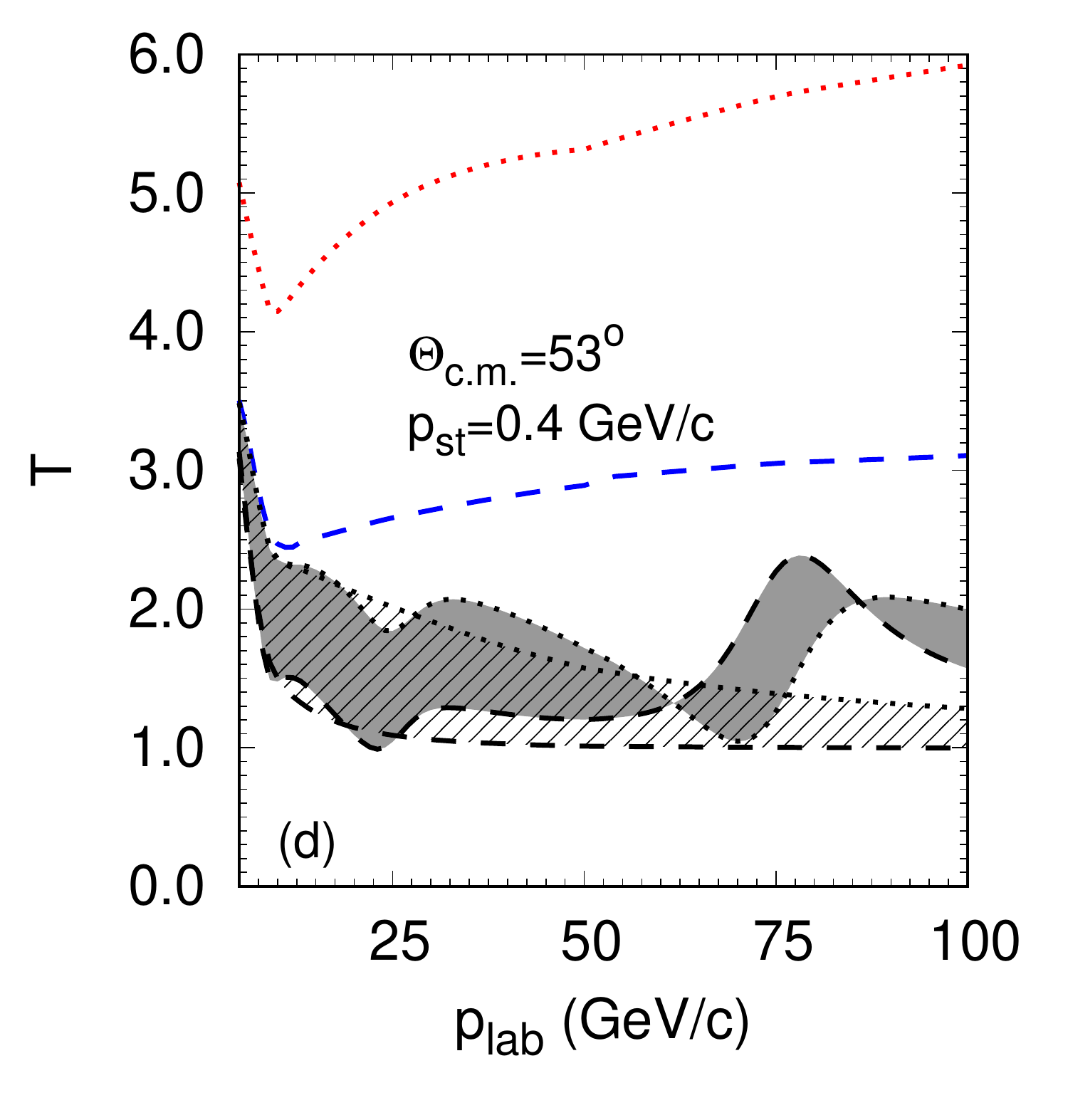}
  \caption{\label{fig:T_vs_plab} Beam momentum dependence of the transparency $T$
    at $\alpha_s=1$, $\phi=180\degree$ for different values of $p_{st}$ and $\Theta_{c.m.}$ as indicated on the panels.
    The hatched band with borders given by the mass denominator of the coherence length,
    $\Delta M^2 = 0.7$ GeV$^2$ (black dashed line) and 3 GeV$^2$ (black dotted line) 
    shows the calculations without filtering out the Landshoff component, i.e. when
    CT influences both the QC and Landshoff parts of rescattering amplitudes. 
    Other notations are the same as in Fig.~\ref{fig:T_180deg}.}
\end{figure}

The beam momentum dependence of the CT effects is better represented in Fig.~\ref{fig:sig_vs_plab} 
for differential cross section and in Fig.~\ref{fig:T_vs_plab} for $T$. 
At $p_{st} = 0.2$ GeV/c, the transparency oscillates as a function of $p_{\rm lab}$ which is a consequence of
a filtering effect. For comparison, the transparency is also shown without the filtering effect, i.e. when the Landshoff part
of the amplitude is treated like the QC part so that CT affects the entire QC+Landshoff amplitude (this corresponds to the scenario when
the Landshoff process results in the PLC formation). In this case, the oscillations disappear
and $T$ shows up a monotonic increase with beam momentum, as expected.

At $p_{\rm lab}=30\div60$ GeV/c, deviations from the IA ($T=1$) at smaller transverse momentum, $p_{st} = 0.2$ GeV/c,
are significant even with CT. However, at larger transverse momentum, $p_{st} = 0.4$ GeV/c, the calculation with CT becomes close
to the one in the IA. This is mostly because the absorptive effect of the pure GEA calculation at small $p_{st}$'s is stronger than the
enhancement effect at large $p_{st}$'s, given the double rescattering contribution is included
(cf. Fig.~\ref{fig:T_vs_plab}a,b).

Reducing the c.m. scattering angle from $90\degree$ to $53\degree$ does not greatly change
the effects of rescattering and CT, however, increases the cross section by 2–3 orders of magnitude.
The differences in $T$ for the two values of $\Theta_{c.m.}$ are significant at
$p_{\rm lab} \ltsim 30$ GeV/c but become quite small at larger $p_{\rm lab}$, in particular for $\Delta M^2=0.7$ GeV$^2$.
(At $p_{\rm lab}=100$ GeV/c the momentum of a slower proton is $p_4=47$ GeV/c
at $\Theta_{c.m.}=90\degree$ and $p_4=18$ GeV/c at $\Theta_{c.m.}=53\degree$.
For $\Delta M^2=0.7$ GeV$^2$ this corresponds to full CT, i.e. to practically full disappearance of the QC part
of the rescattering amplitude. But for $\Delta M^2=3$ GeV$^2$ the slower proton is still sensitive to the PLC expansion.)
This is expected since,
at infinitely large beam momentum, the momenta of outgoing particles  become pure longitudinal and
thus in this limit the rescattering effects do not depend on the transverse momenta of outgoing protons.

\begin{figure}
  \begin{center} 
    \begin{tabular}{ccc}
   \vspace{-1cm}   
   \includegraphics[scale = 0.35]{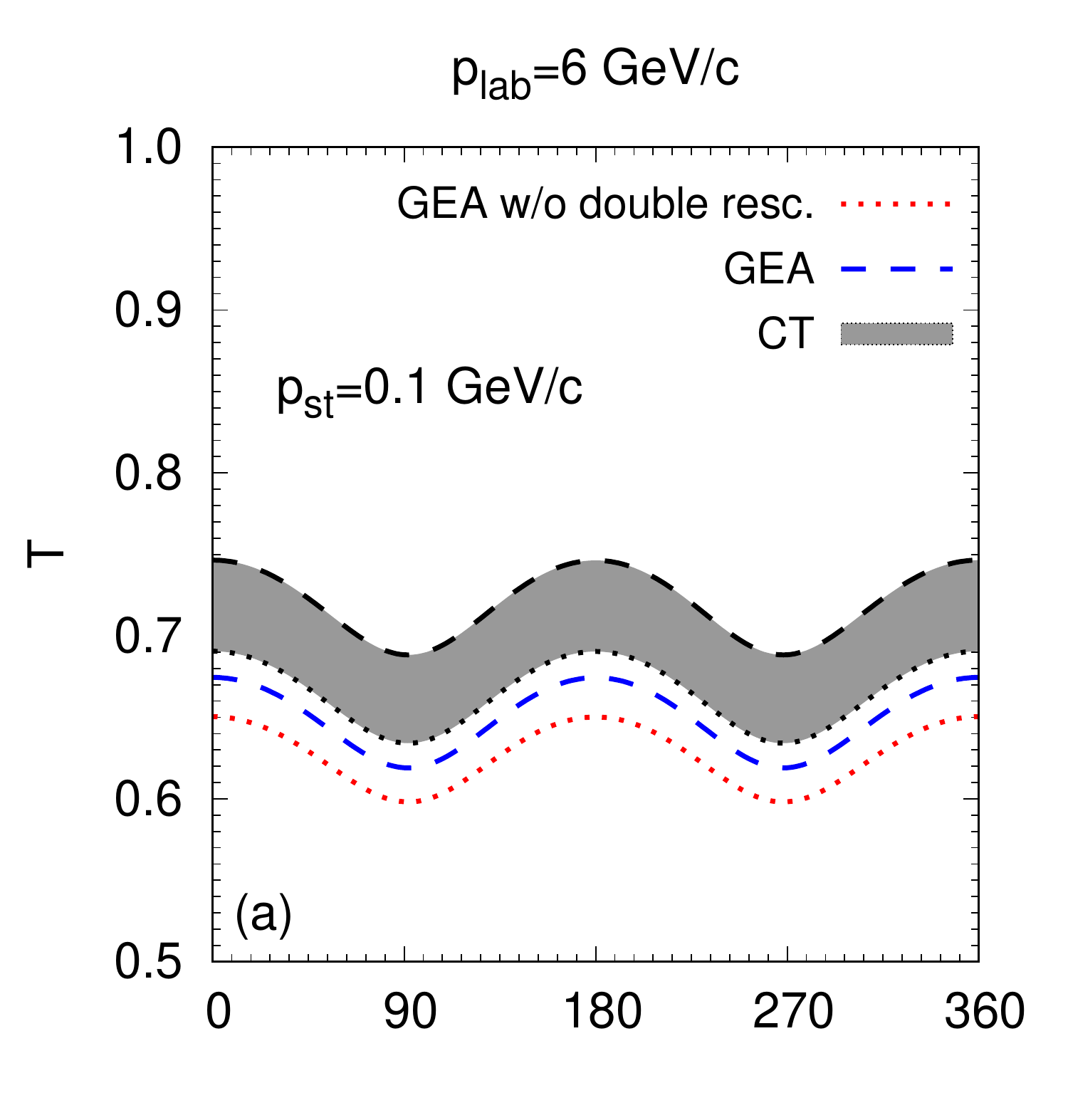} &
   \includegraphics[scale = 0.35]{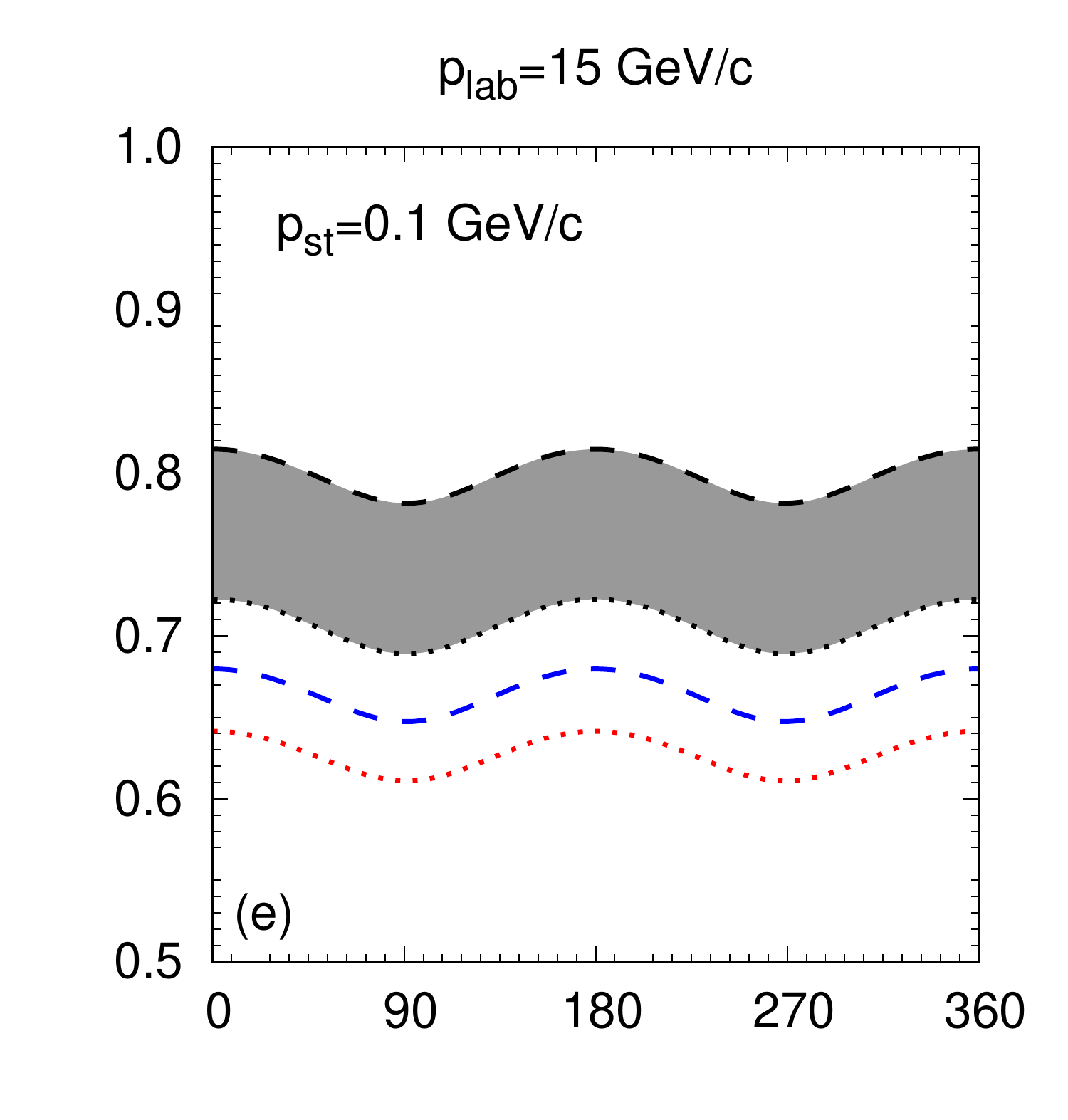} &
   \includegraphics[scale = 0.35]{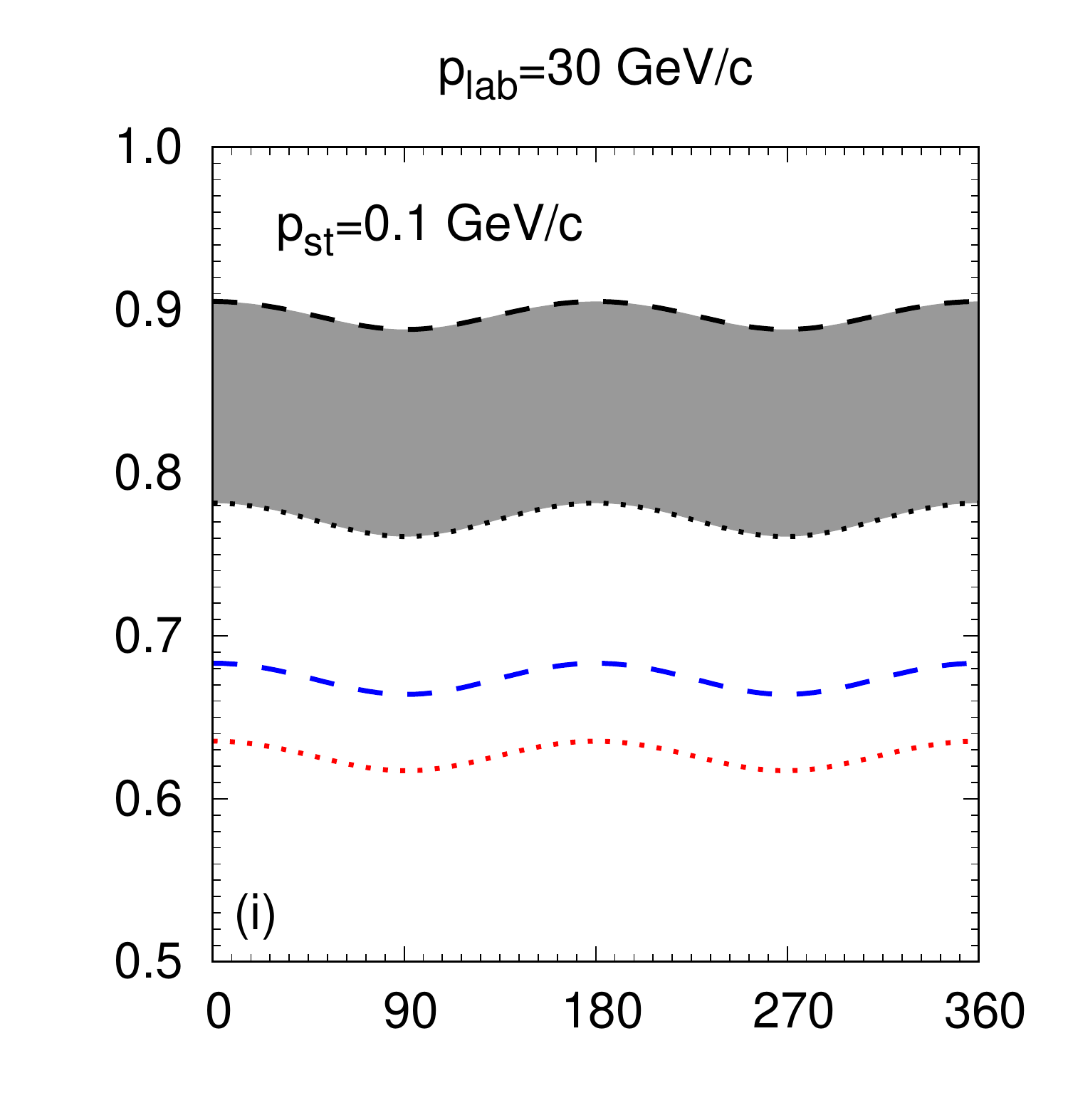} \\
   \vspace{-1cm}
   \includegraphics[scale = 0.35]{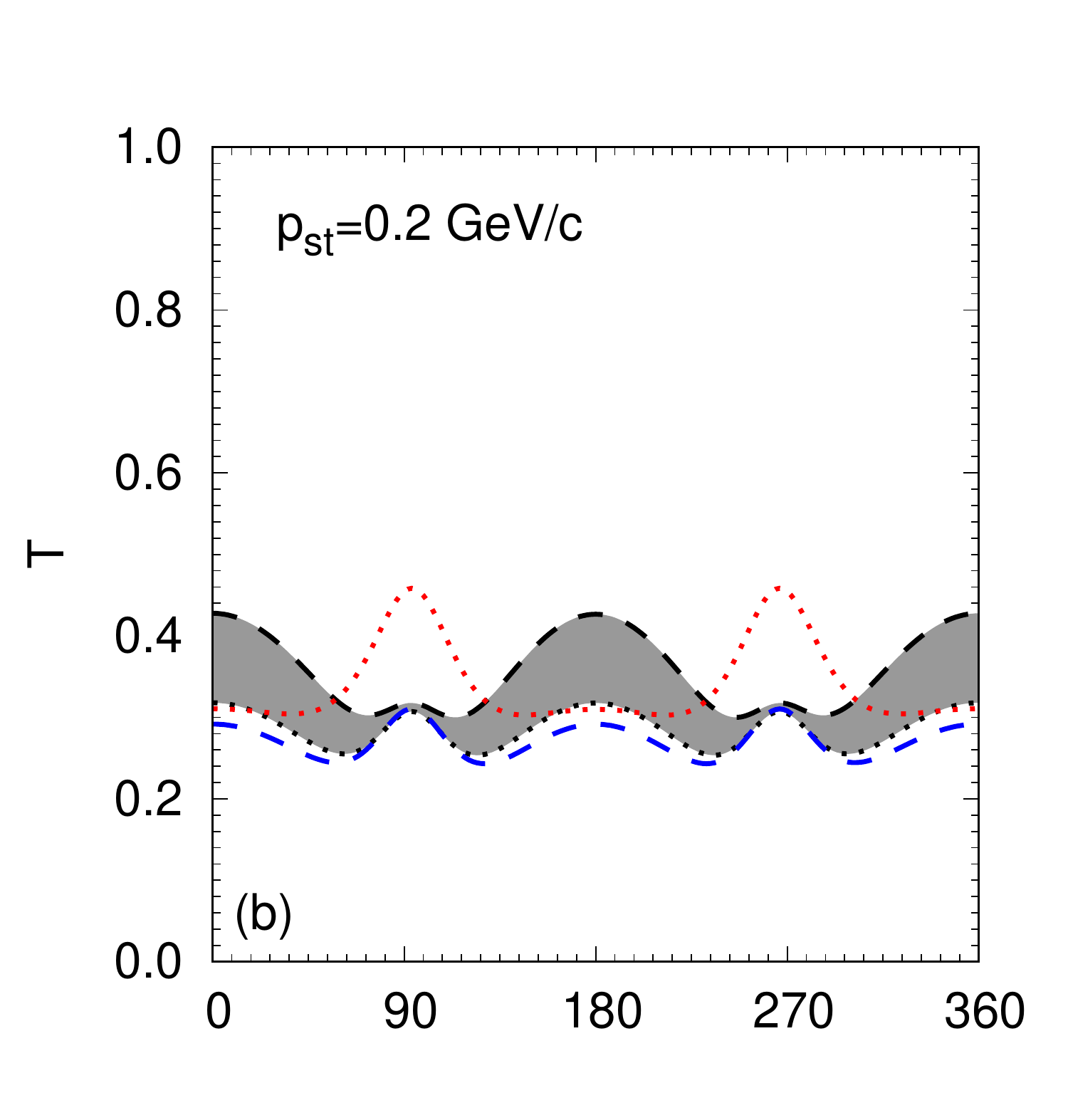} &
   \includegraphics[scale = 0.35]{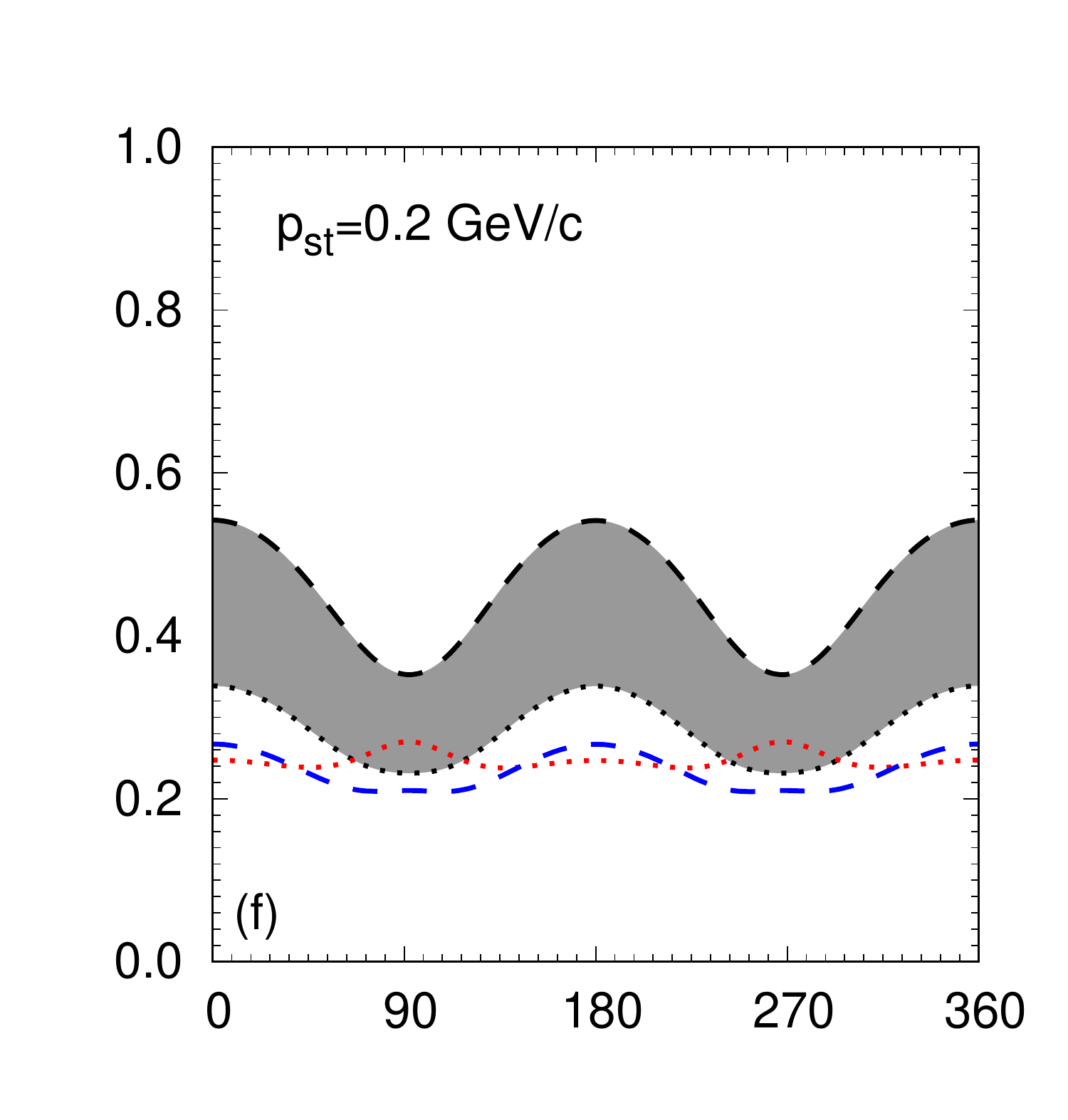} &
   \includegraphics[scale = 0.35]{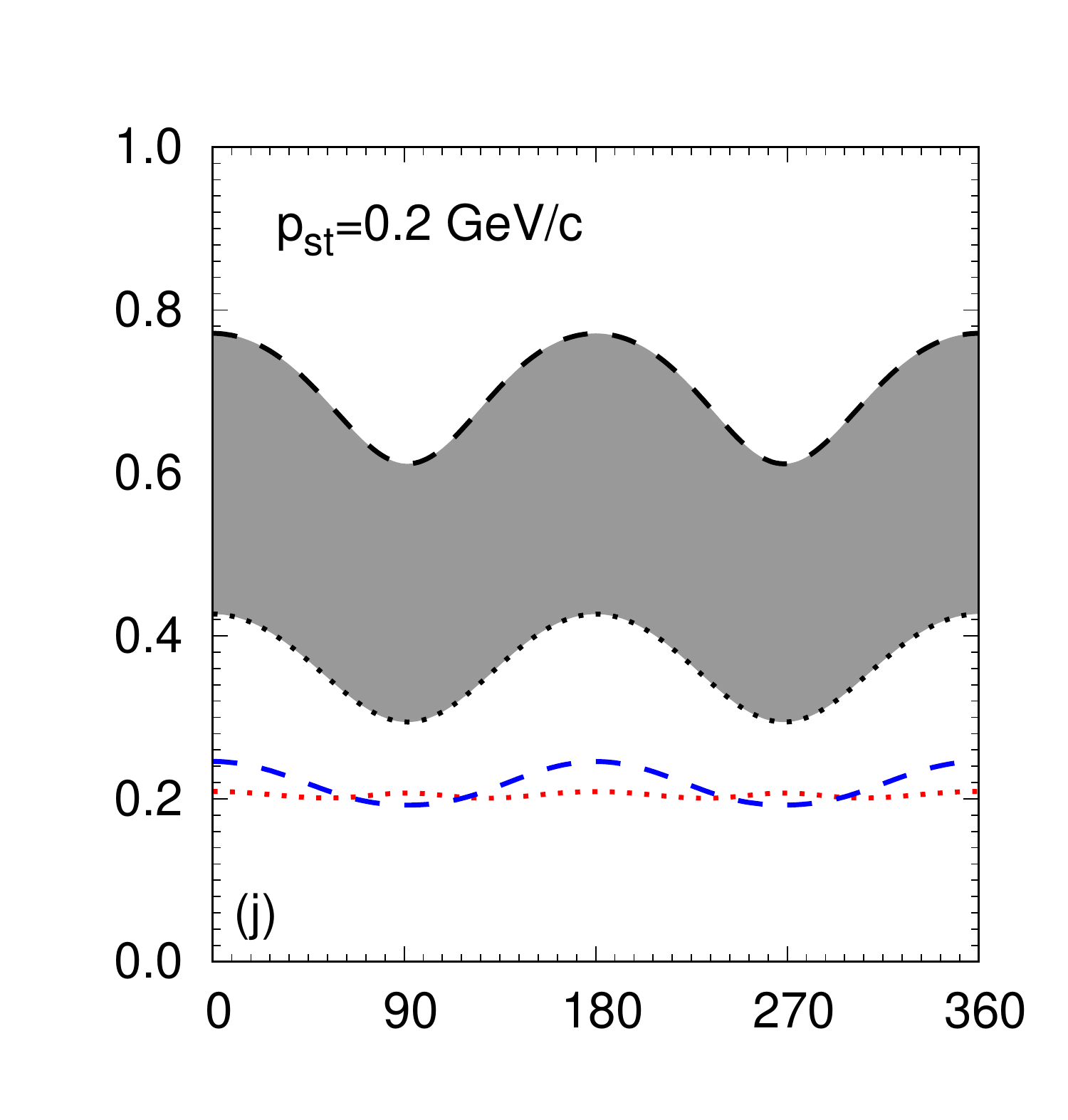} \\
   \vspace{-1cm}
   \includegraphics[scale = 0.35]{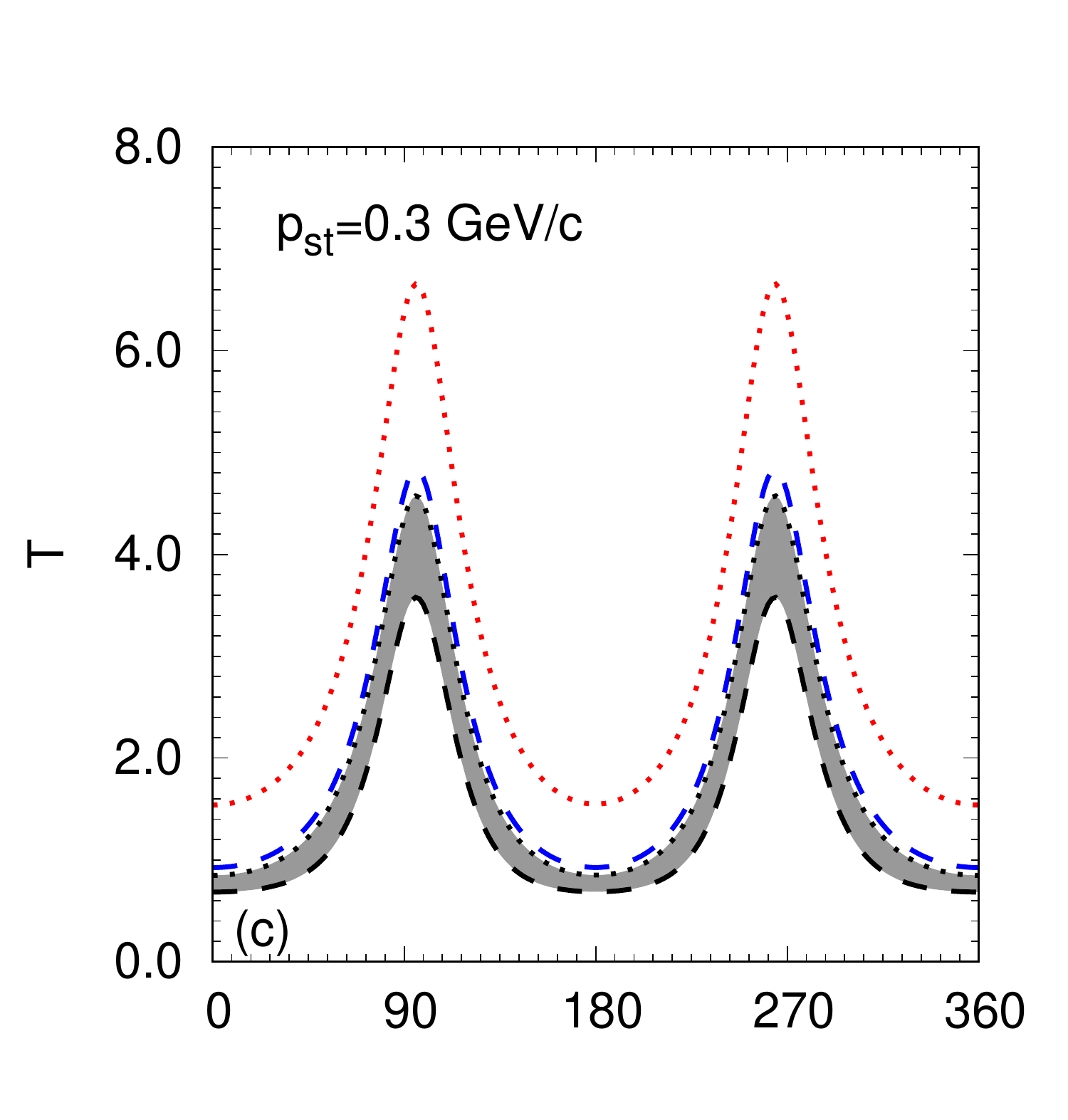} &
   \includegraphics[scale = 0.35]{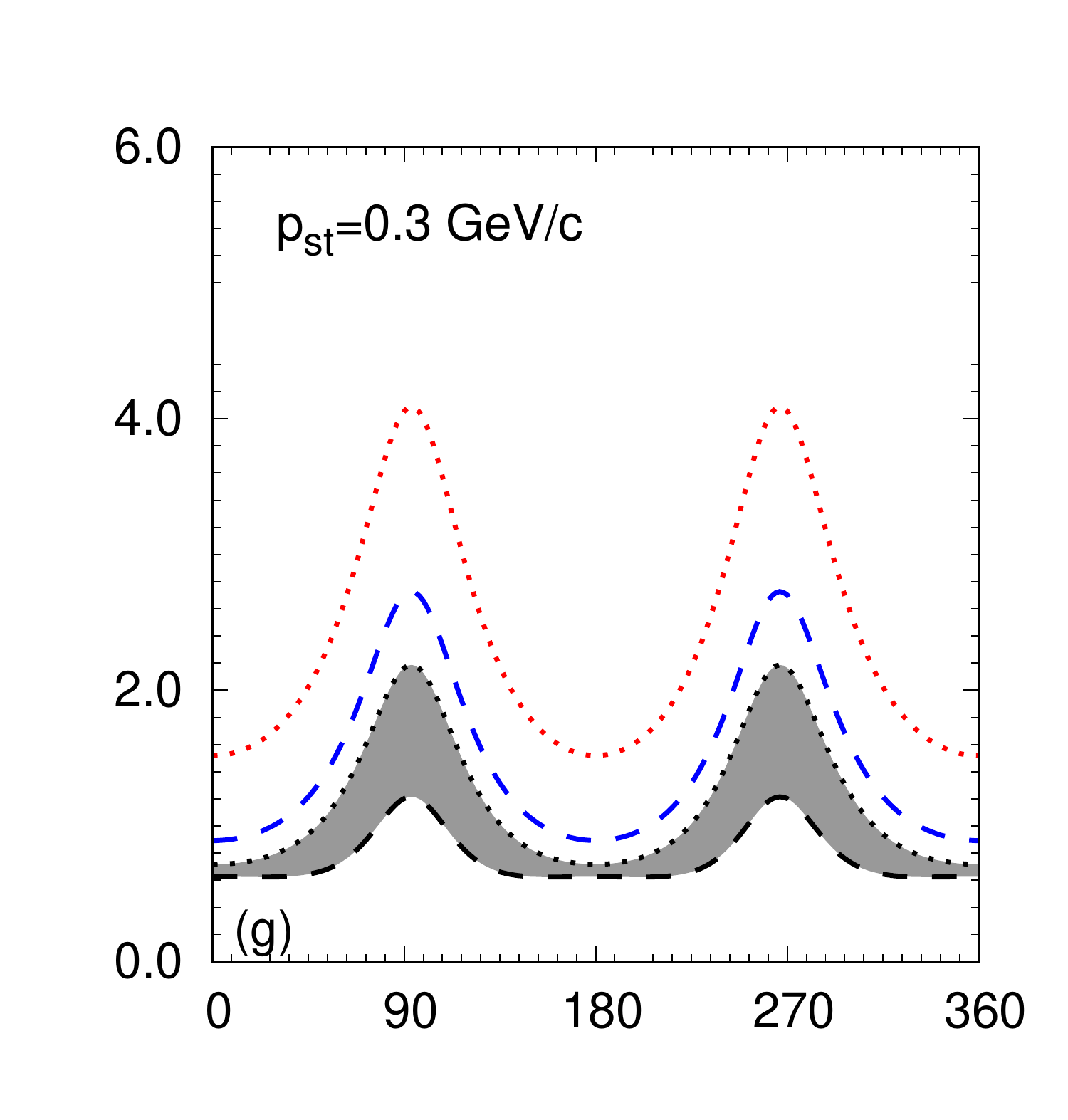} &
   \includegraphics[scale = 0.35]{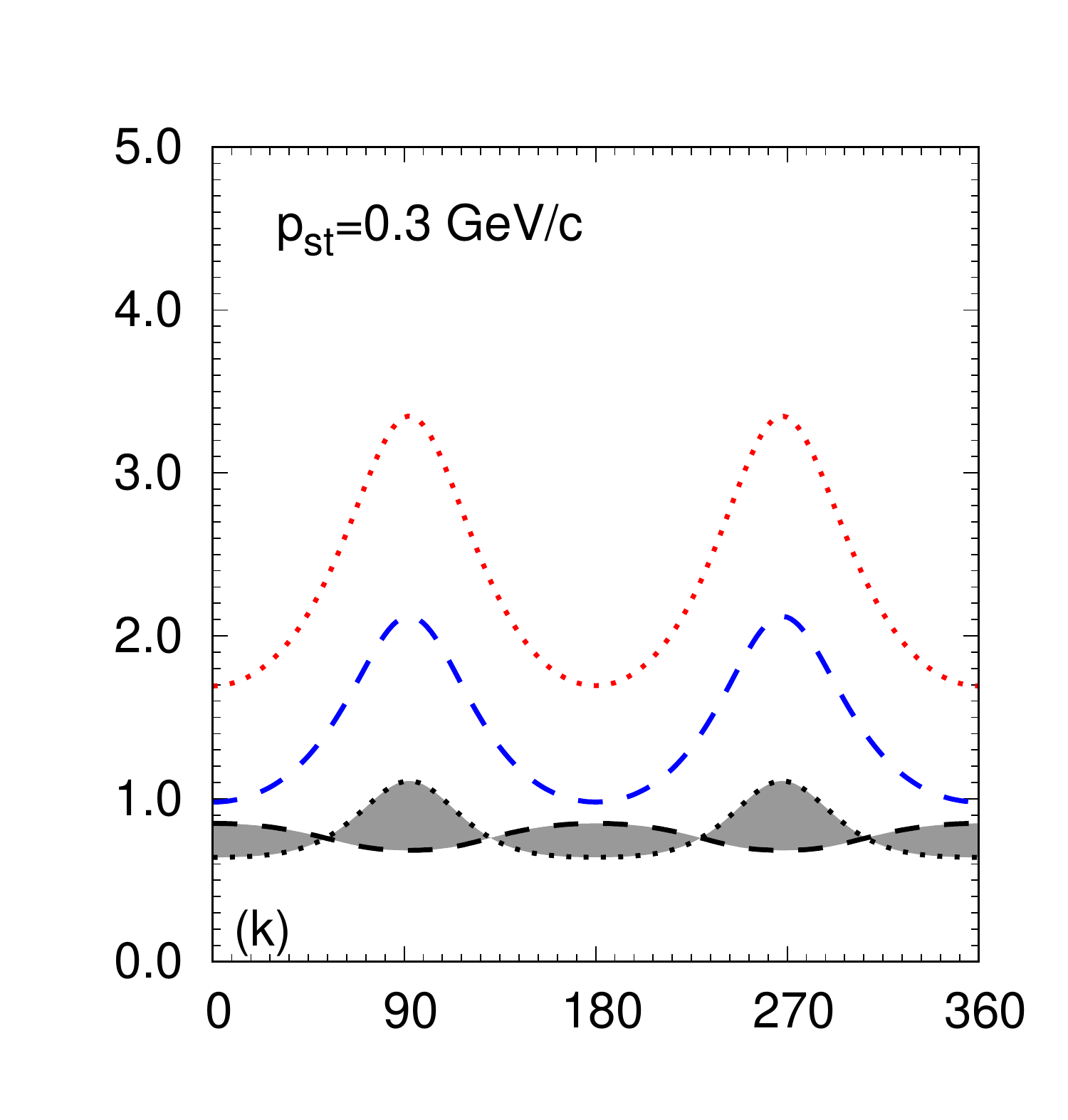} \\
   \includegraphics[scale = 0.35]{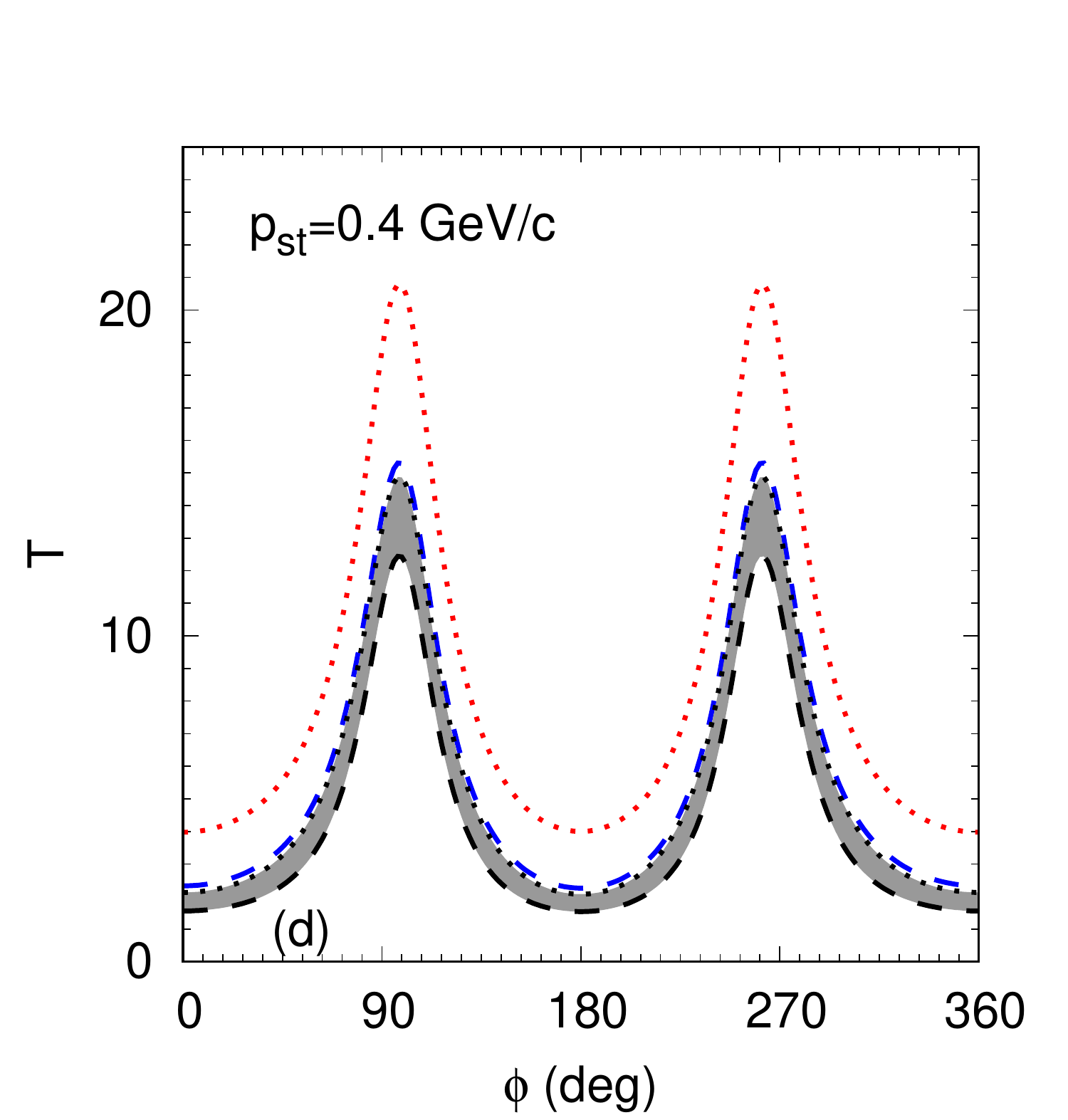} &
   \includegraphics[scale = 0.35]{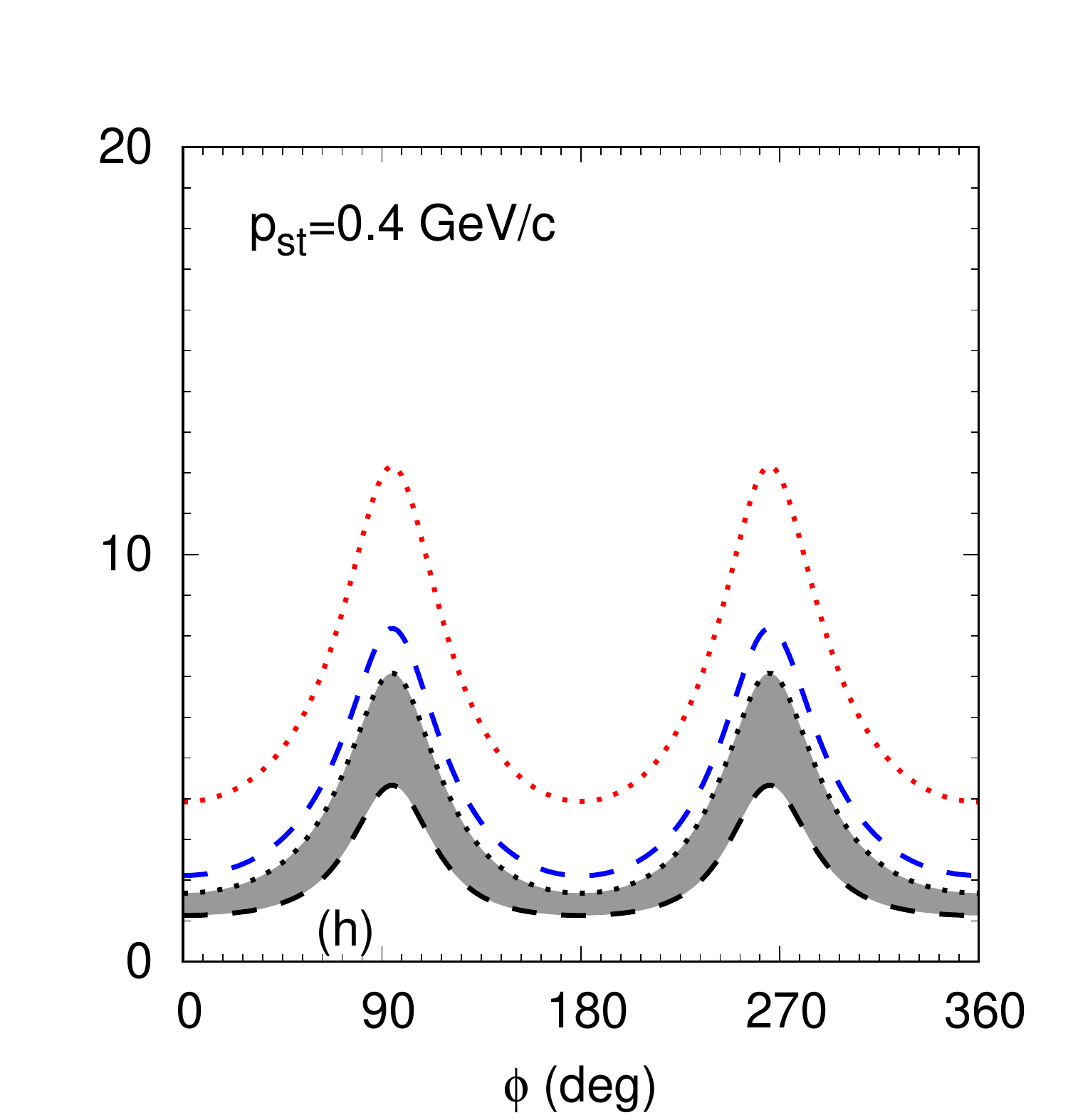} &
   \includegraphics[scale = 0.35]{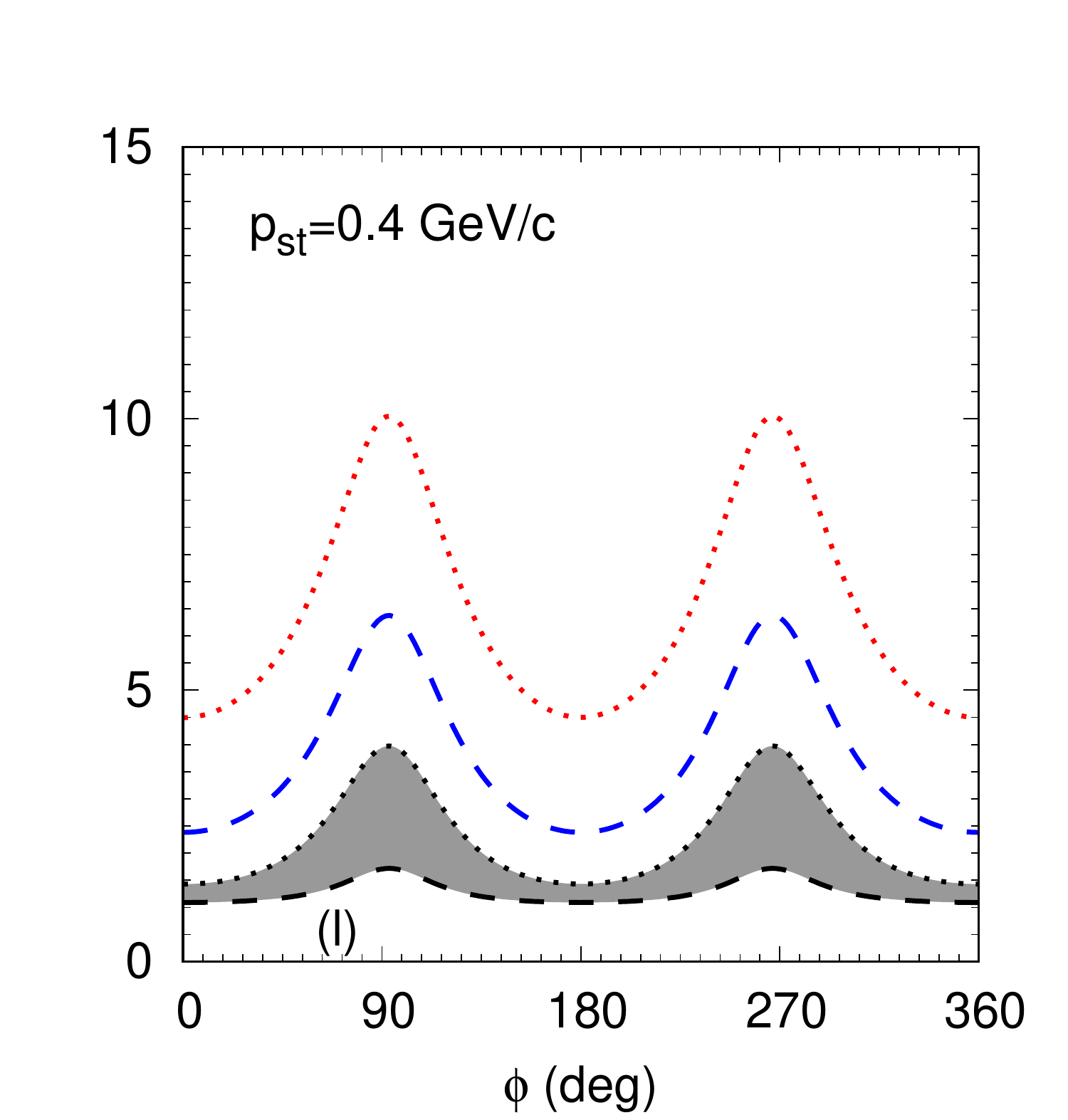} \\
   \end{tabular}
 \end{center}   
 \caption{\label{fig:T_6&15&30gevc_phiDep} The transparency $T$ for $pd \to ppn$ at $p_{\rm lab}=6$ GeV/c (left column),
   $p_{\rm lab}=15$ GeV/c (middle column), and $p_{\rm lab}=30$ GeV/c (right column) 
   as a function of relative azimuthal angle between the scattered proton and spectator neutron
   for $p_{st}=0.1, 0.2, 0.3$ and 0.4 GeV/c (from top to bottom panels).
   Line notations are the same as in Fig.~\ref{fig:T_180deg}.}
\end{figure}

\begin{figure}
  \begin{center} 
   \begin{tabular}{ccc}
   \vspace{-1cm}   
   \includegraphics[scale = 0.35]{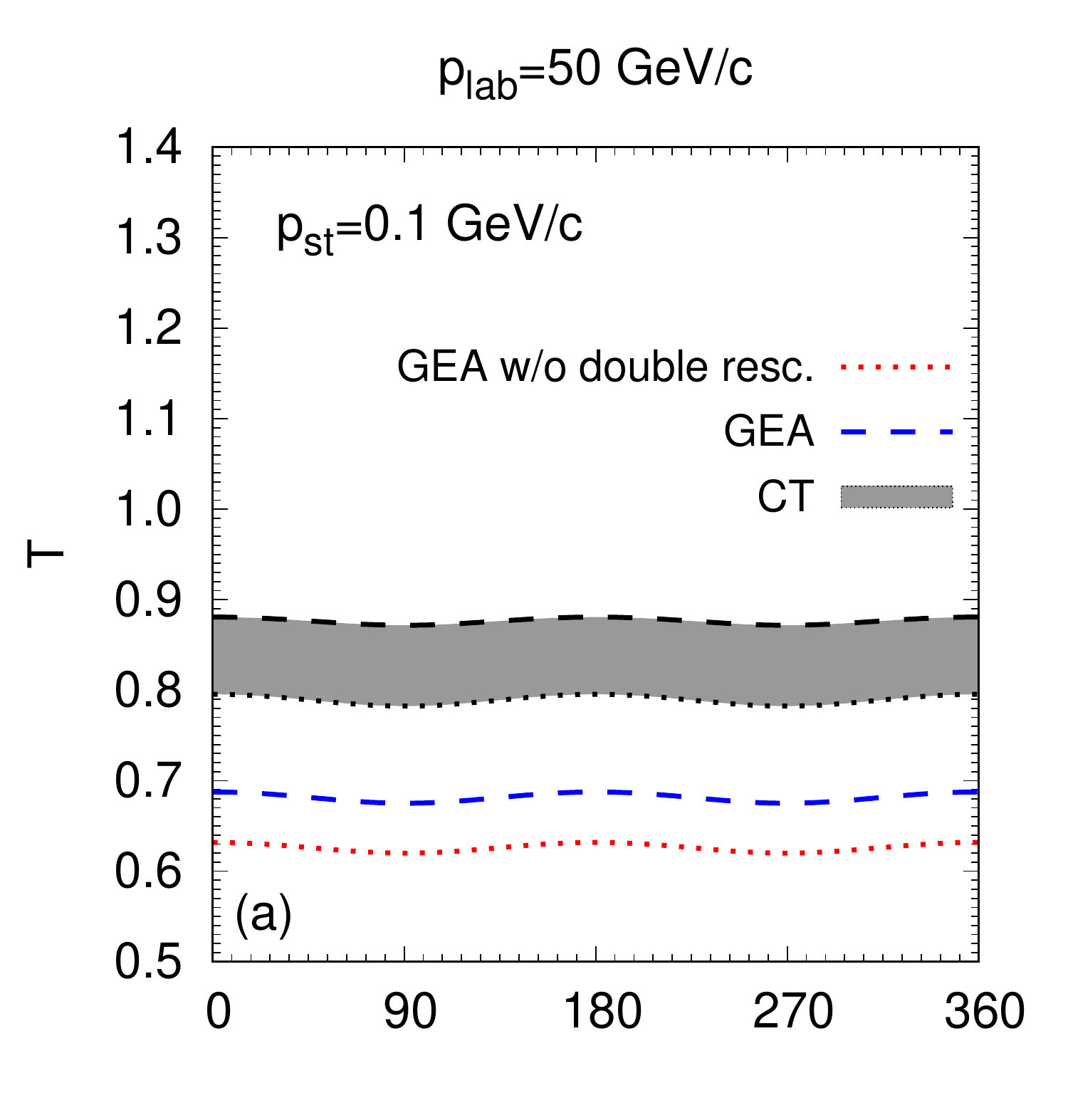} &
   \includegraphics[scale = 0.35]{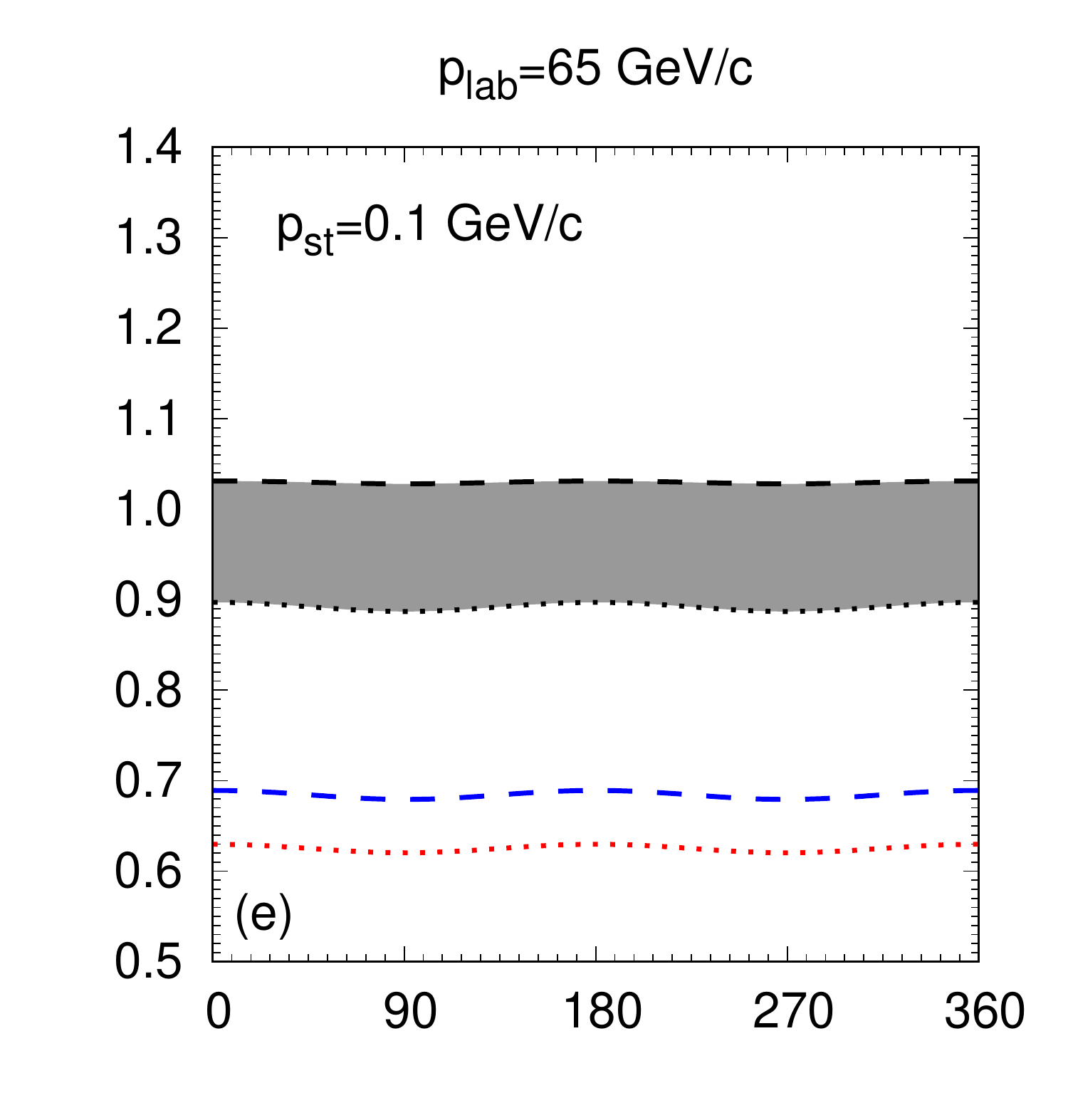} &   
   \includegraphics[scale = 0.35]{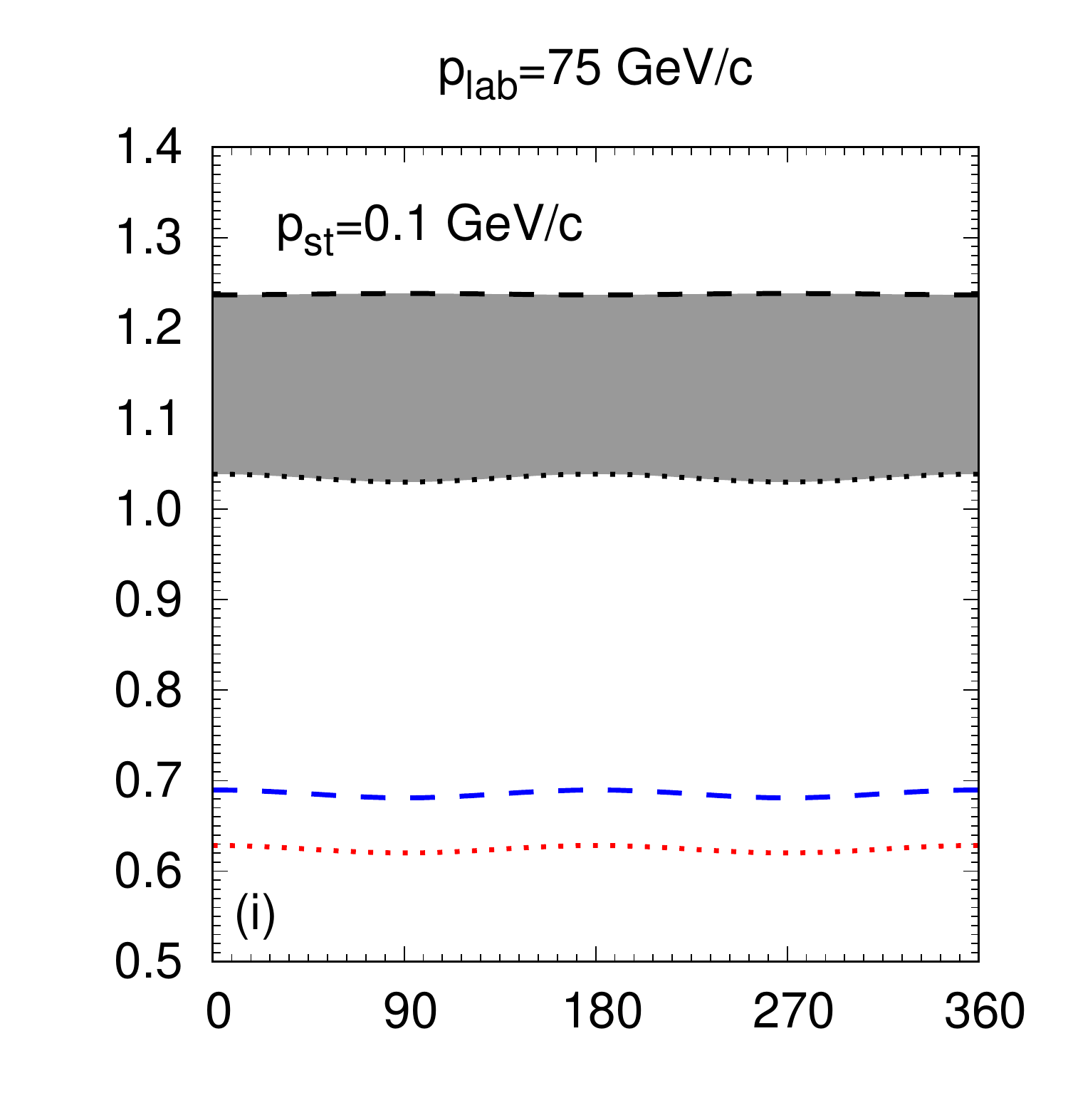} \\
   \vspace{-1cm}
   \includegraphics[scale = 0.35]{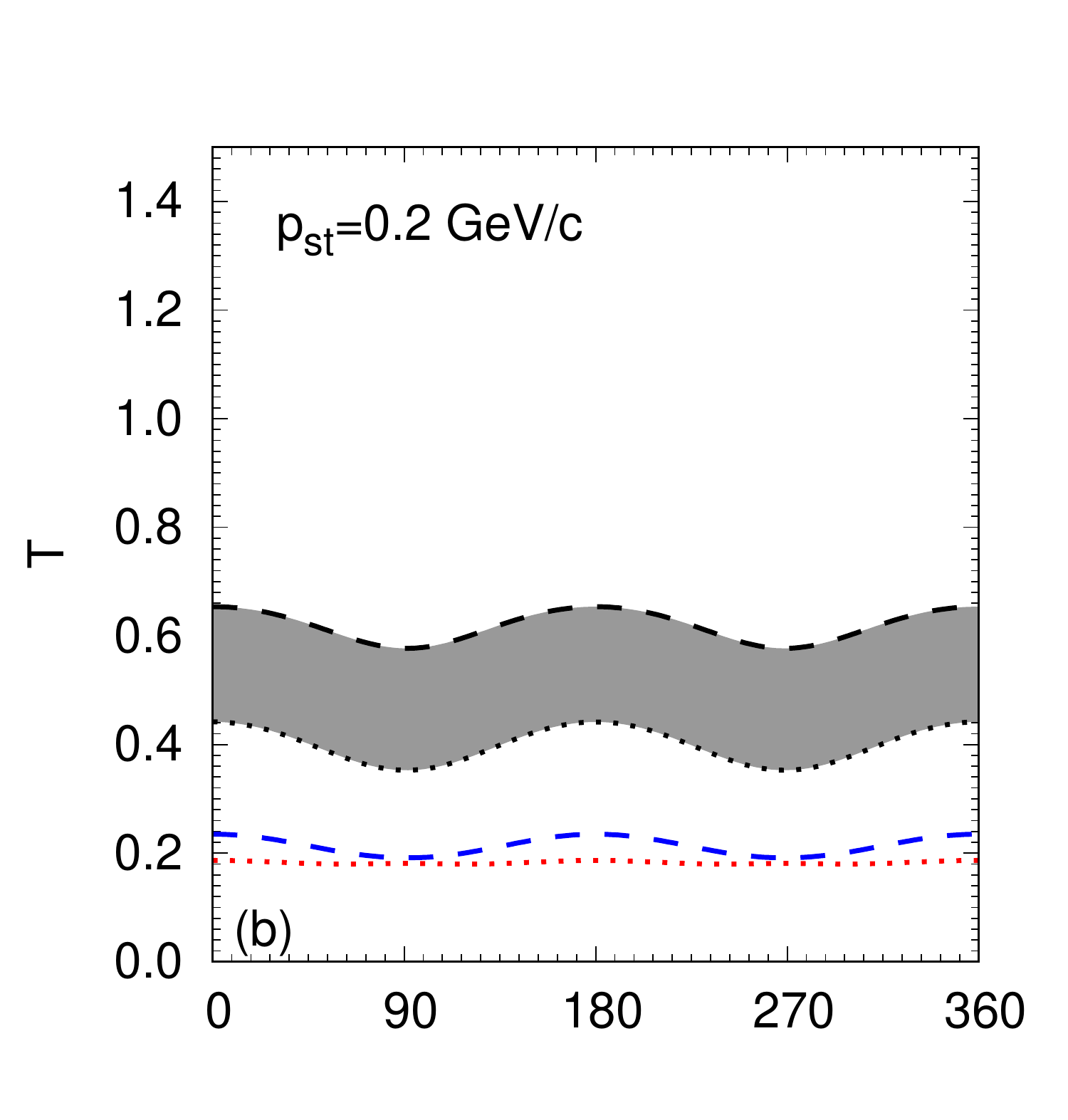} &
   \includegraphics[scale = 0.35]{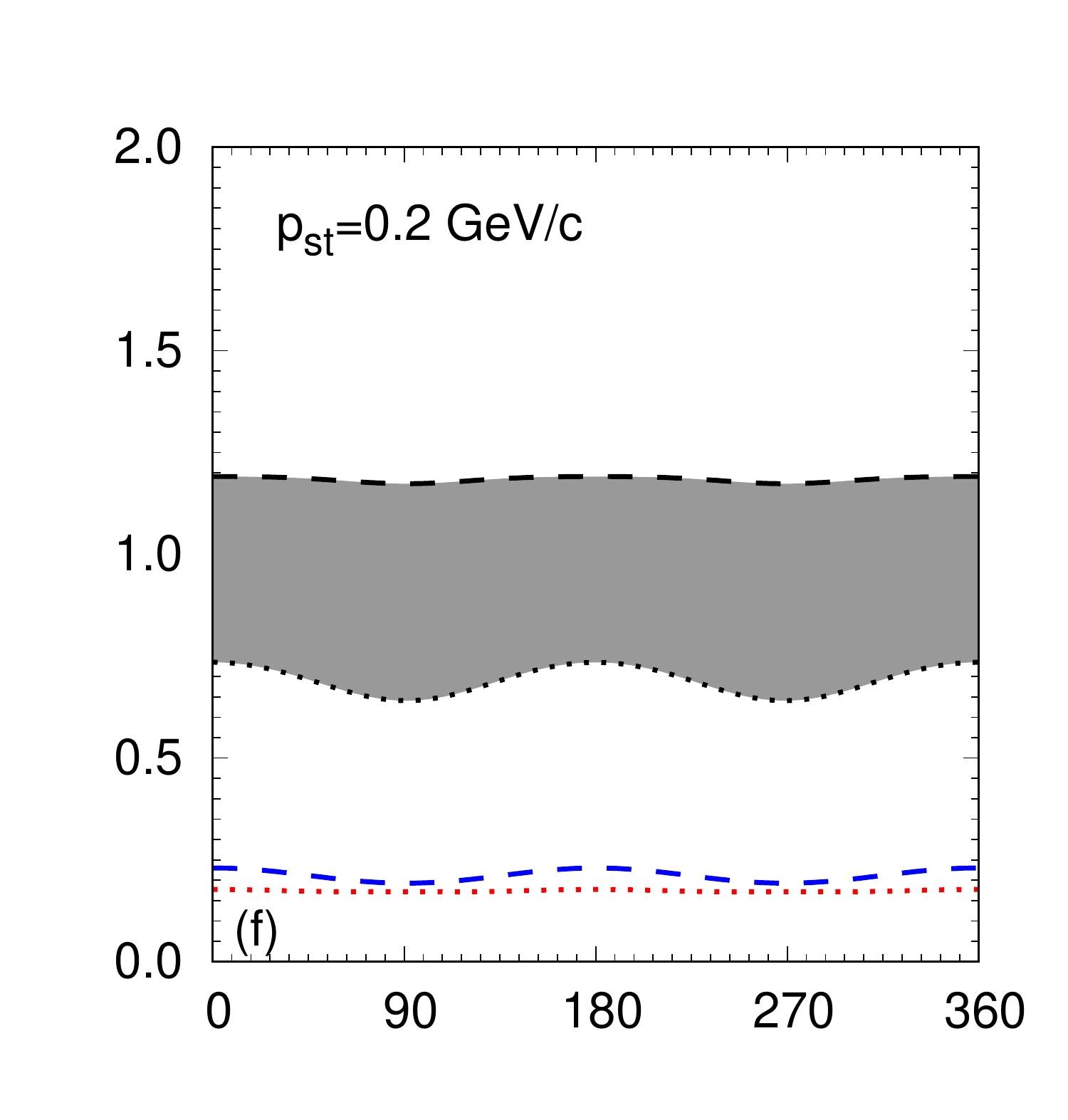} &
   \includegraphics[scale = 0.35]{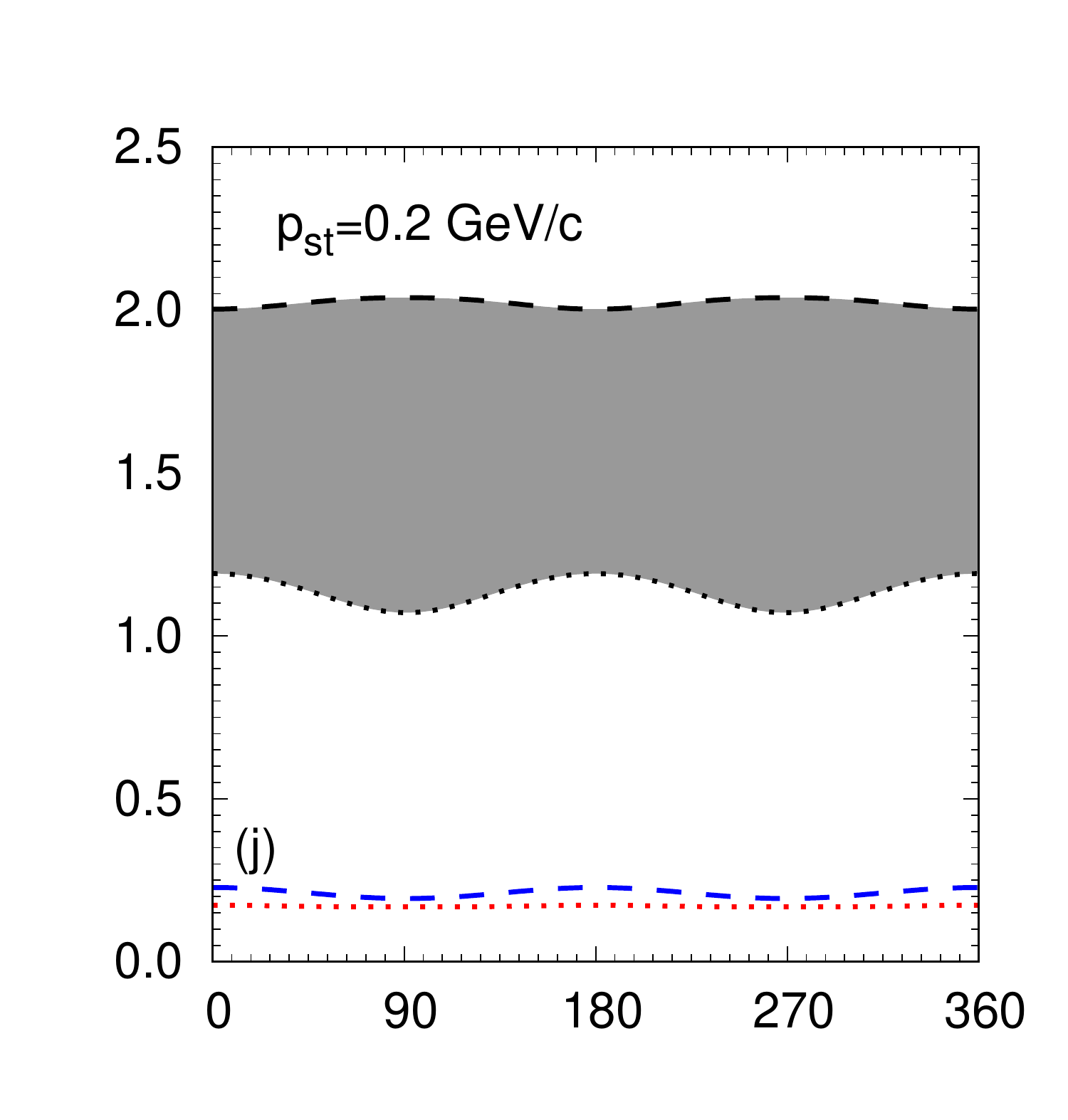} \\
   \vspace{-1cm}
   \includegraphics[scale = 0.35]{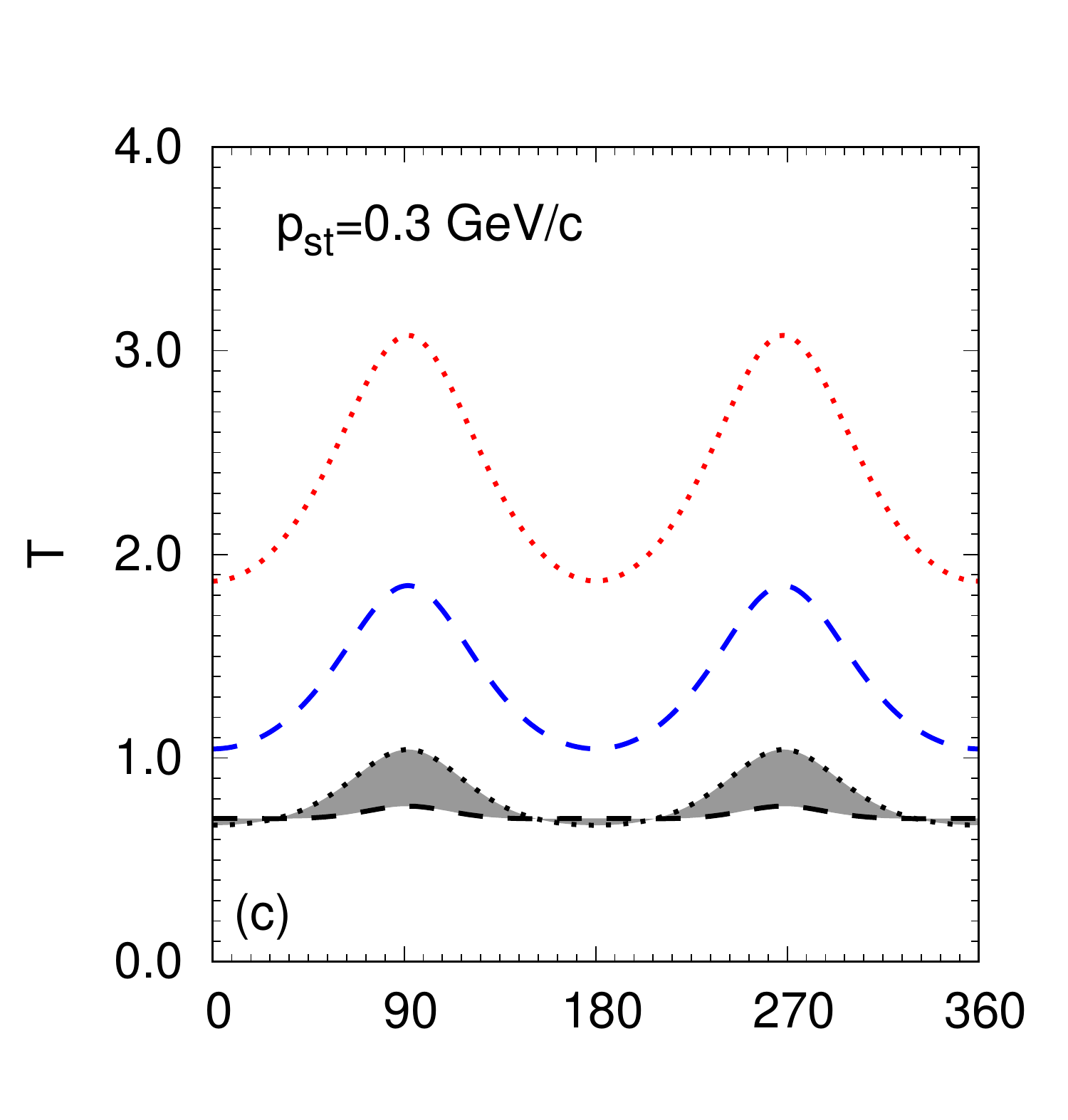} &
   \includegraphics[scale = 0.35]{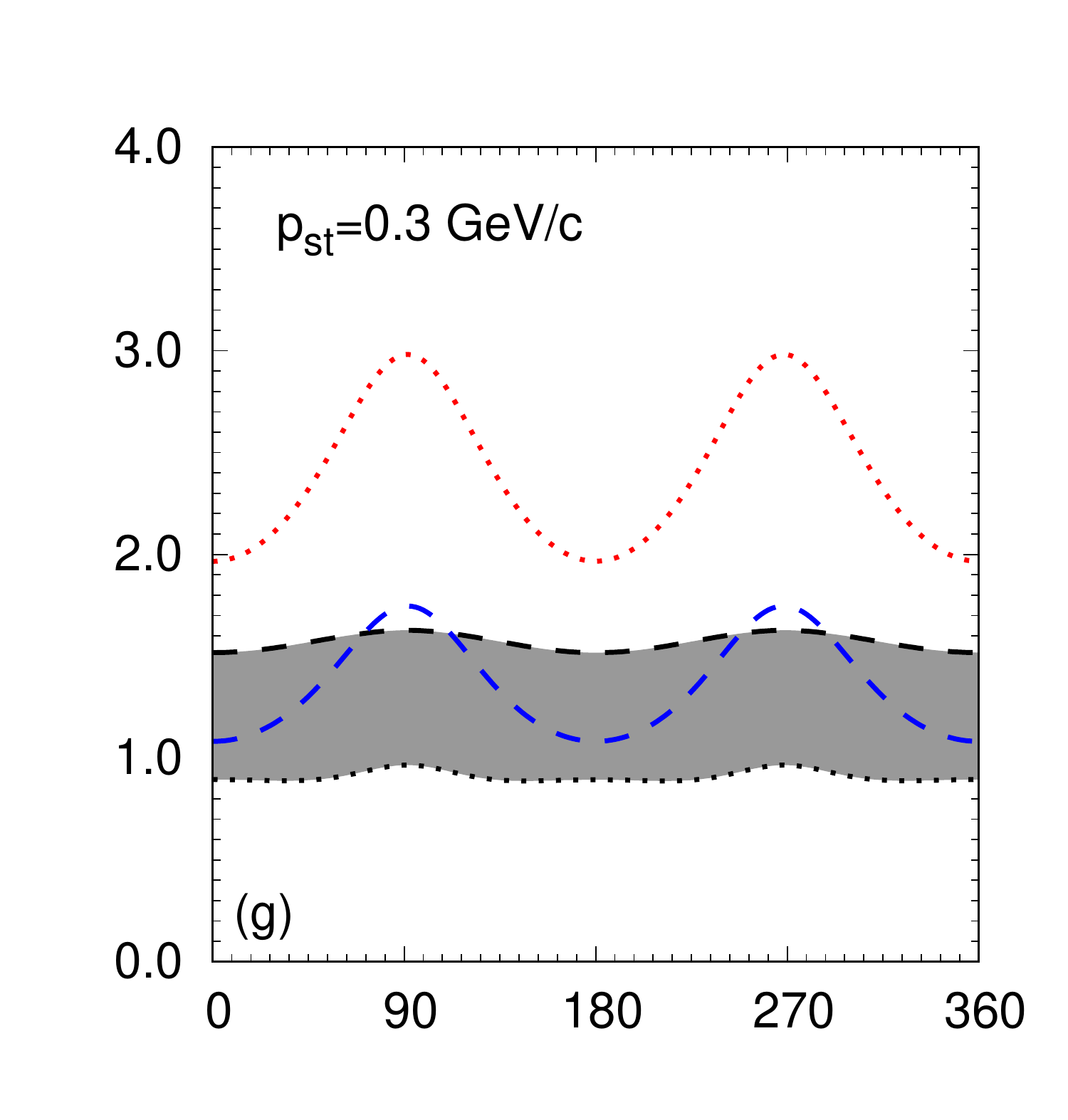} &
   \includegraphics[scale = 0.35]{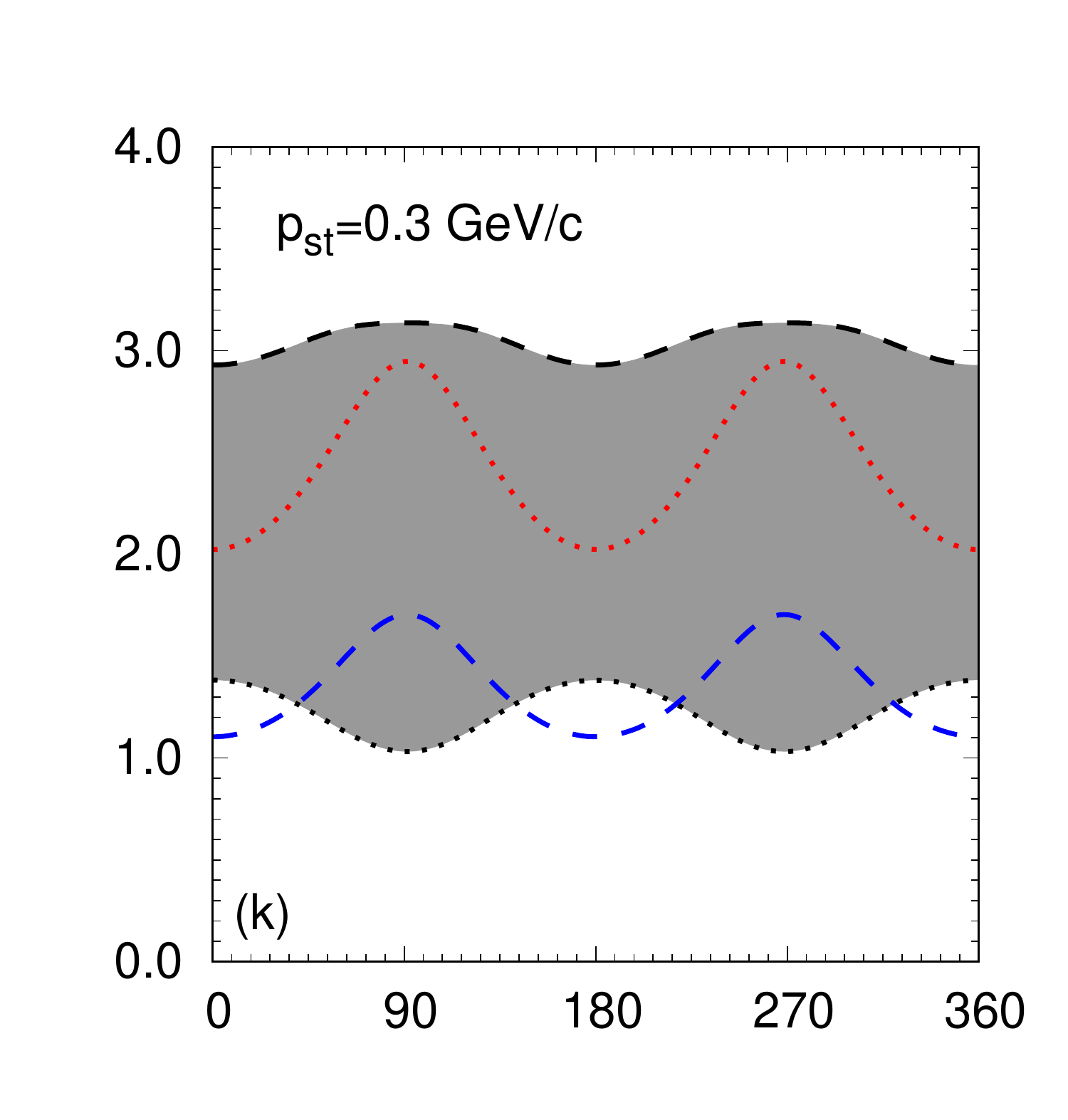} \\
   \includegraphics[scale = 0.35]{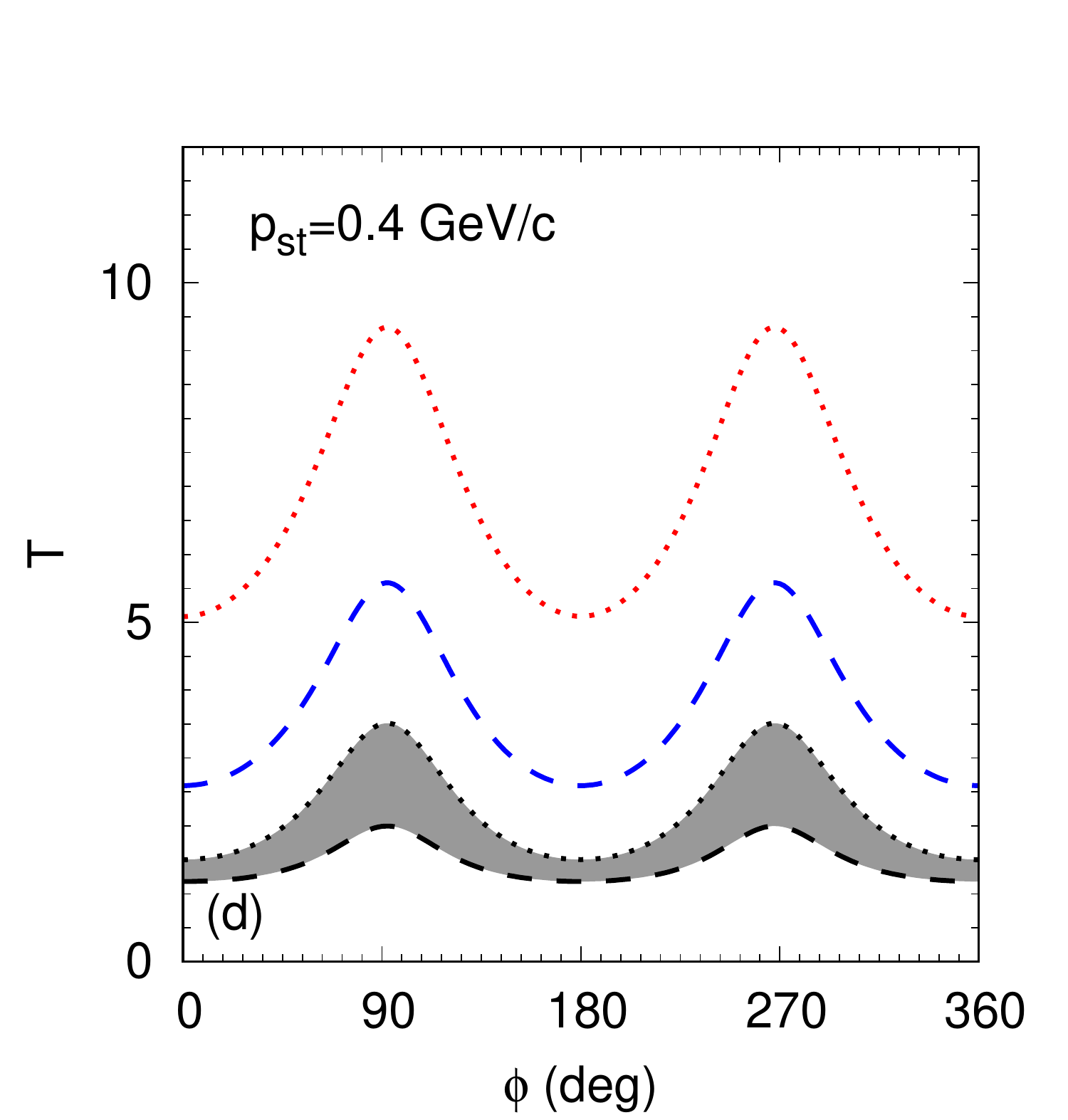} &
   \includegraphics[scale = 0.35]{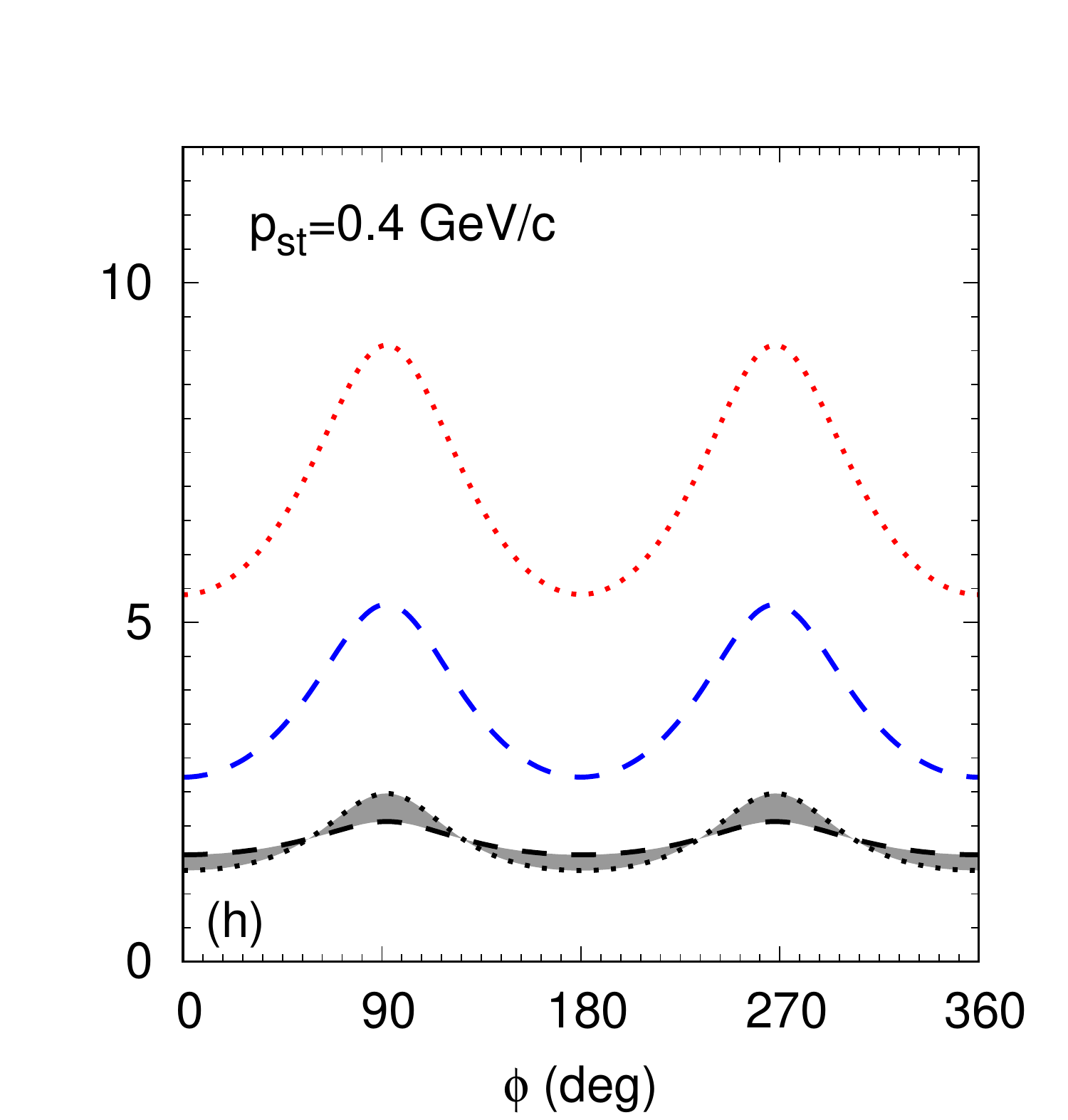} &
   \includegraphics[scale = 0.35]{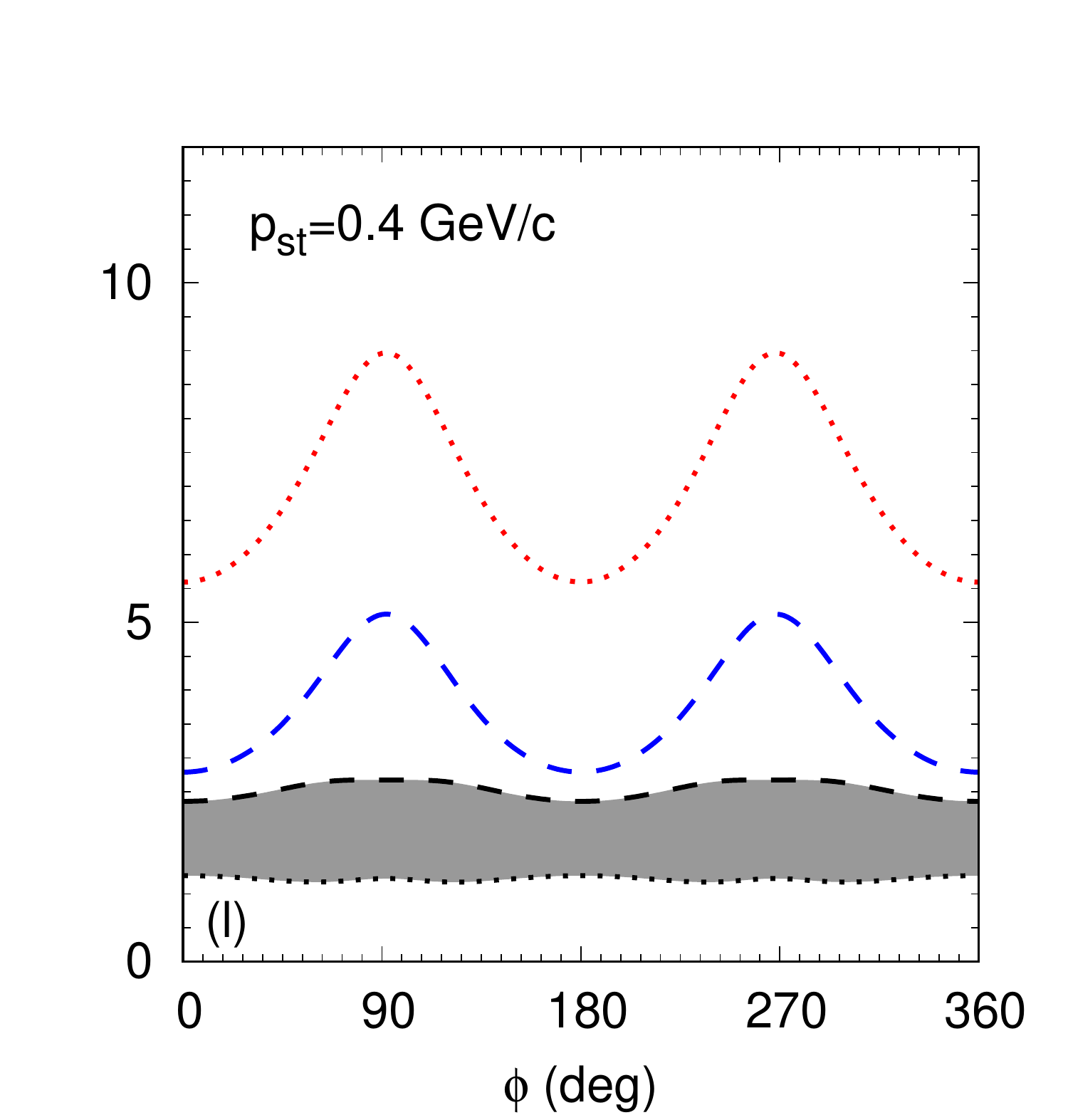} \\
   \end{tabular}
 \end{center}   
 \caption{\label{fig:T_50&65&75gevc_phiDep} The transparency $T$ for $pd \to ppn$ at $p_{\rm lab}=50$ GeV/c (left column),
   $p_{\rm lab}=65$ GeV/c (middle column), and $p_{\rm lab}=75$ GeV/c (right column) 
   as a function of relative azimuthal angle between the scattered proton and spectator neutron
   for $p_{st}=0.1, 0.2, 0.3$ and 0.4 GeV/c (from top to bottom panels).
   Line notations are the same as in Fig.~\ref{fig:T_180deg}.}
\end{figure}

Rescattering significantly affects the dependence of the transparency on the relative azimuthal angle $\phi$ between the scattered proton
and the spectator neutron, as shown in Figs.~\ref{fig:T_6&15&30gevc_phiDep} and \ref{fig:T_50&65&75gevc_phiDep}.
Note that, in the IA, the four-differential cross section can be considered as $\phi$-independent with an accuracy of about $10-20\%$.
Hence, $T$ reflects the azimuthal dependence of the cross section (not shown).
At $p_{\rm lab}=6$ and 15 GeV/c, transparencies calculated in the GEA  are rather close to those shown in Fig. 7 of Ref. \cite{Frankfurt:1996uz},
although we obtained larger maximum values at $p_{st}=0.4$ GeV/c, in particular, at the lowest beam momentum.
Transparencies at 6 and 15 GeV/c with CT effects are also quite close to those of Ref. \cite{Frankfurt:1996uz},
despite the fact that the latter do not include separation to the PLC and large-size configurations
and are calculated in a narrower range, $\Delta M^2=0.7\div1.1$ GeV$^2$.

The azimuthal dependence of $T$ is mostly determined by the rescattering amplitudes (b),(c) in Fig.~\ref{fig:diagr}.
It can be seen from Eq.(\ref{M^(b)_coord}) that the phase of the integrand is minimized when $\bvec{p}_s = \bvec{k}_t$
where $\bvec{k}_t \perp \bvec{p}_4$.
Since in the kinematics with $\alpha_s=1$ the spectator momentum is almost orthogonal to the beam axis, we conclude
that the phase is minimized and, hence, the rescattering amplitude is largest by absolute value for $\phi=90\degree$ and $270\degree$.
(Note that the beam momentum and the momenta of the outgoing protons are almost coplanar.)
In contrast, at $\phi=0\degree$ and $180\degree$, the spectator momentum always has a component along $\bvec{p}_4$ which leads
to the quickly oscillating phase as a function of $\tilde z$ resulting in the suppressed rescattering amplitude.
At small $p_{st}$'s, where the main effect of the rescattering amplitude is due to its interference with the IA amplitude,
this leads to the increased (reduced) absorption at $\phi=90\degree$ and $270\degree$ ($\phi=0\degree$ and $180\degree$).
Similarly, at large $p_{st}$'s, when the IA amplitude is small, the square of the rescattering amplitude leads to an increased (decreased)
yield at $\phi=90\degree$ and $270\degree$ ($\phi=0\degree$ and $180\degree$).

The influence of double rescattering amplitudes (g) and (h) of Fig.~\ref{fig:diagr} is quite modest at $p_{st} \ltsim 0.2$ GeV/c,
but becomes stronger at larger transverse momentum of the spectator neutron.
These amplitudes have to be taken into account at $p_{st} \gtsim 0.3$ GeV/c
since in their absence $T$ will be overestimated by $\sim 50\%$.
These observations agree with Ref. \cite{Frankfurt:1996uz}.

In calculations with CT, the azimuthal dependence of $T$ saturates between $p_{\rm lab}=30$ and 50 GeV/c,
in accordance with $p_{st}$-dependence, and begins to change between 50 and 75 GeV/c towards isotropy.

\subsection{Polarized deuteron}
\label{pol}

\begin{figure}
 \begin{center} 
   \begin{tabular}{cc}
  \includegraphics[scale = 0.40]{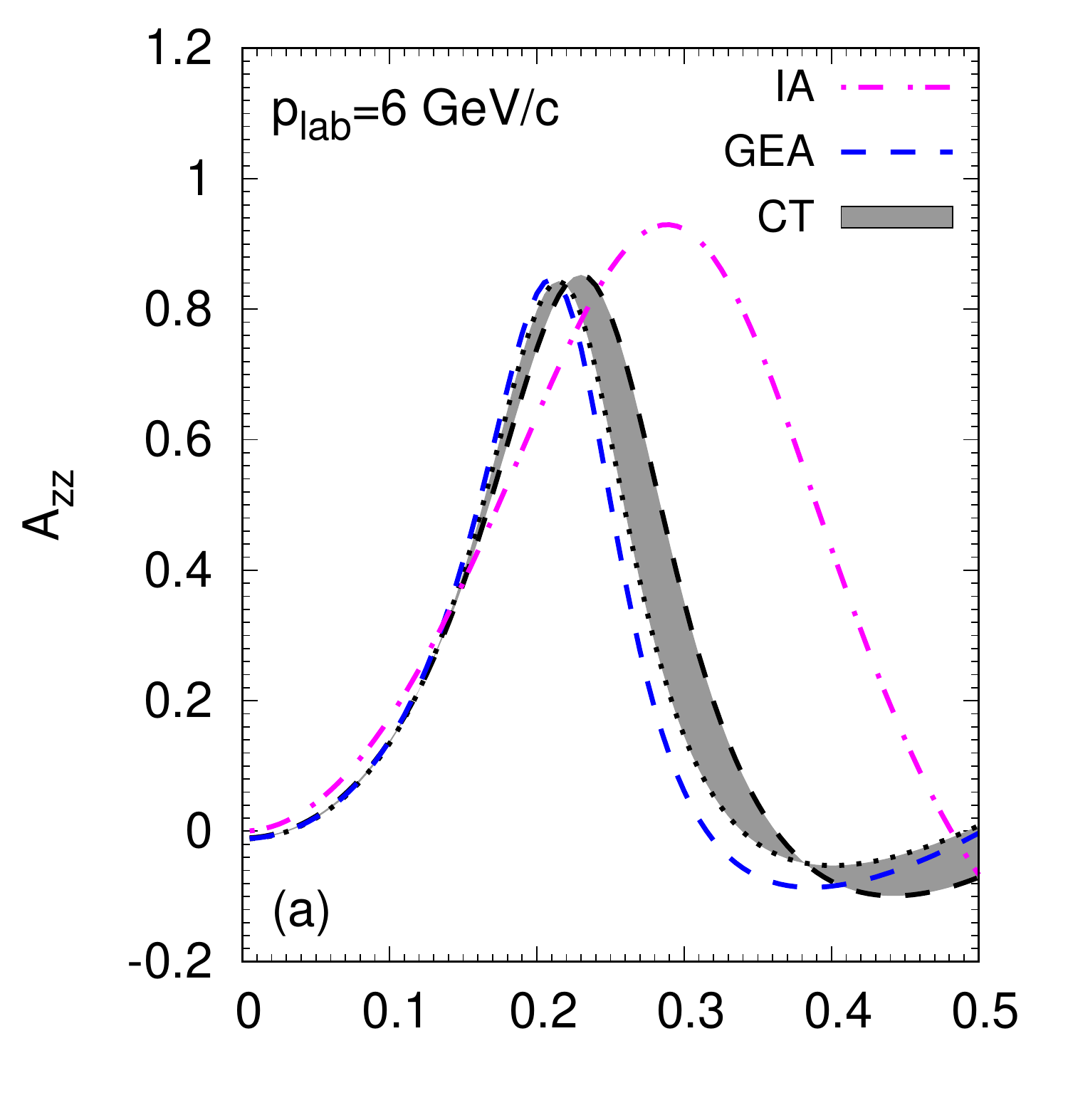} &
  \includegraphics[scale = 0.40]{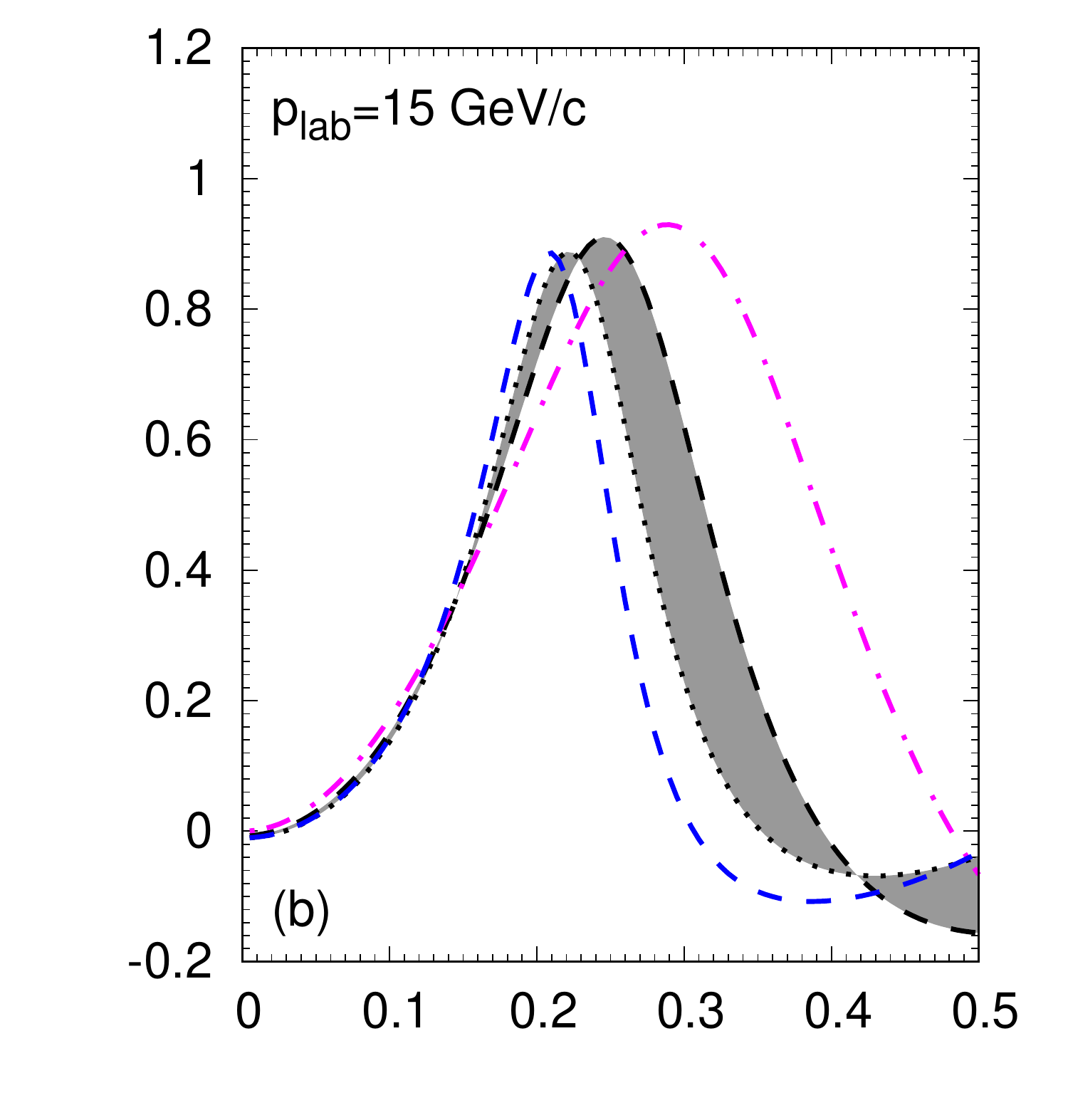}\\
  \includegraphics[scale = 0.40]{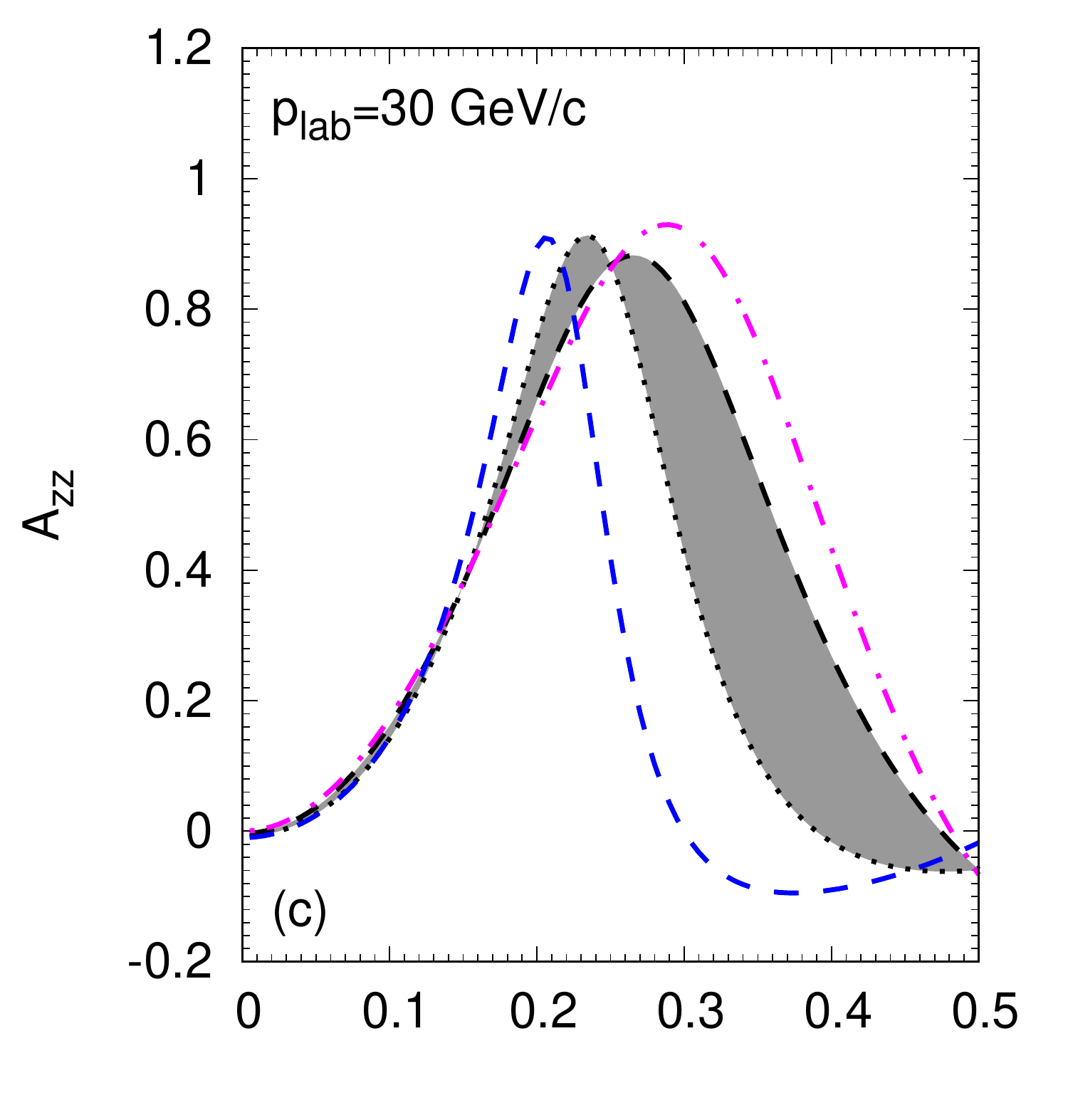} &
  \includegraphics[scale = 0.40]{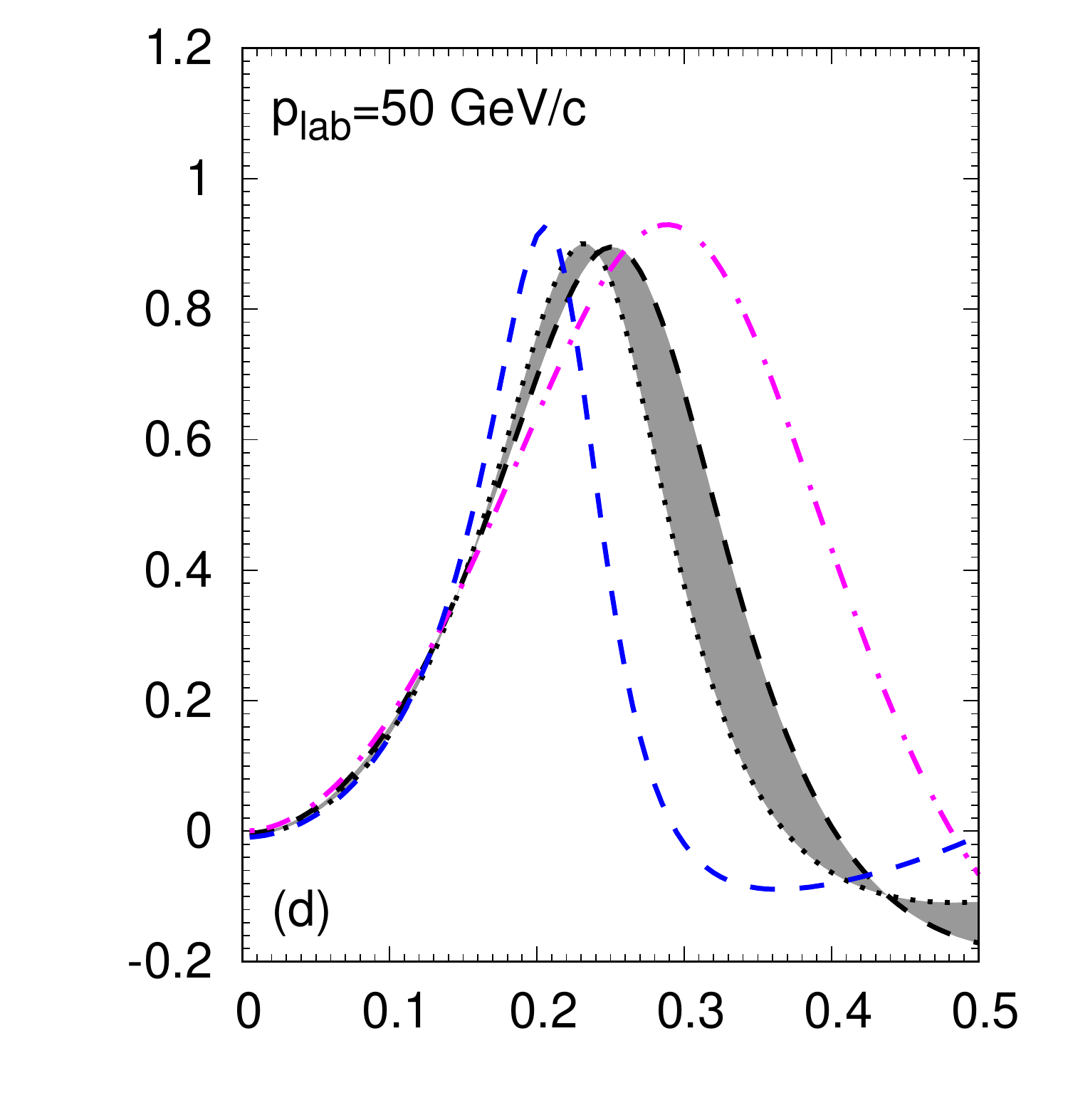} \\
  \includegraphics[scale = 0.40]{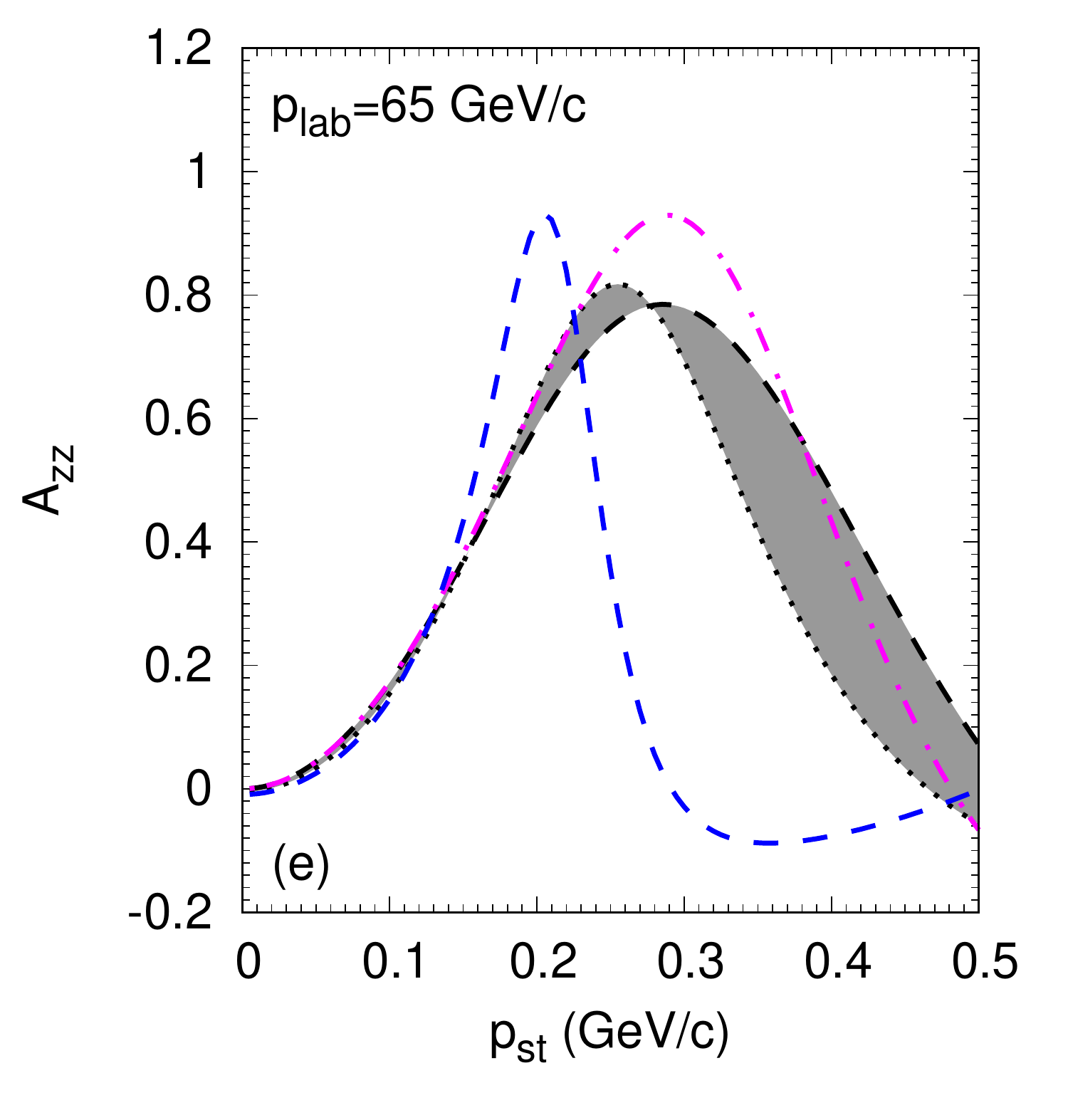} &
  \includegraphics[scale = 0.40]{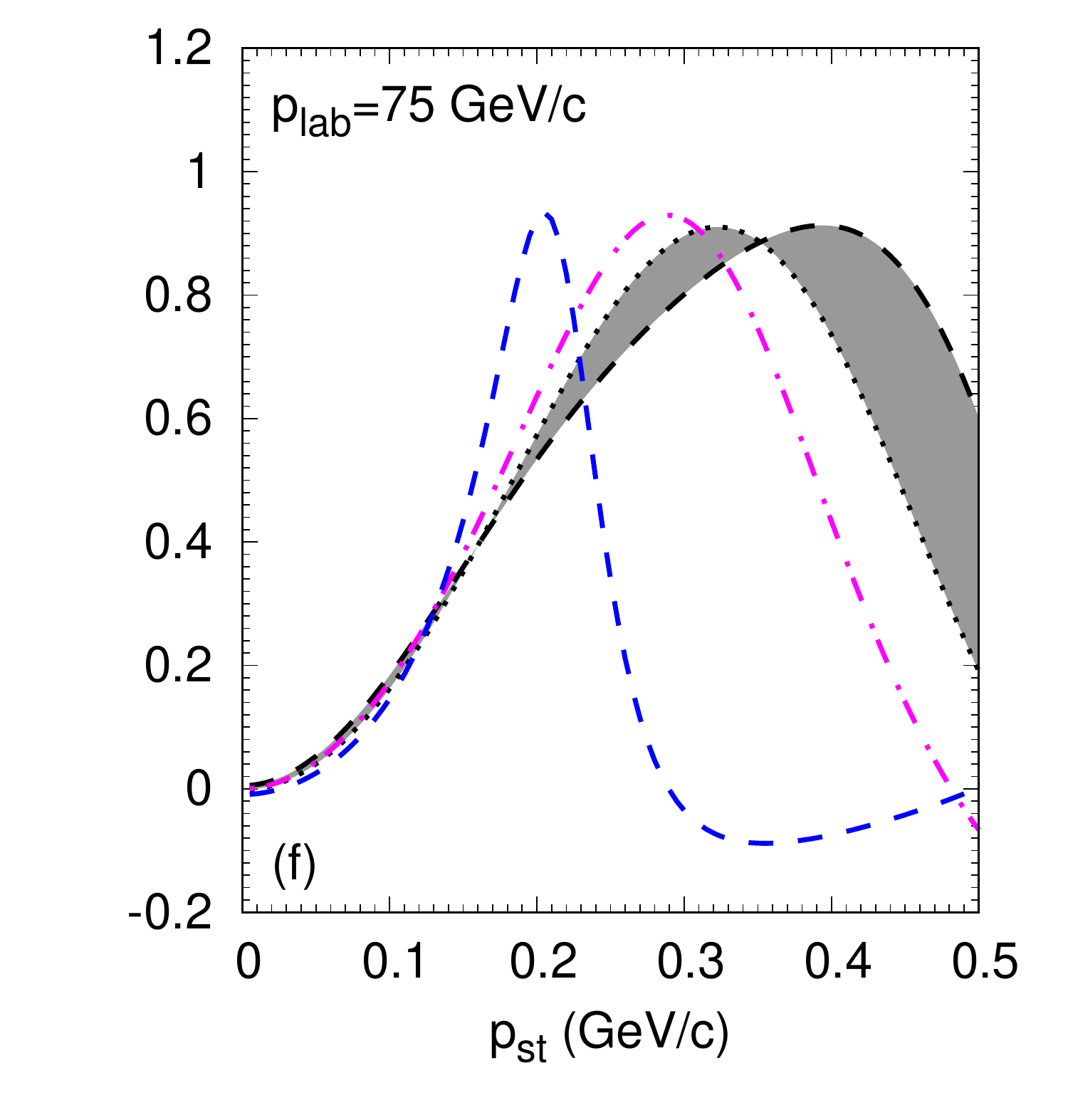}\\ 
   \end{tabular}
 \end{center}
  \caption{\label{fig:Azz_180deg} Tensor analyzing power, Eq.(\ref{A_zz}), for $pd \to ppn$ as a function
    of transverse momentum of spectator neutron. The kinematics with $\alpha_s=1$, $\phi=180\degree$,
    and $\Theta_{c.m.}=90\degree$ is chosen.
    Different panels correspond to different values of beam momentum as indicated.
    Line notations are the same as in Fig.~\ref{fig:sig_180deg}.}
\end{figure}

Fig.~\ref{fig:Azz_180deg} shows the tensor analyzing power $A_{zz}$ vs the spectator transverse momentum.
In the IA, $A_{zz}$ can be calculated from Eq.(\ref{A_zz^IA}) that does not depend neither on $p_{\rm lab}$ nor on $\phi$ and
-- since $\alpha_s=1$ is set -- is fully determined by the transverse momentum of the spectator.
This results in a peak at $p_{st}=0.3$ GeV/c. The ISI/FSI introduced in the framework of the GEA
lead to the pronounced change of the $p_{st}$-dependence by shifting the peak down to $p_{st}=0.2$ GeV/c
and reducing its width. However, the shape of the $p_{st}$-dependence varies with beam momentum very weakly.
In contrast, in the calculation with CT, the function $A_{zz}(p_{st})$ strongly varies with the beam momentum.
The strongest sensitivity to the ISI/FSI and CT is visible in the interval of $p_{st}=0.3-0.4$ GeV/c
where the GEA predicts negative values of $A_{zz} \simeq -0.1$ while the IA gives a large positive $A_{zz}$.

In Ref. \cite{Frankfurt:1994kt}, the deuteron polarization effects in $d(e,e^\prime p)n$ reaction with $Q^2 =1-10$ GeV$^2$
have been studied theoretically. This $Q^2$ range with simultaneous requirement
of $x=1$ selects the produced proton in the momentum range between 1 and 6 GeV/c and the neutron with momentum perpendicular
to the momentum of a virtual photon. The asymmetry calculated in Ref. \cite{Frankfurt:1994kt}
\footnote{In the notation of Ref. \cite{Frankfurt:1994kt}, $A_{zz} \equiv A_d$ is the asymmetry and
$A_{zz}^{IA} = \rho_{20}^d(p_s,p_s)/\rho_d(p_s,p_s)$ is the ratio of the polarized and unpolarized density matrices.}
for the neutron production at $90\degree$ 
in the Glauber approximation falls substantially below the asymmetry calculated in the plane-wave IA for neutron
momenta between 0.3 and 0.4 GeV/c, see Fig.~7a,b in Ref. \cite{Frankfurt:1994kt} which is in a qualitative agreement
with our Fig.~\ref{fig:Azz_180deg}. 

\begin{figure}
 \begin{center} 
   \begin{tabular}{ccc}
   \vspace{-1cm}
   \includegraphics[scale = 0.35]{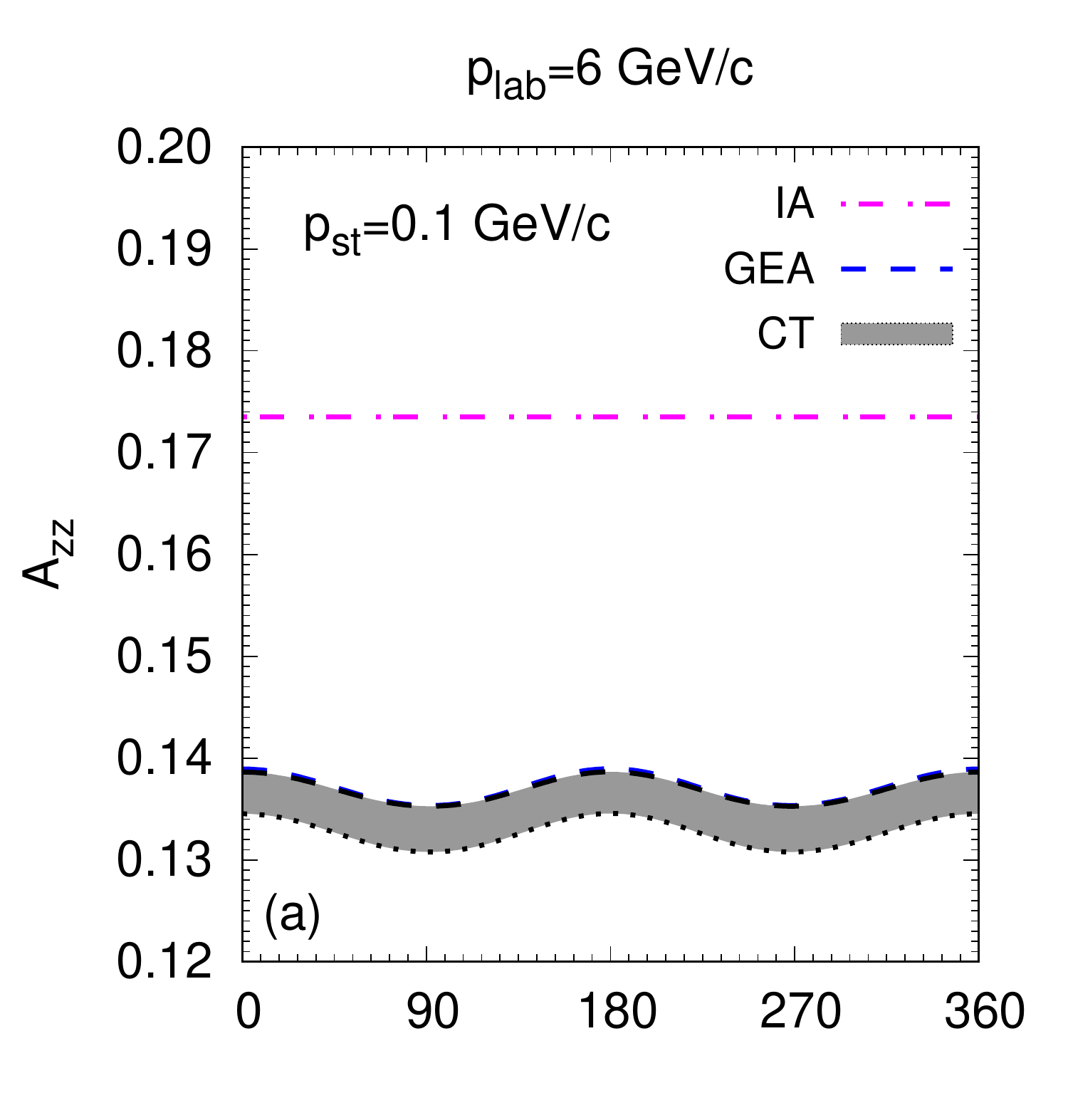} &
   \includegraphics[scale = 0.35]{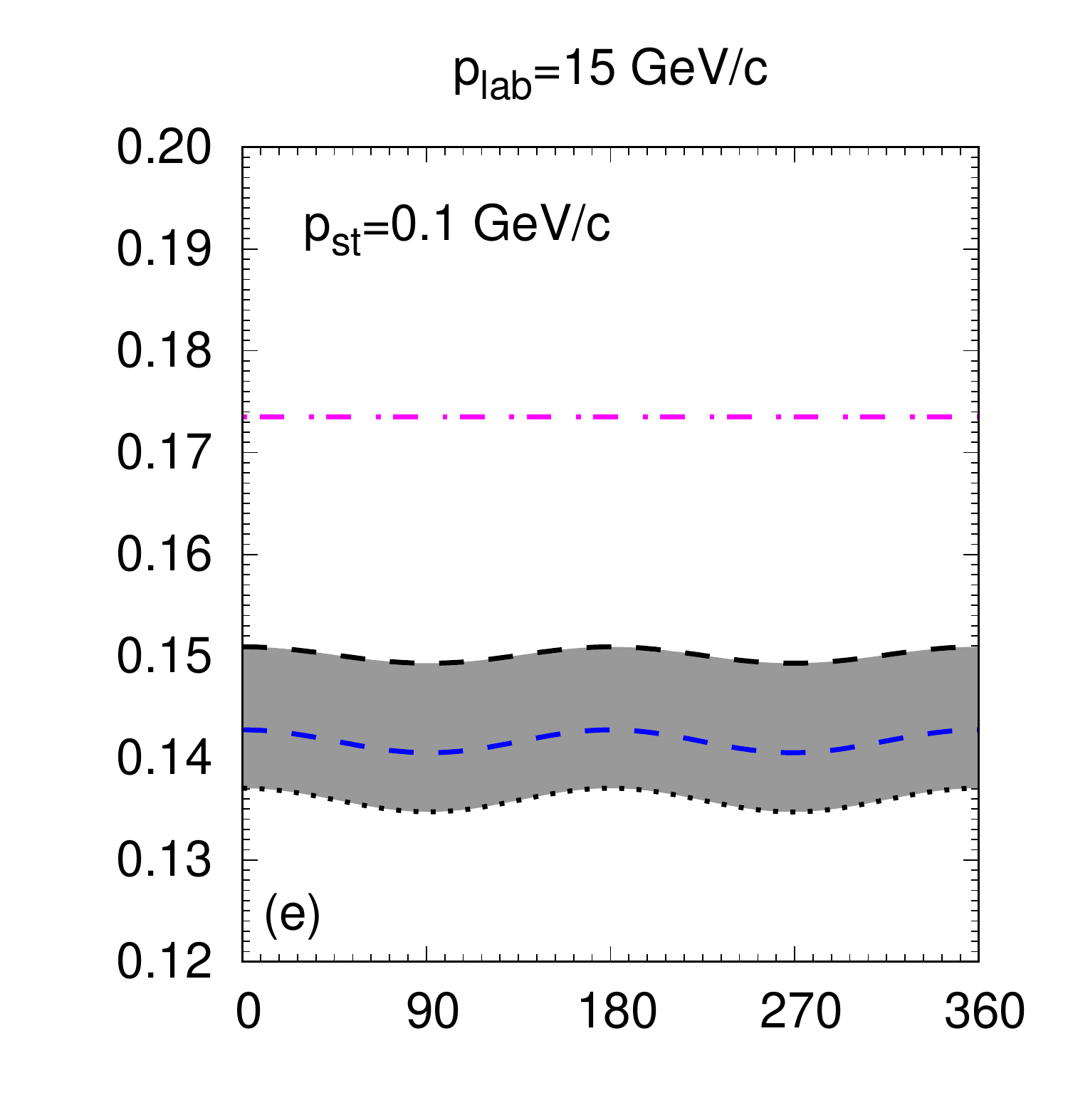} &
   \includegraphics[scale = 0.35]{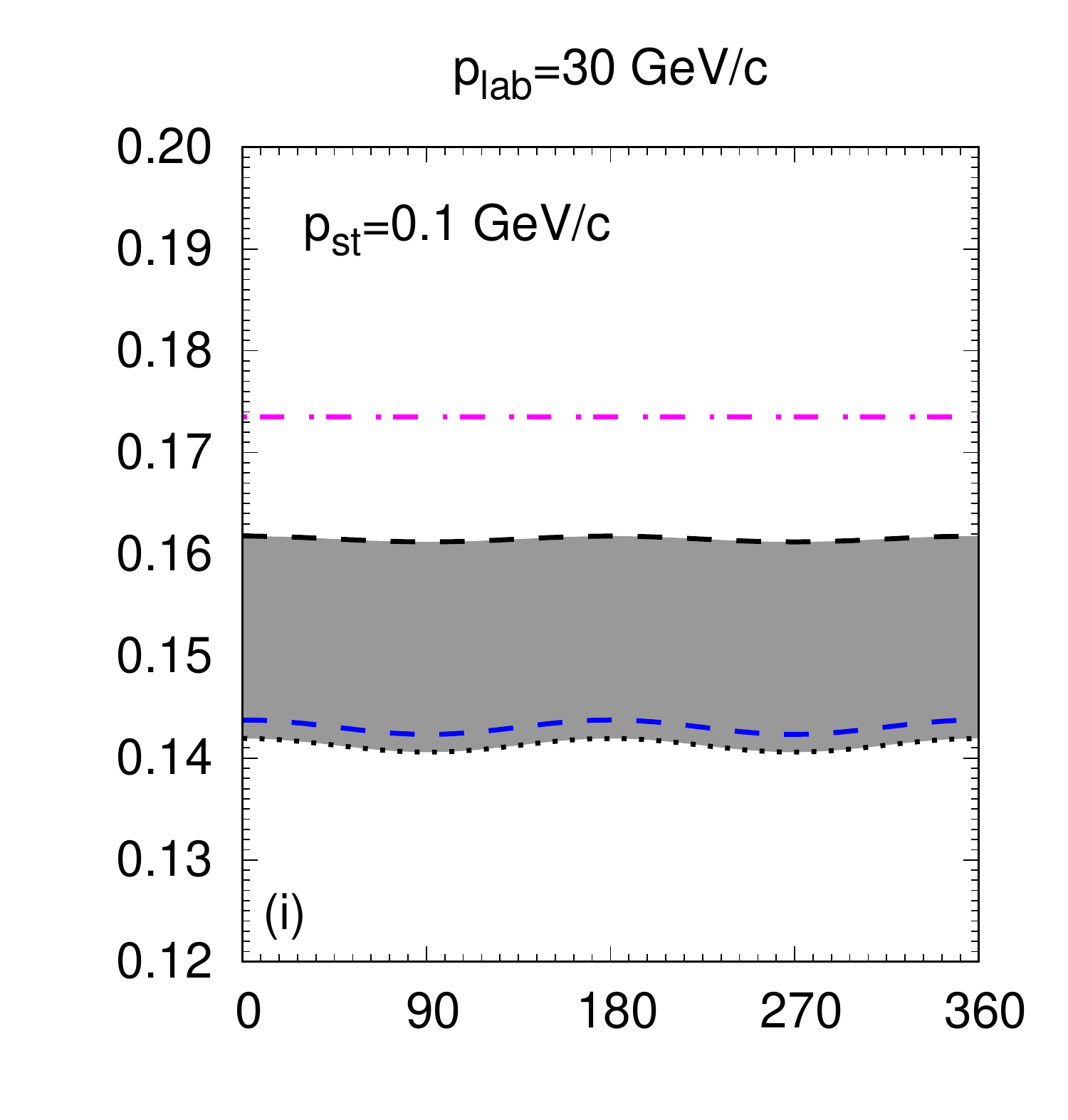} \\
   \vspace{-1cm}
   \includegraphics[scale = 0.35]{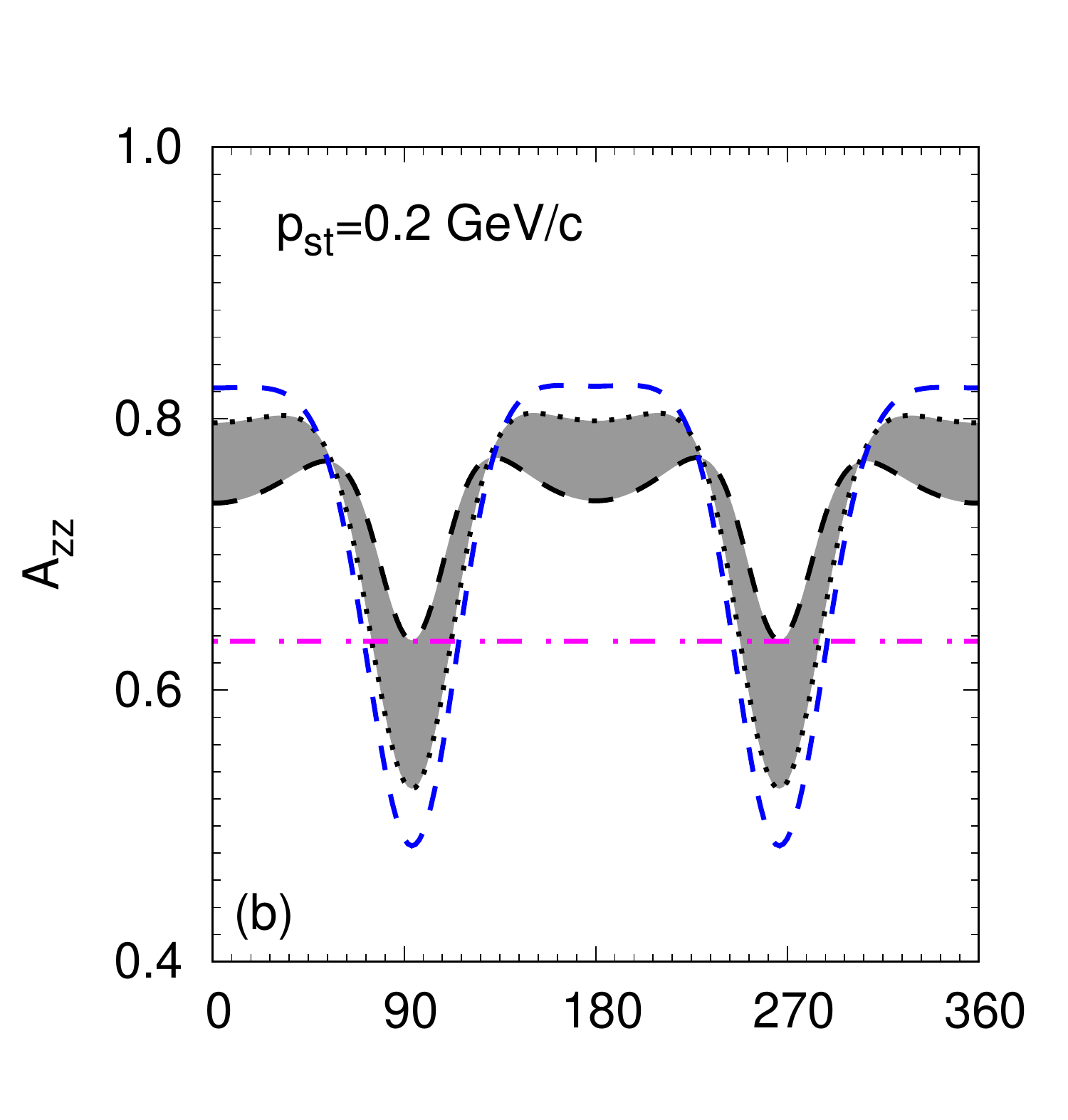} &
   \includegraphics[scale = 0.35]{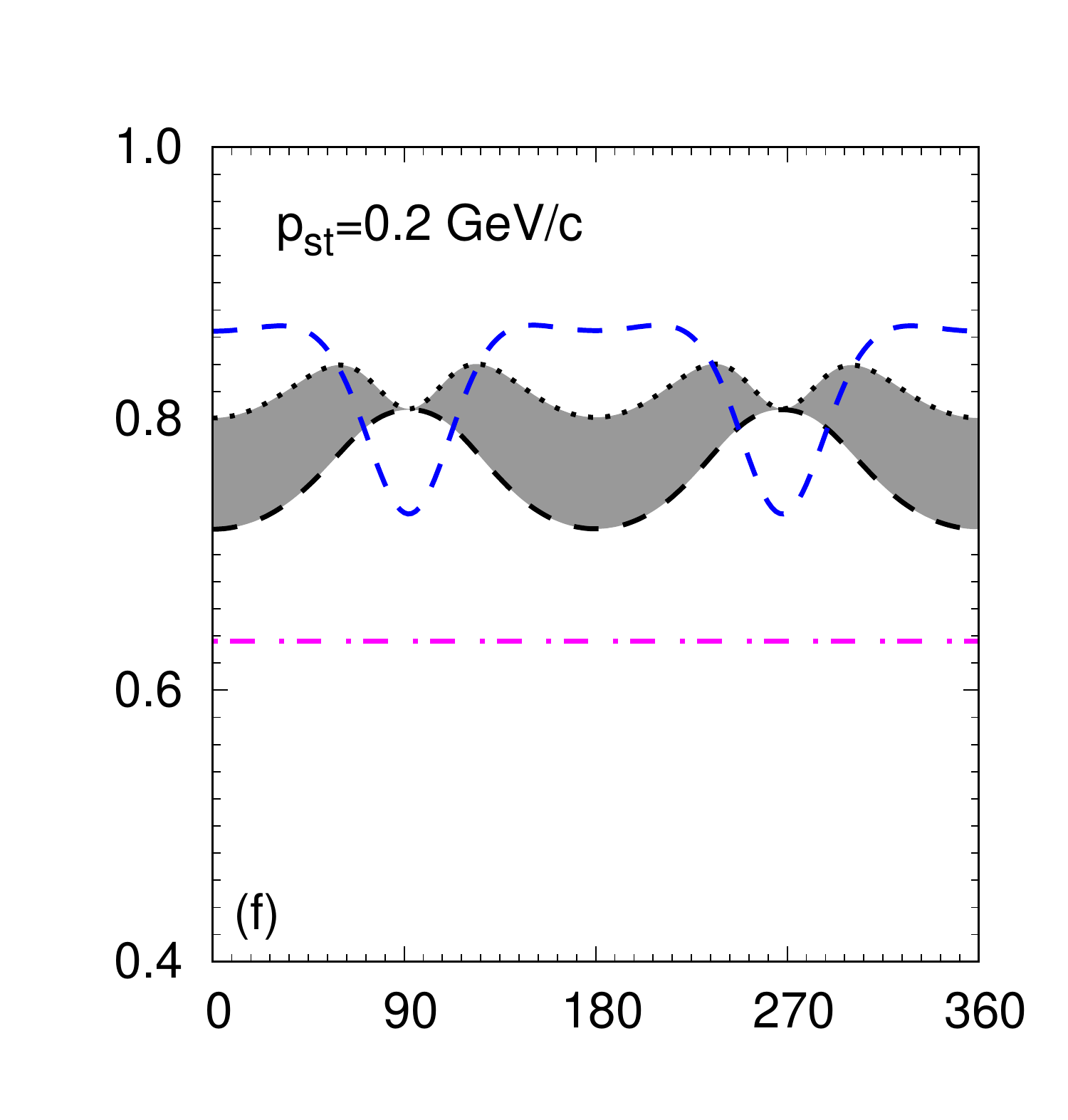} &
   \includegraphics[scale = 0.35]{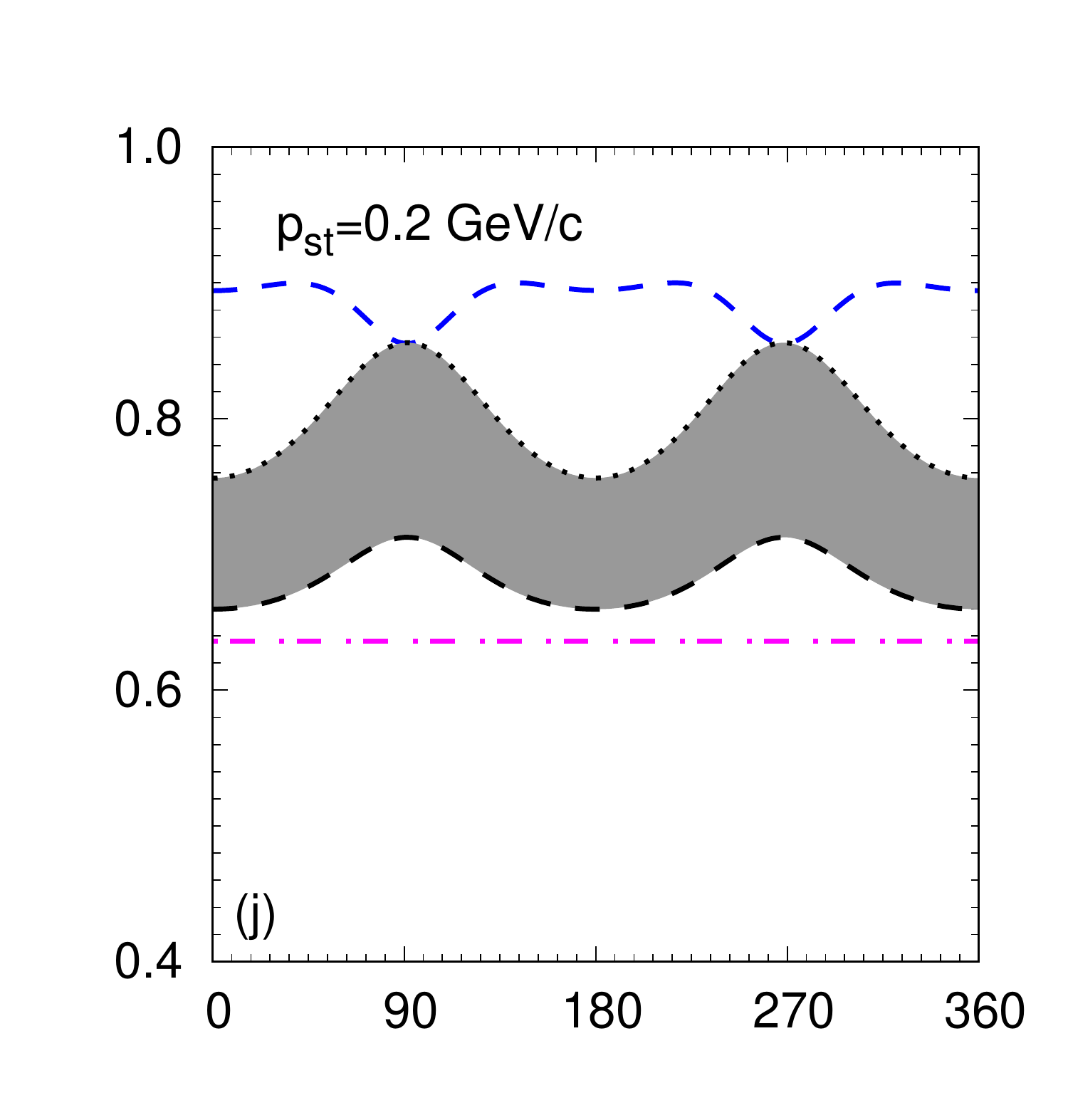} \\  
   \vspace{-1cm}
   \includegraphics[scale = 0.35]{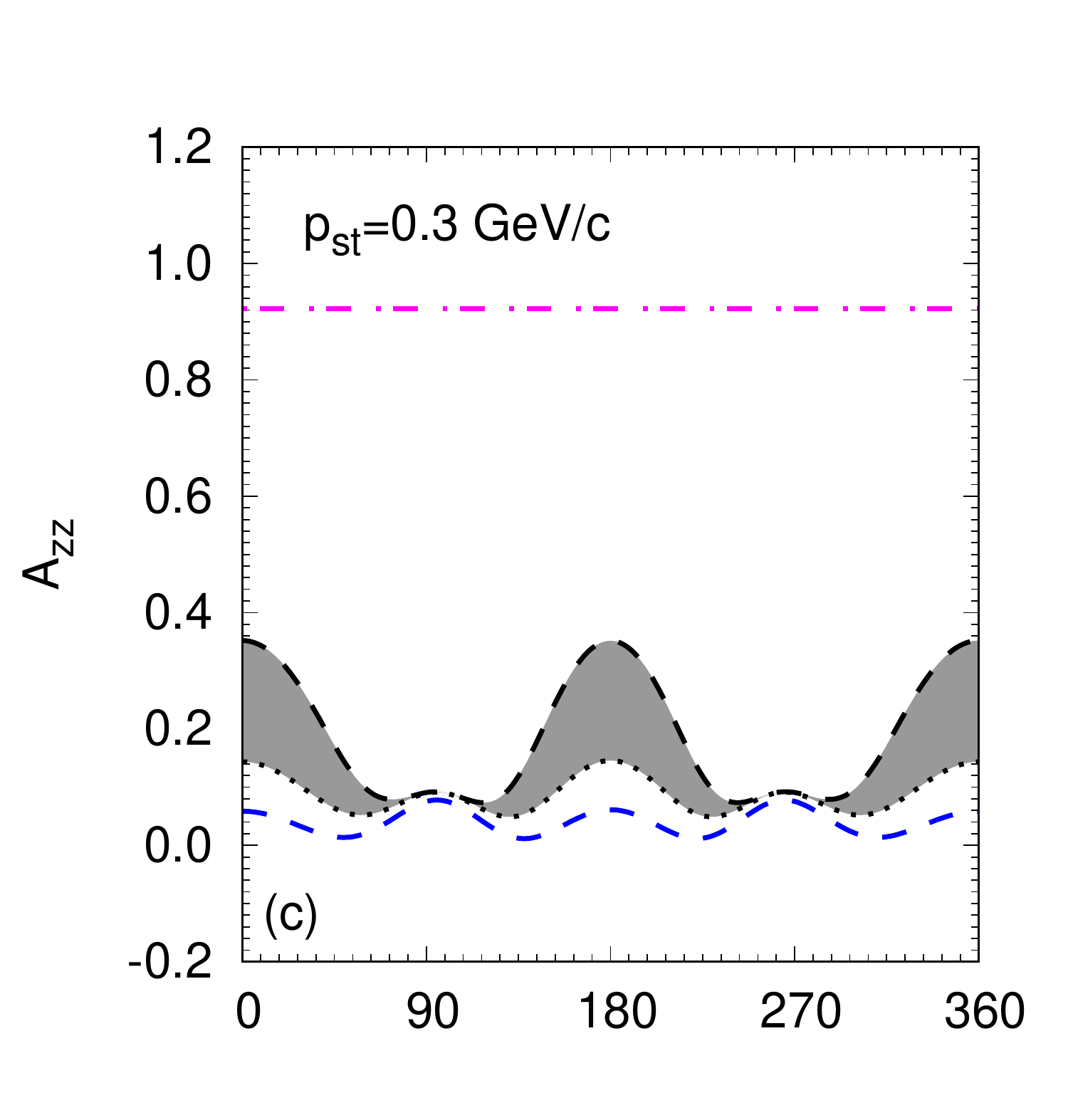} &
   \includegraphics[scale = 0.35]{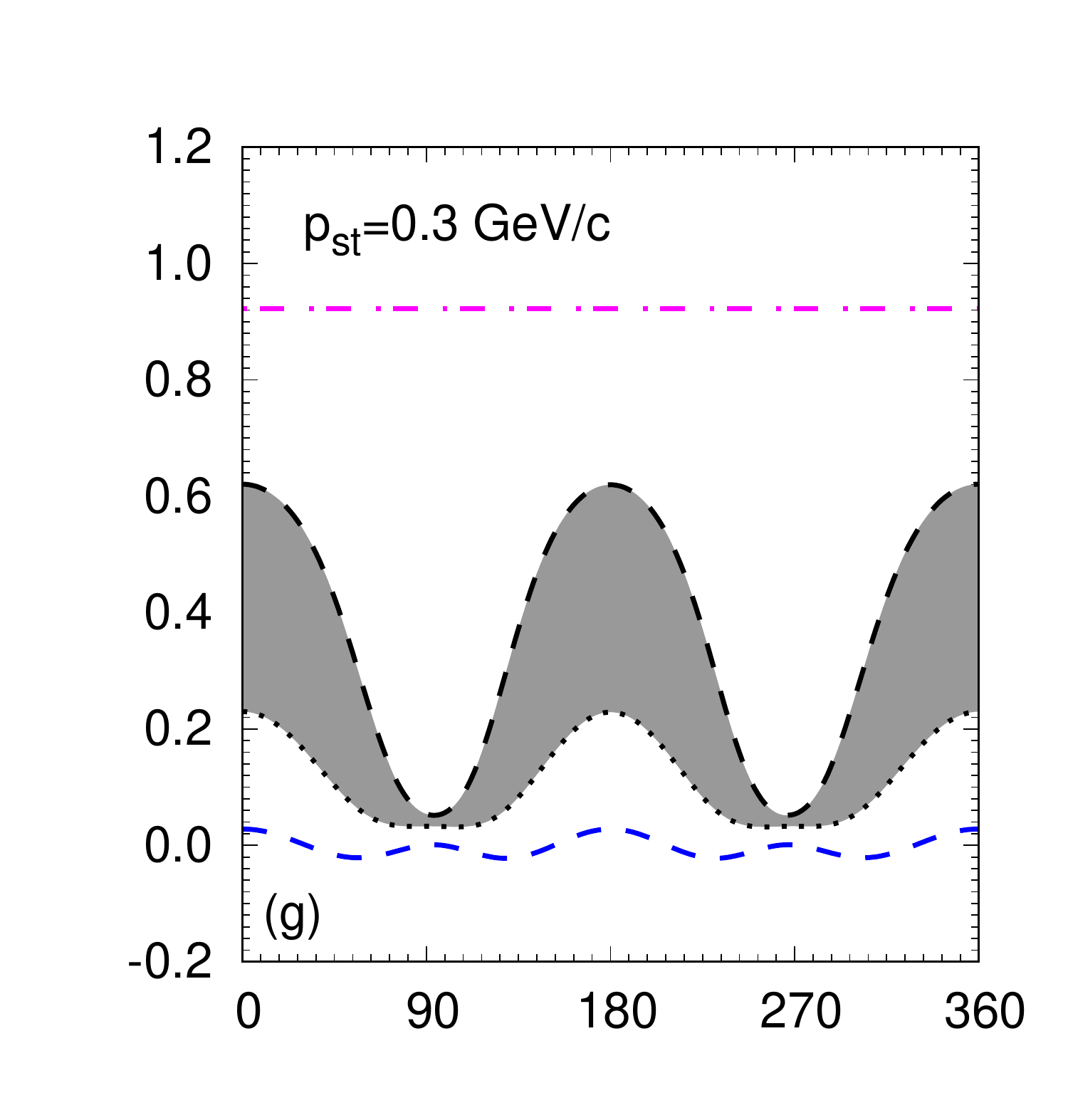} &
   \includegraphics[scale = 0.35]{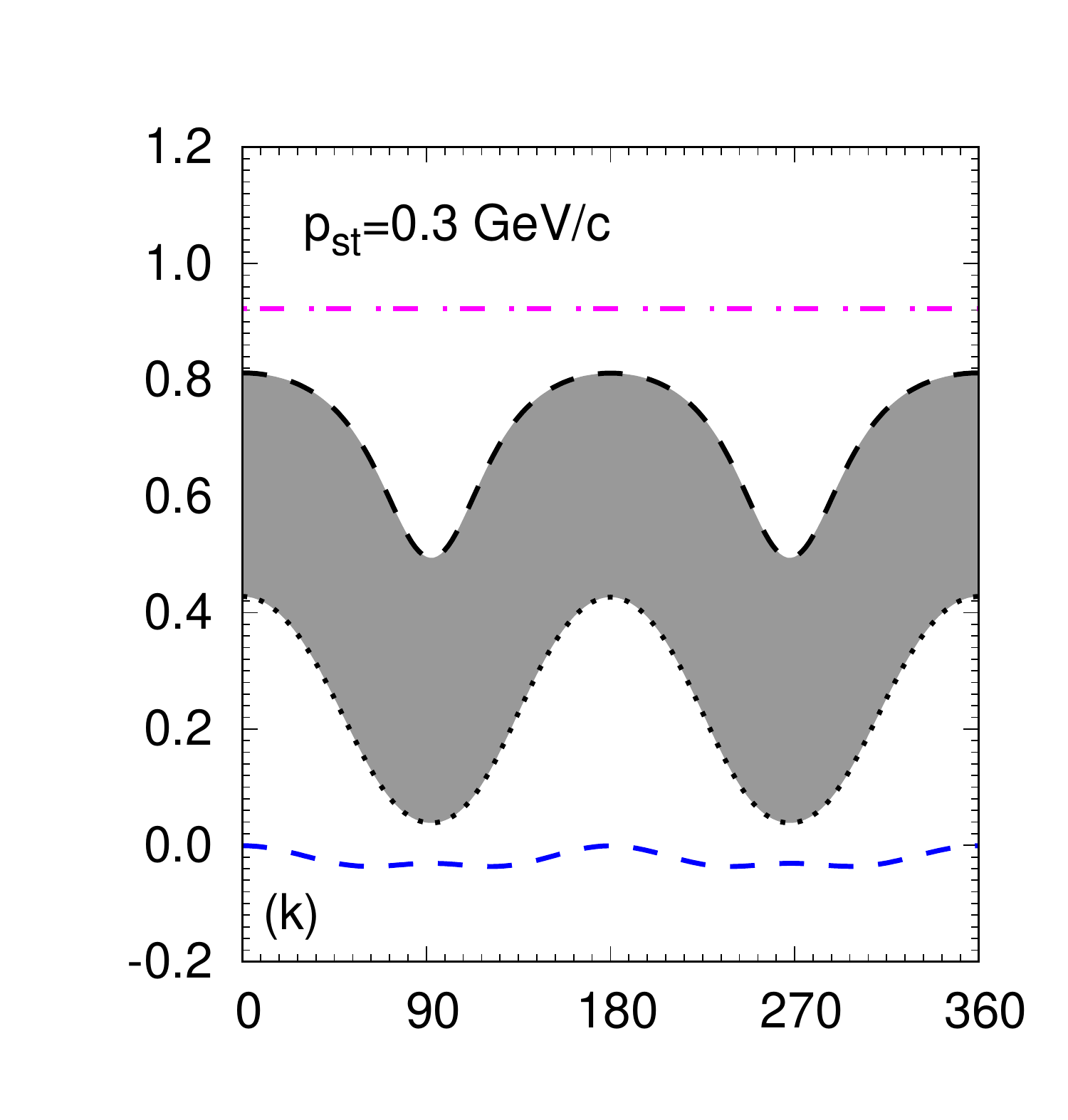} \\
   \includegraphics[scale = 0.35]{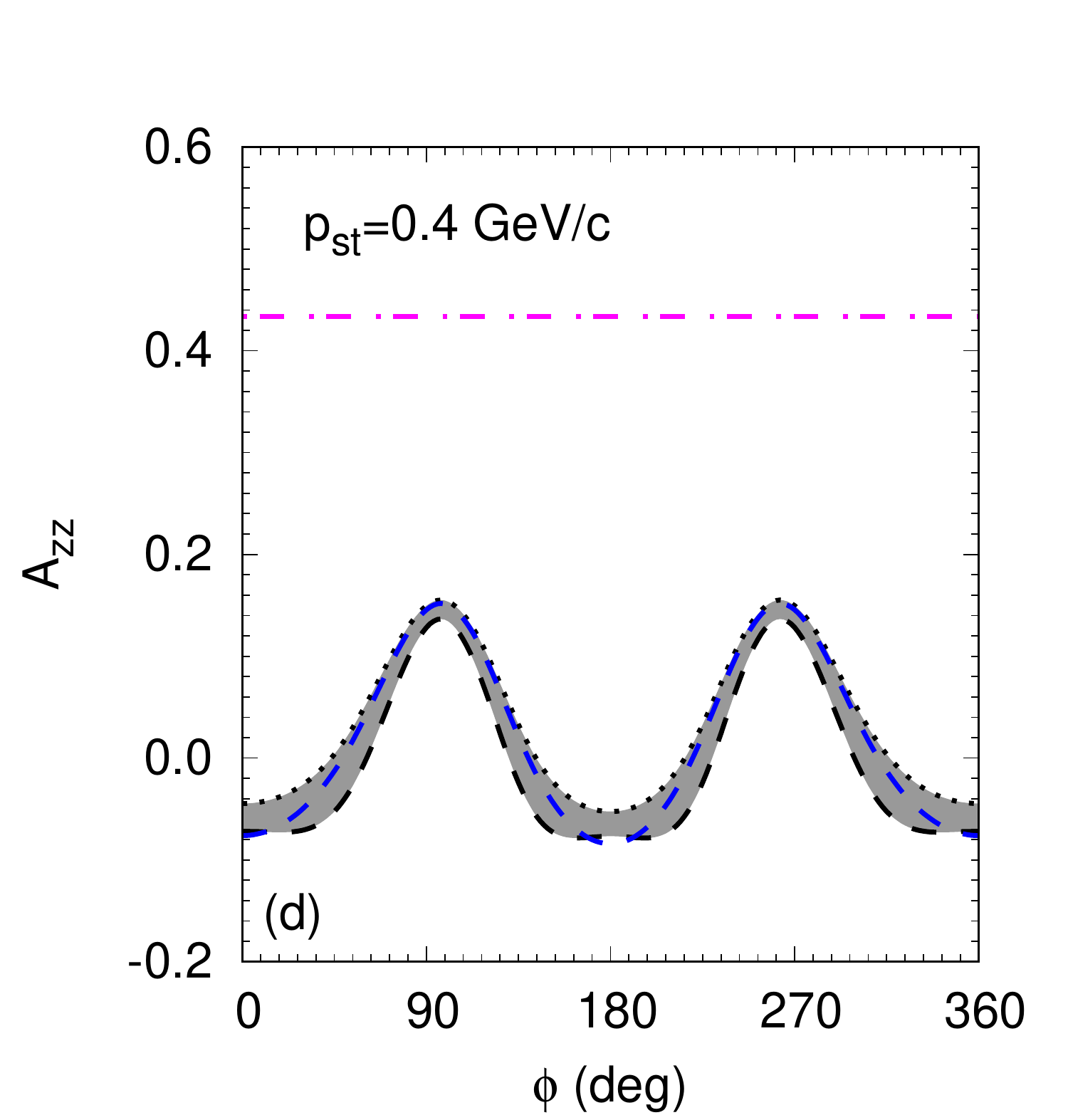} &
   \includegraphics[scale = 0.35]{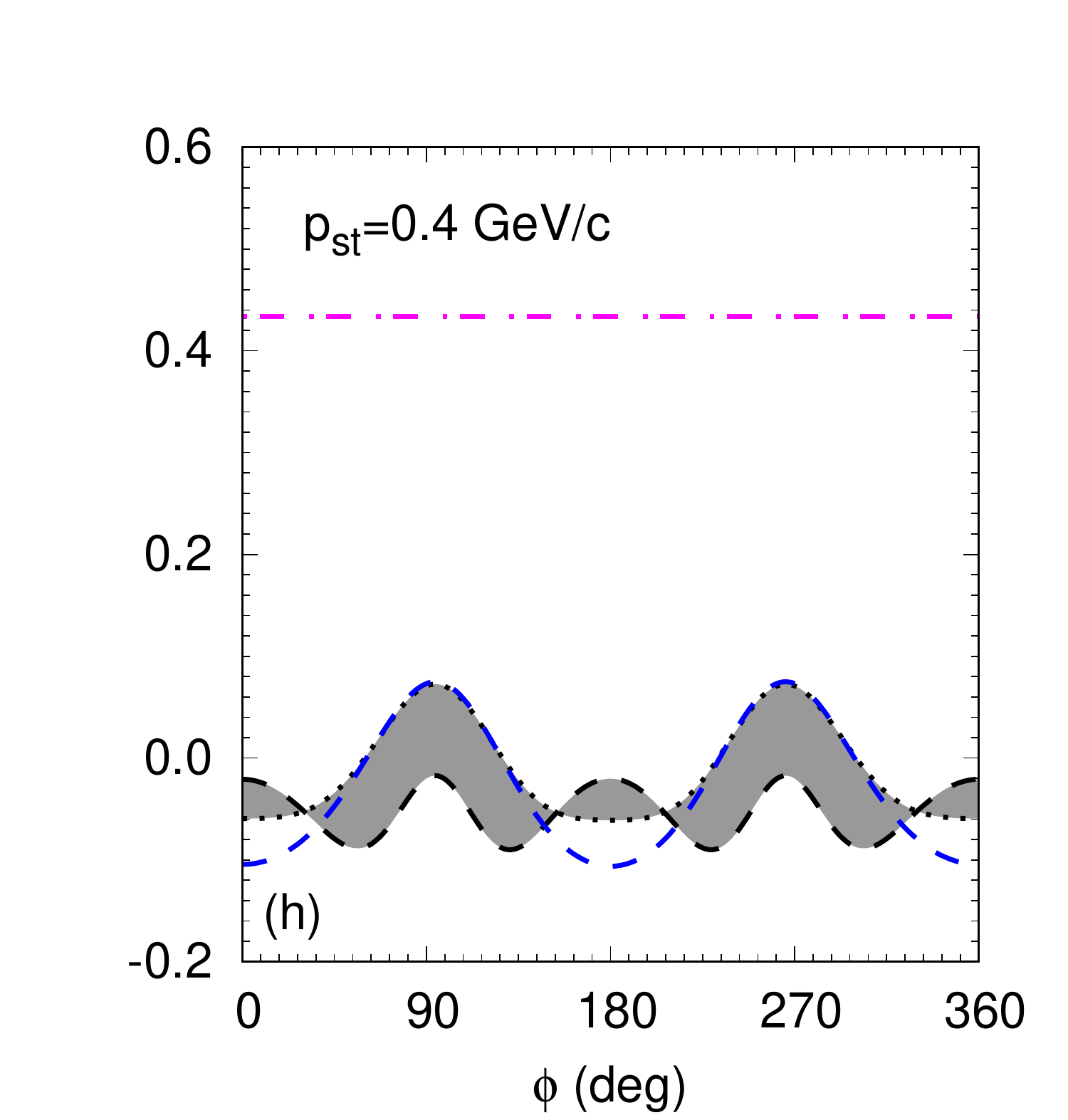} &
   \includegraphics[scale = 0.35]{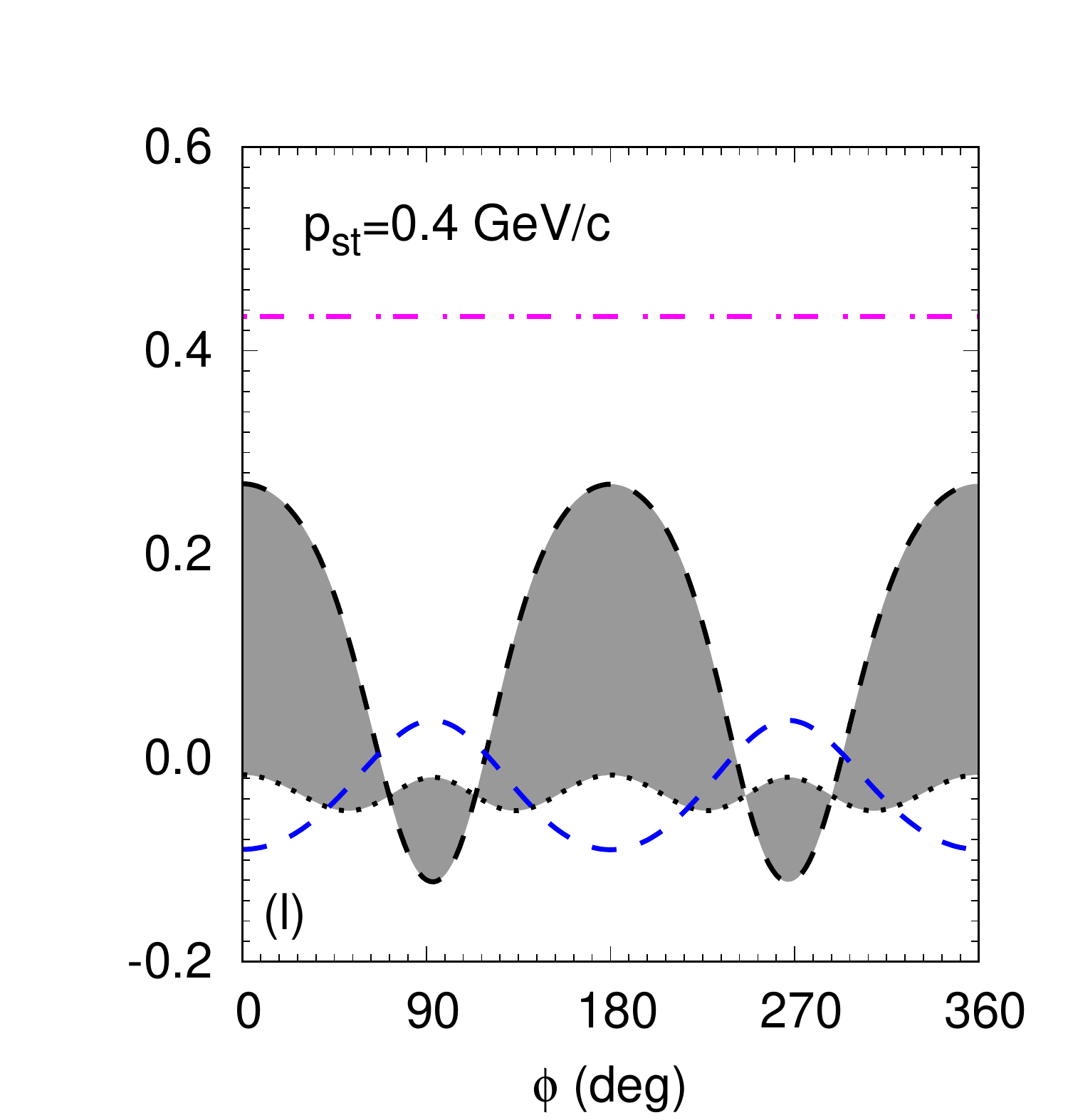} \\
   \end{tabular}
 \end{center}
 \caption{\label{fig:Azz_6&15&30gevc_phiDep} The tensor analyzing power $A_{zz}$ 
   for $pd \to ppn$ at $p_{\rm lab}=6$ GeV/c (left column),
   $p_{\rm lab}=15$ GeV/c (middle column), and $p_{\rm lab}=30$ GeV/c (right column) 
   as a function of relative azimuthal angle between the scattered proton and spectator neutron
   for $p_{st}=0.1, 0.2, 0.3$ and 0.4 GeV/c (from top to bottom panels).
   Line notations are the same as in Fig.~\ref{fig:sig_180deg}.}
\end{figure}

\begin{figure}
 \begin{center} 
   \begin{tabular}{ccc}
   \vspace{-1cm}
   \includegraphics[scale = 0.35]{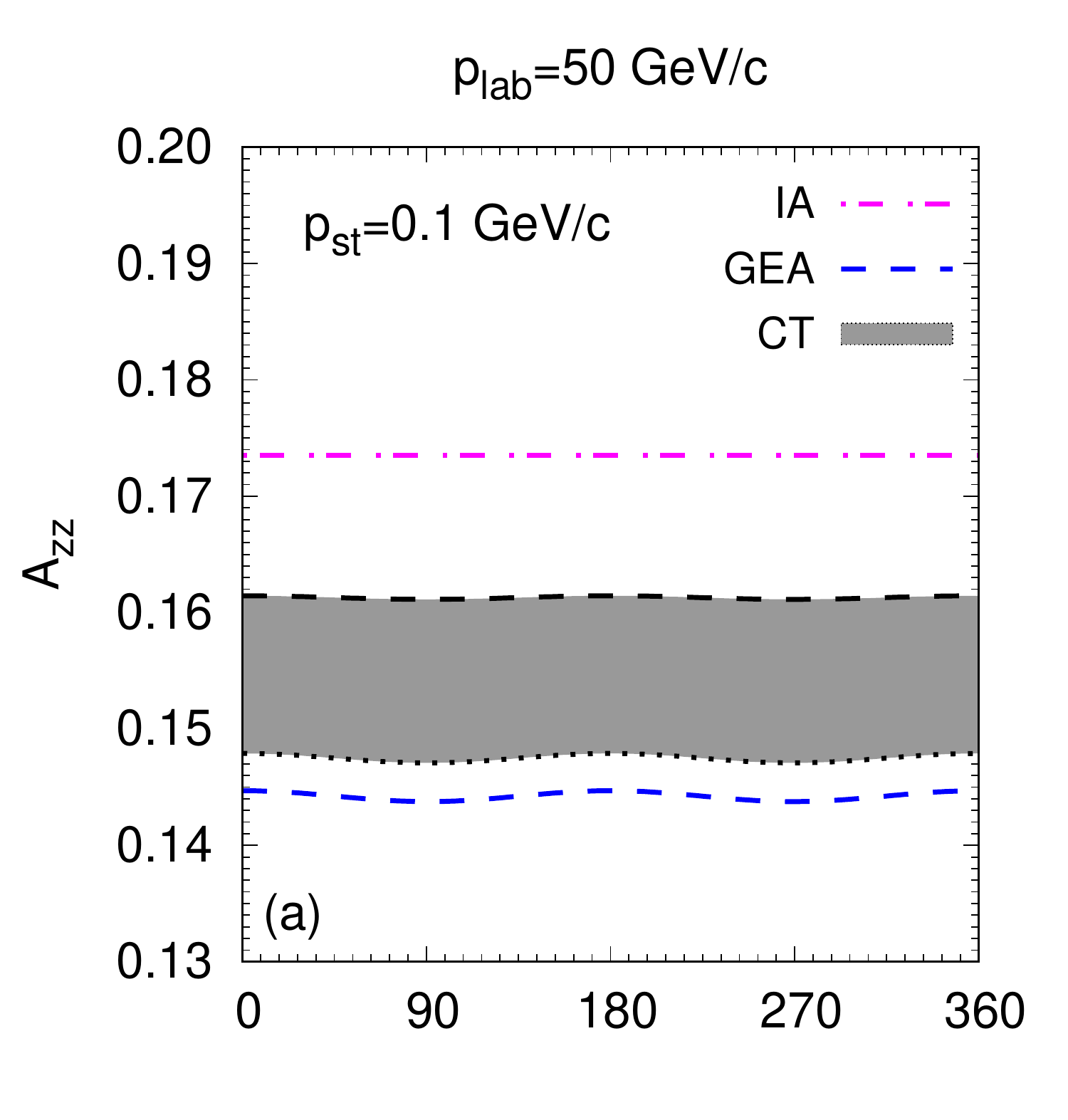} &
   \includegraphics[scale = 0.35]{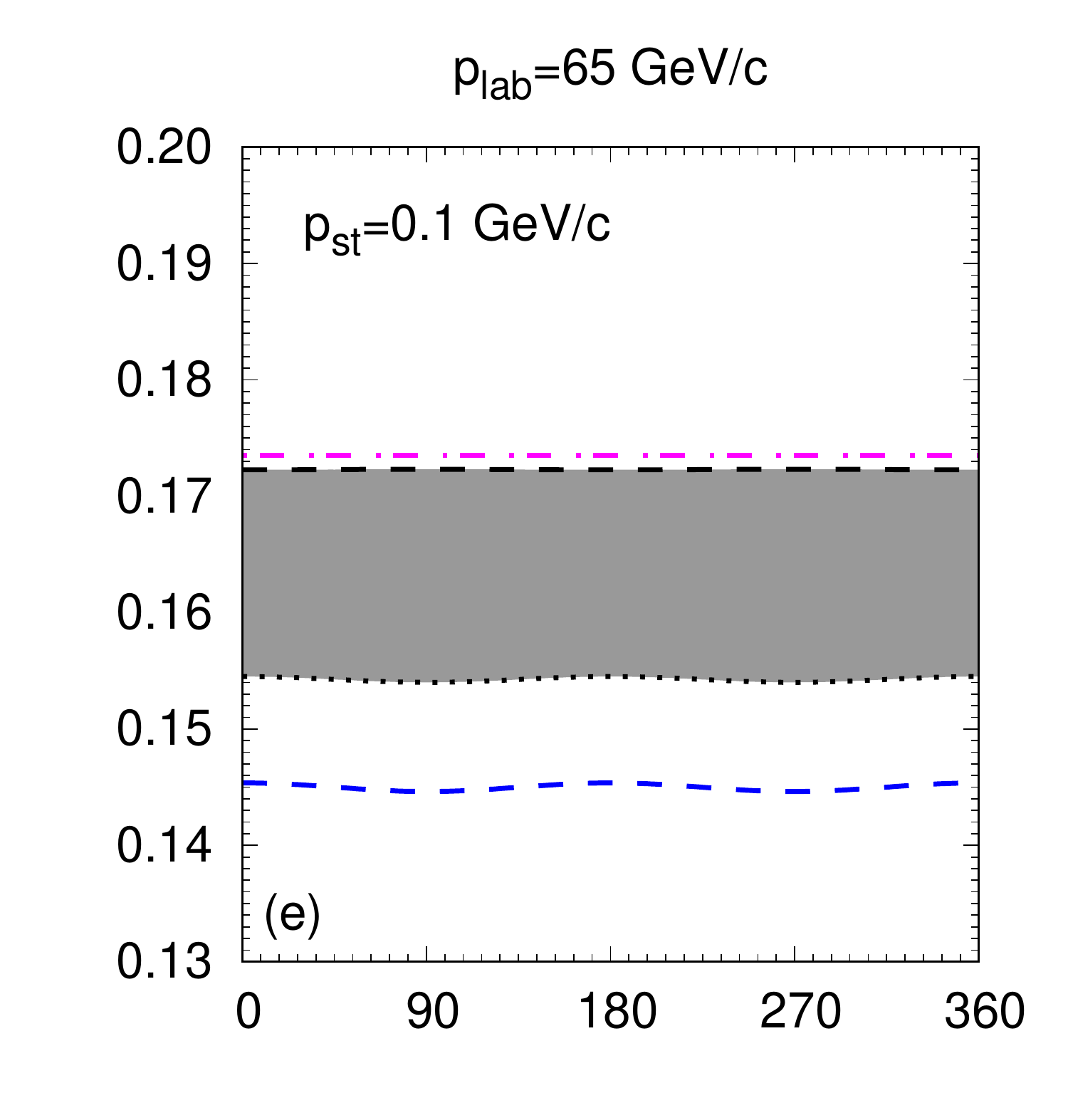} &
   \includegraphics[scale = 0.35]{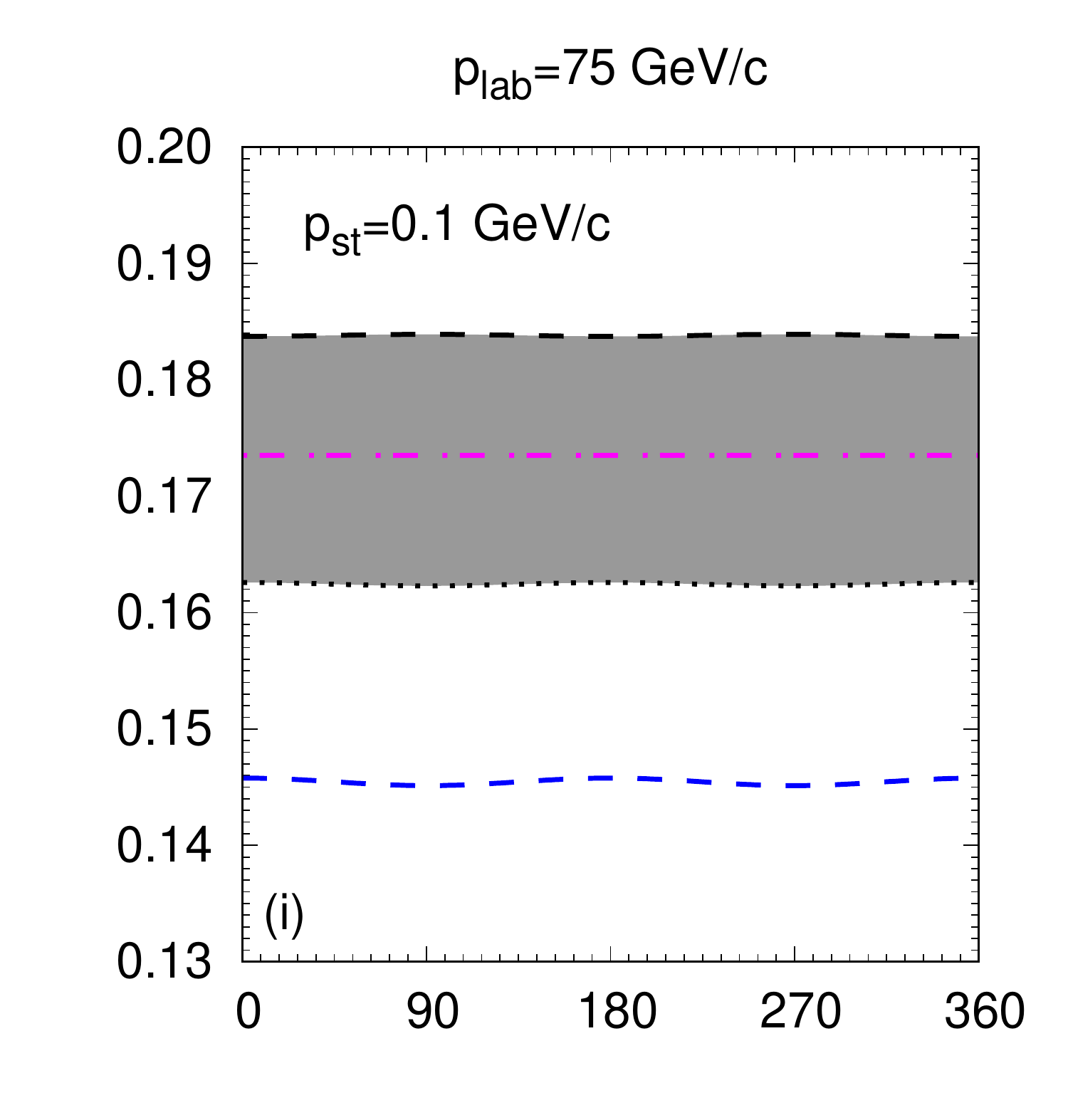} \\
   \vspace{-1cm}
   \includegraphics[scale = 0.35]{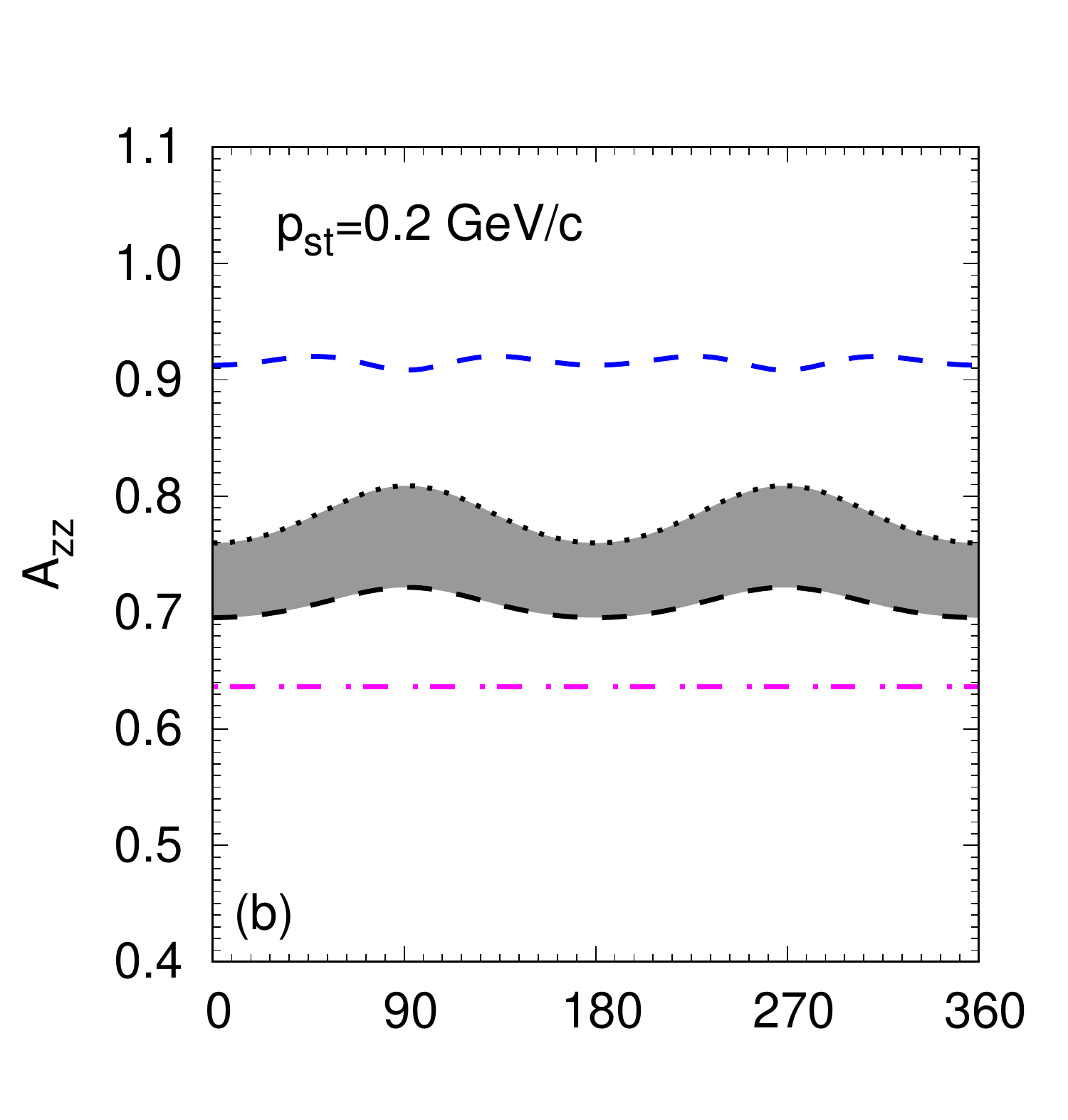} &
   \includegraphics[scale = 0.35]{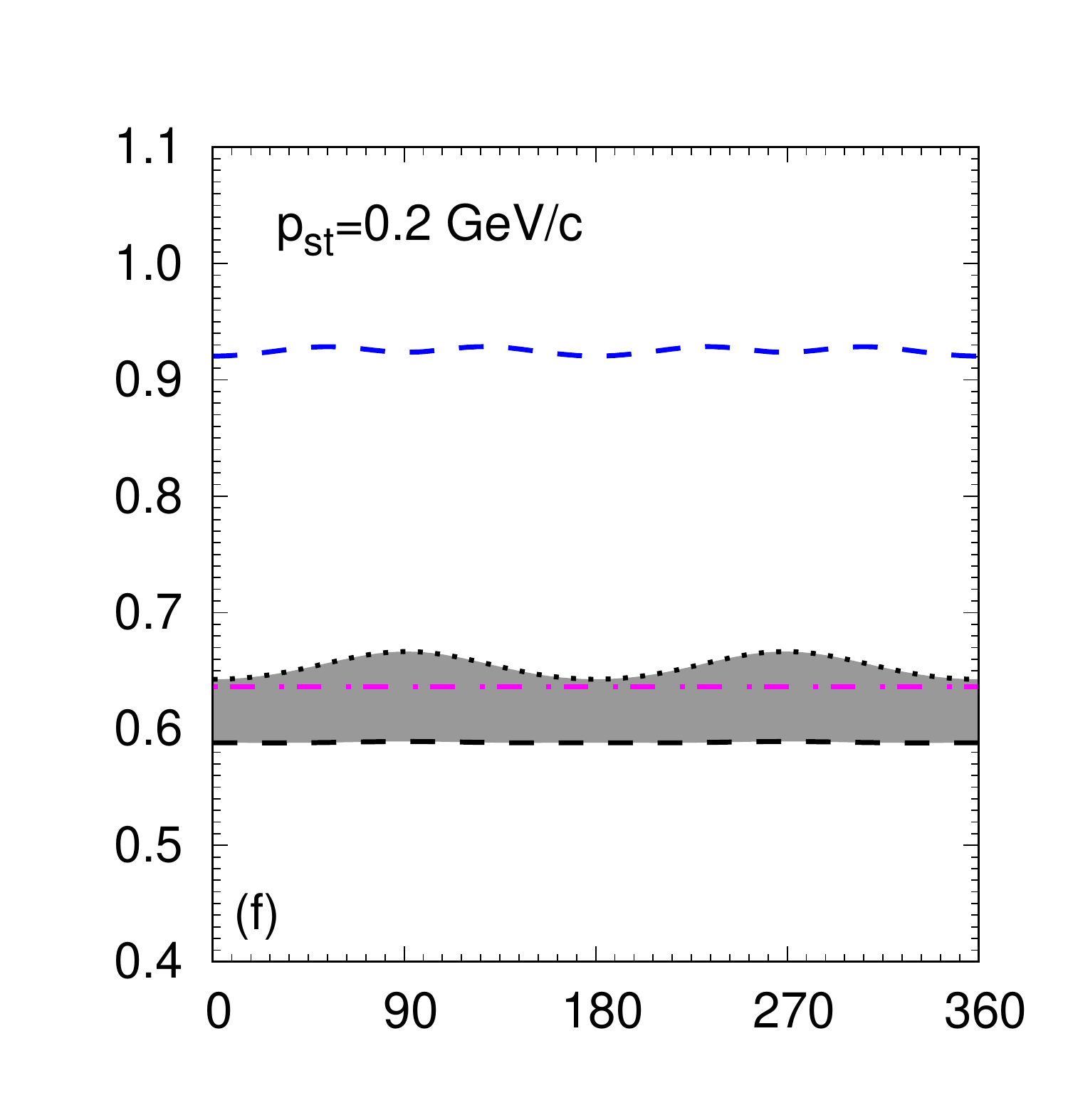} &
   \includegraphics[scale = 0.35]{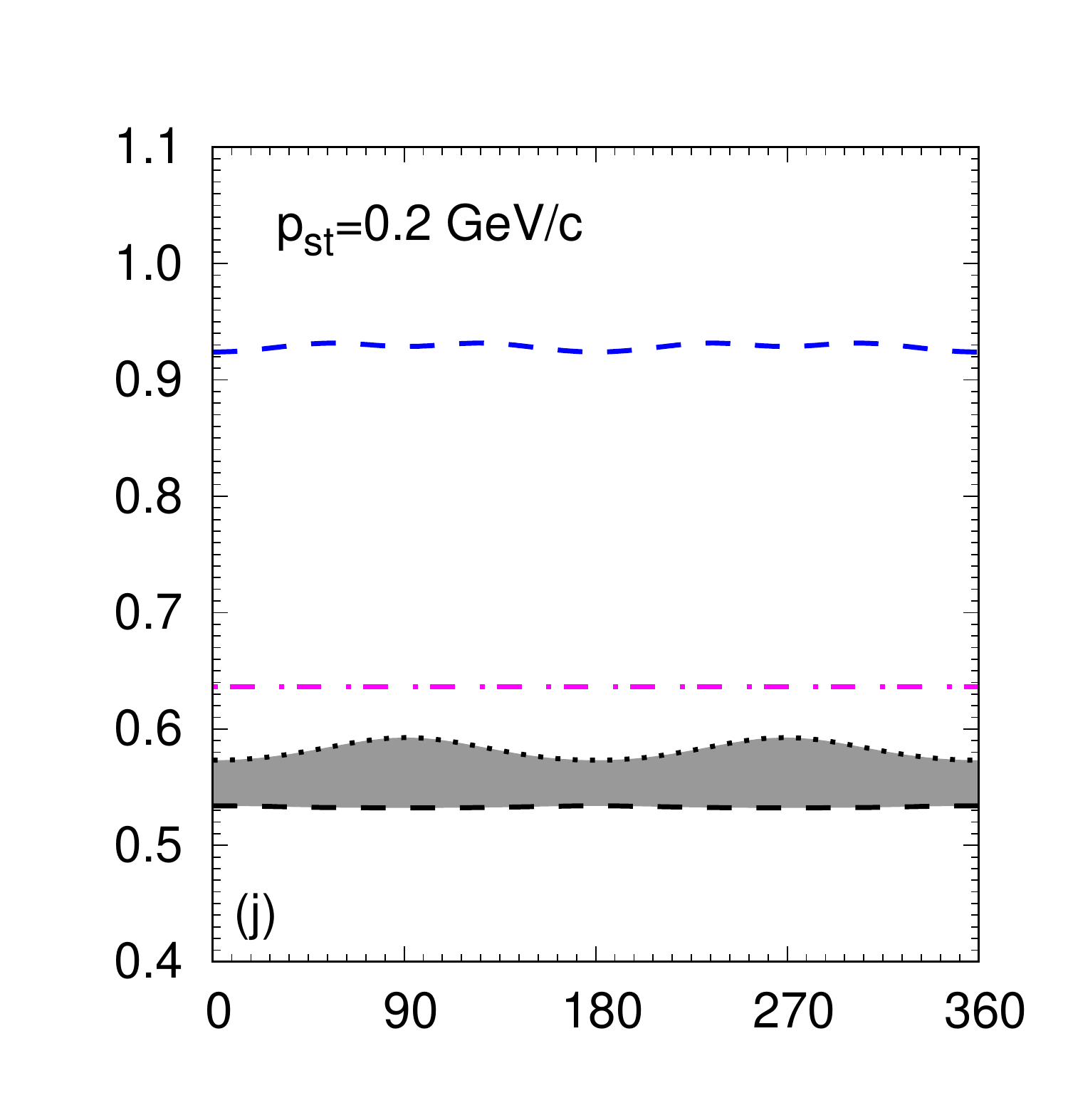} \\  
   \vspace{-1cm}
   \includegraphics[scale = 0.35]{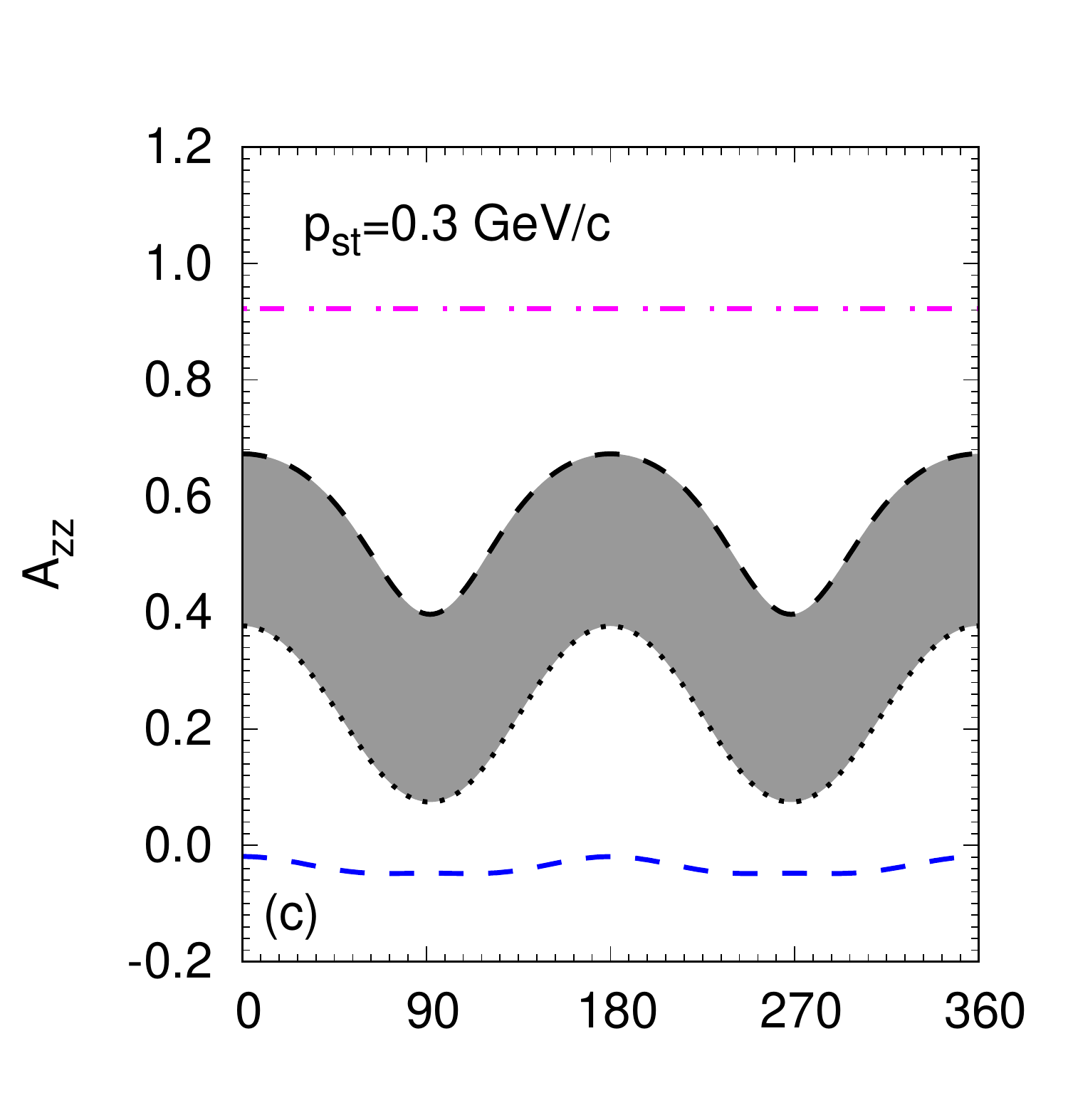} &
   \includegraphics[scale = 0.35]{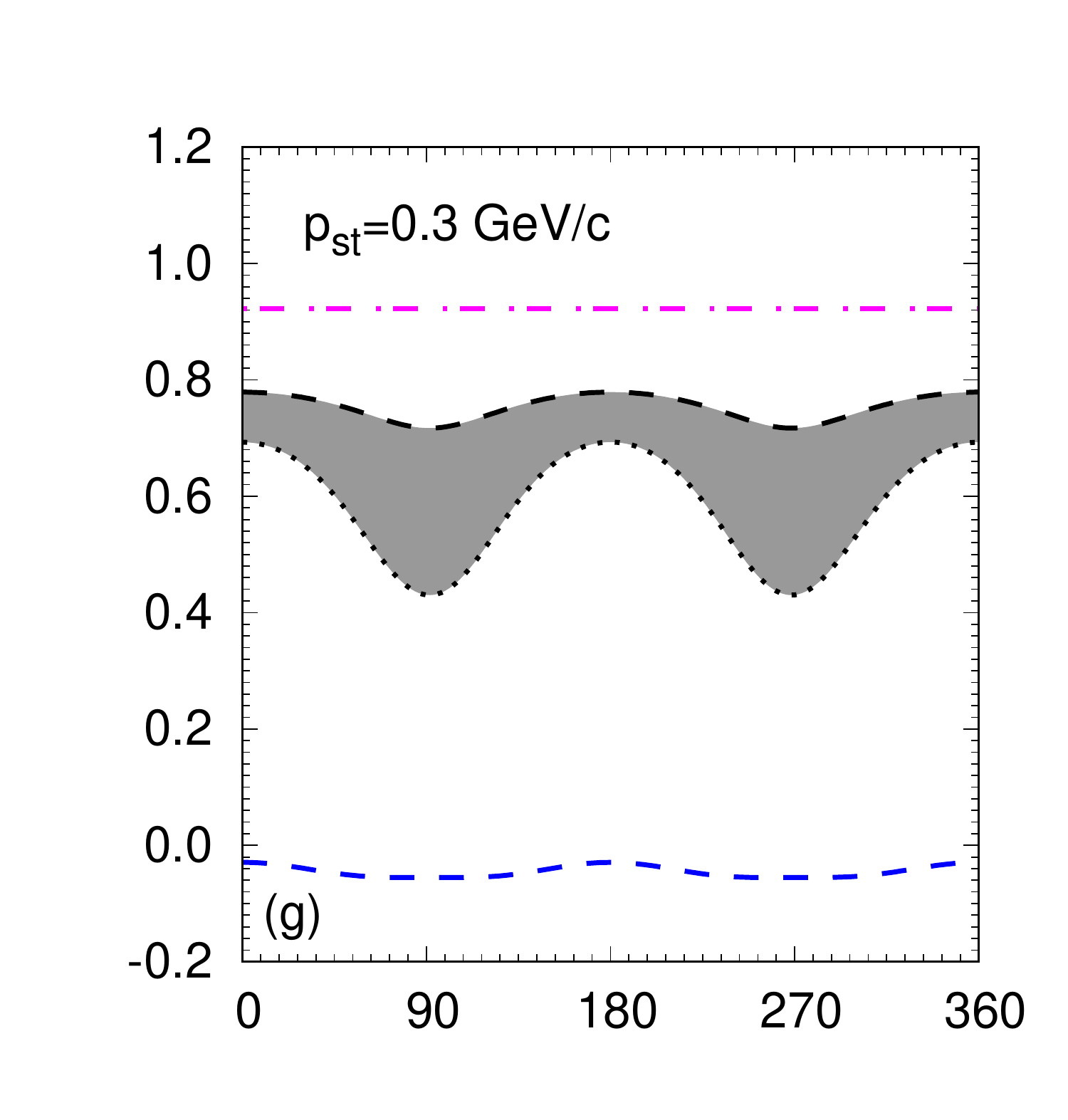} &
   \includegraphics[scale = 0.35]{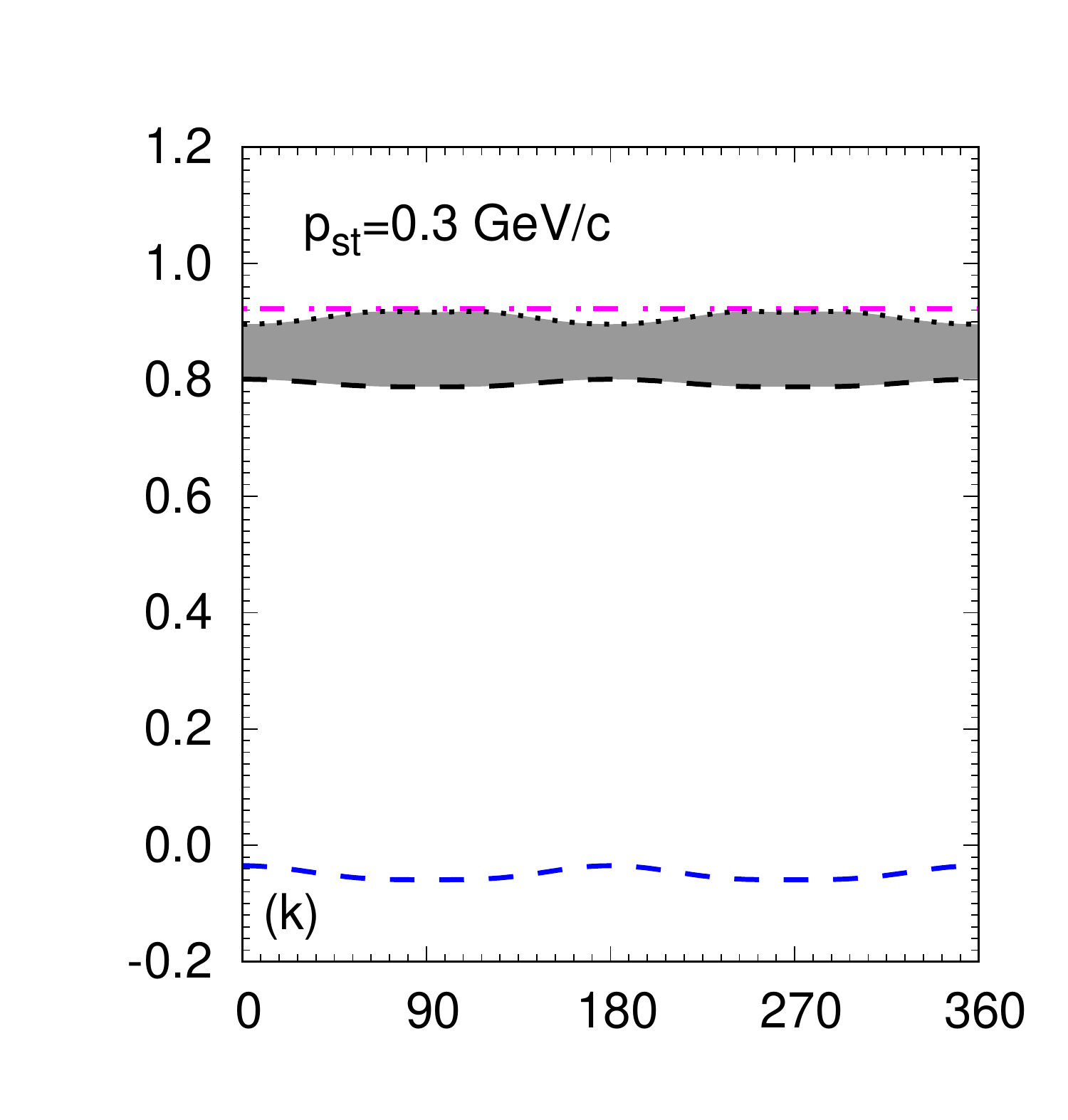} \\
   \includegraphics[scale = 0.35]{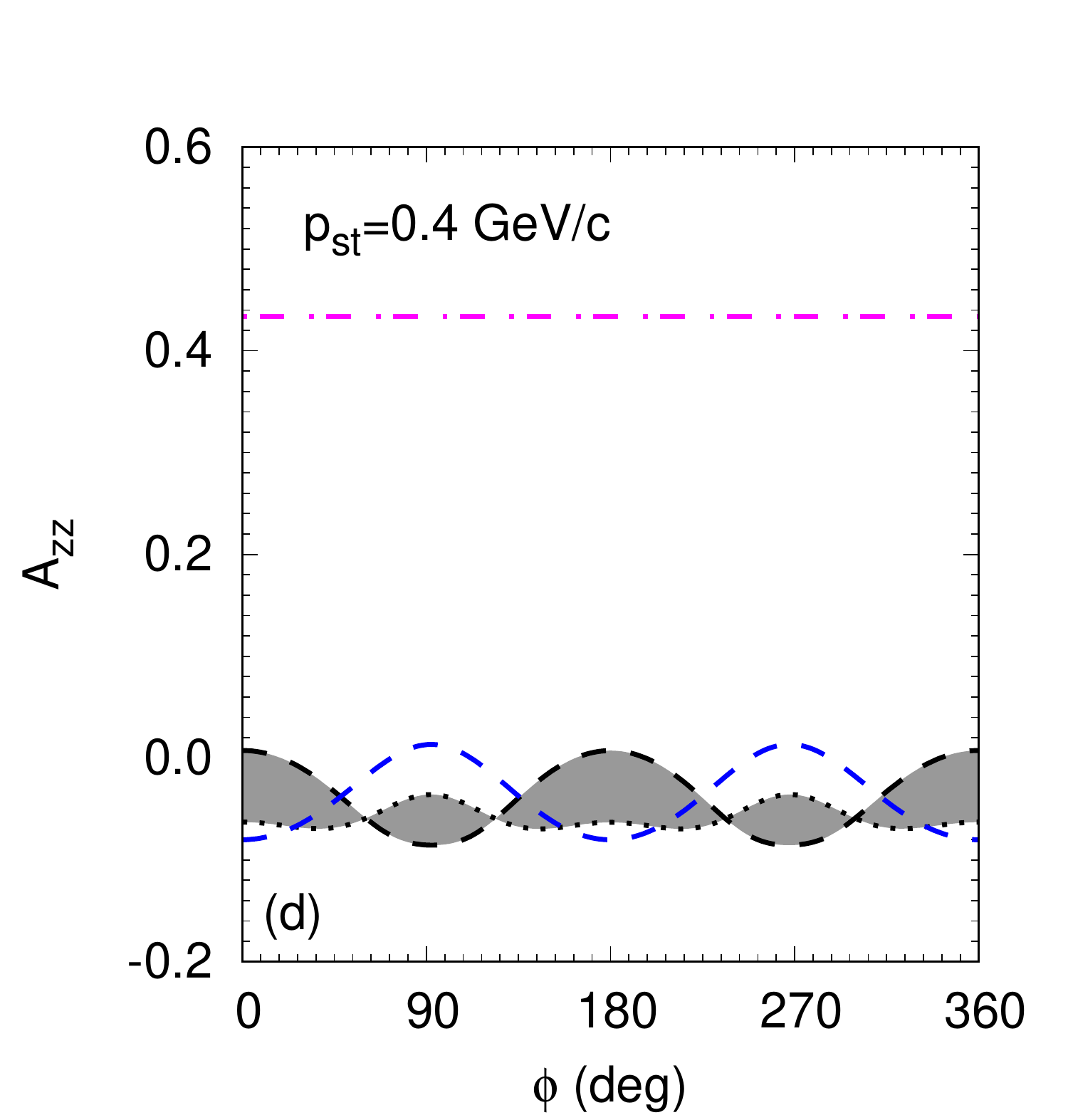} &
   \includegraphics[scale = 0.35]{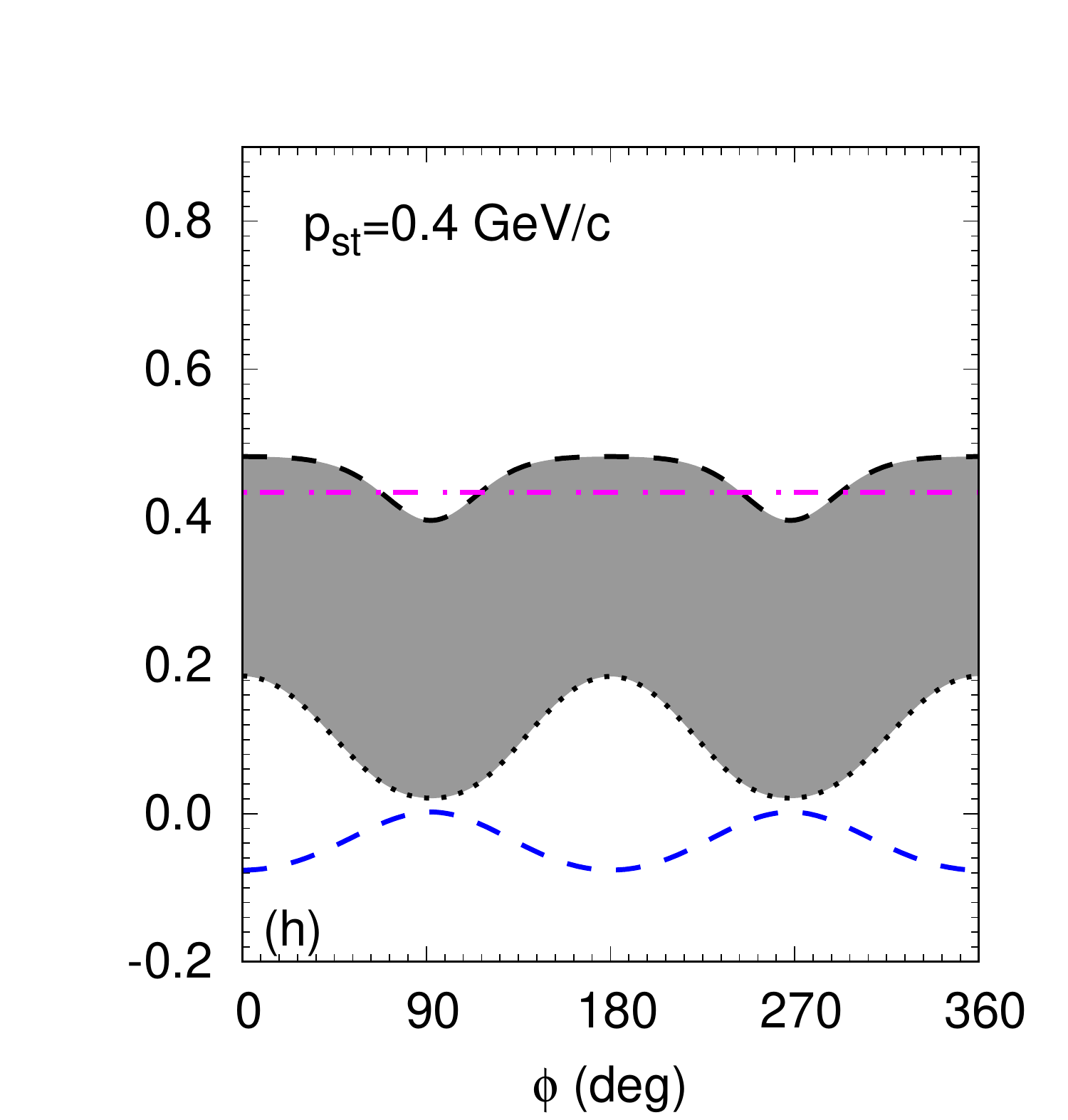} &
   \includegraphics[scale = 0.35]{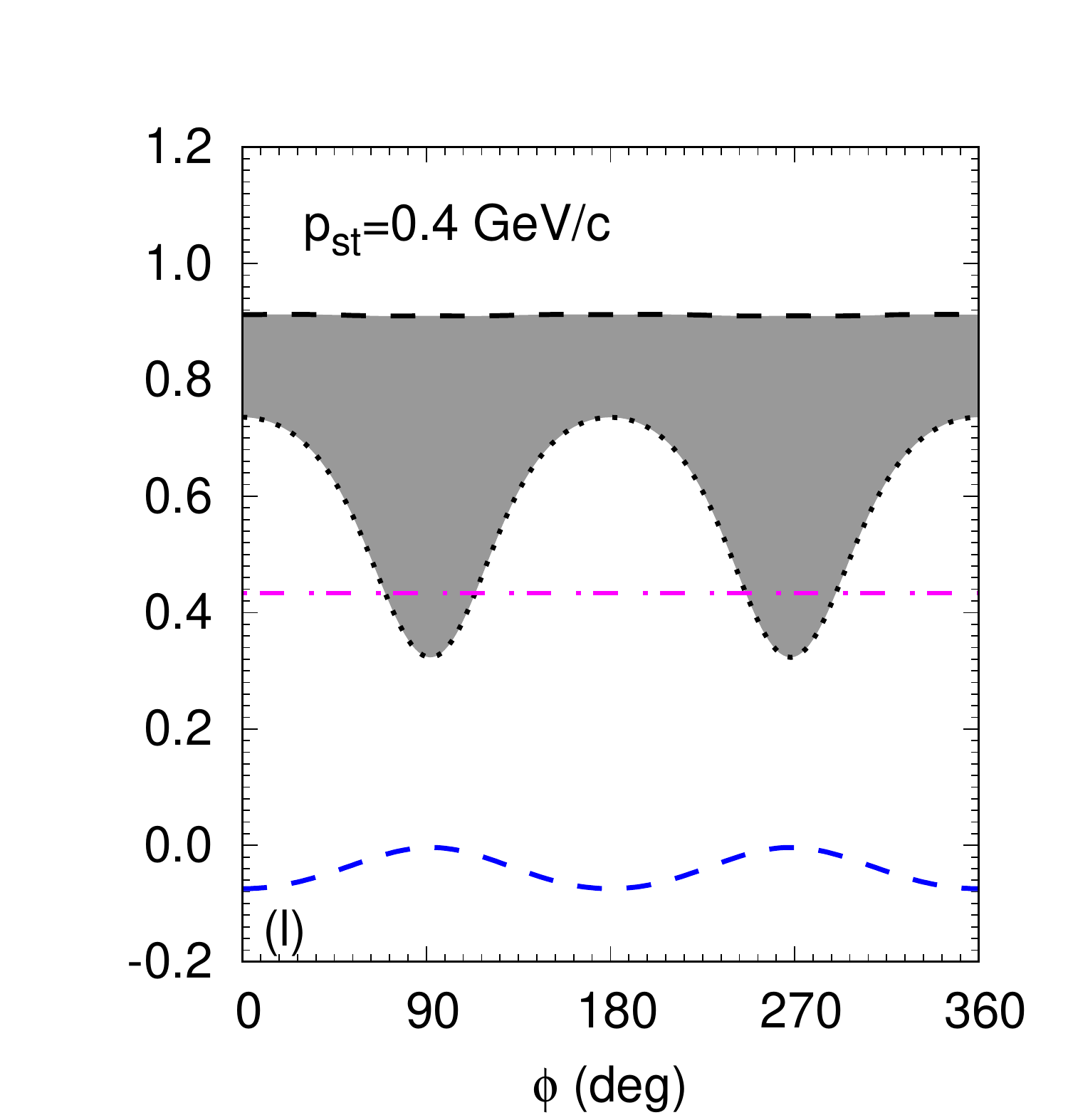} \\
   \end{tabular}
 \end{center}
 \caption{\label{fig:Azz_50&65&75gevc_phiDep} Same as in Fig.~\ref{fig:Azz_6&15&30gevc_phiDep}
   but at $p_{\rm lab}=50$ GeV/c (left column) $p_{\rm lab}=65$ GeV/c (middle column)
   and $p_{\rm lab}=75$ GeV/c (right column).}
\end{figure}

The azimuthal dependence of $A_{zz}$ is shown in Figs.~\ref{fig:Azz_6&15&30gevc_phiDep},\ref{fig:Azz_50&65&75gevc_phiDep}.
For small transverse momenta of the spectator, the GEA calculation gives minima,
and for large -- maxima at $\phi=90\degree$ and $270\degree$.
The $\phi$-dependence of $A_{zz}$ does not change much with the beam momentum in the GEA calculation,
especially at large beam momenta where the produced protons move practically forward while the parameters
of the soft $pn$ scattering amplitude, Eq.(\ref{M_el}), do not vary much.
These properties of $A_{zz}$ are quite similar to those of $T$.

CT changes the $\phi$-dependence of $A_{zz}$ at $p_{\rm lab} \geq 15$ GeV/c drastically.
This can be explained by the D-state dominance in the tensor analyzing power since the distance between neutron and proton
is in this case smaller than for S-state and, hence, the expansion of the PLC is less pronounced which favors CT effects.
In contrast, the transparency is governed by the S-state of the DWF which is characterized by larger distances in the deuteron
and, hence, the expansion of the PLC becomes substantial.

The most pronounced difference between the GEA- and CT calculations is visible at $p_{st}=0.3$ GeV/c, where the former gives
practically constant $A_{zz} \simeq 0$ while the latter produces strong variations
of $A_{zz}$ with $\phi$.
These variations are especially strong at $p_{\rm lab}=15-30$ GeV/c and gradually disappear
towards higher beam momenta.

\subsection{Event rate at NICA-SPD}
\label{rates}

CT effects manifest themselves at finite transverse momenta of the spectator neutron, where the cross section is quite small.
Thus, it is necessary to evaluate the possibility of their experimental study.
As an example, let us take $p_{\rm lab}=30$ GeV/c or $\sqrt{s_{NN}}=7.6$ GeV, which is quite high but still in the accessible energy range
at the first stage of NICA-SPD.
From Fig.~\ref{fig:sig_180deg} one can estimate the differential cross section of about $10^{-6}$ $\mu{\rm b}/{\rm GeV}^4$
at $\Theta_{c.m.}=90\degree$ and $p_{st} = 0.2$ GeV/c.
In the ranges $\Delta\alpha_s=0.2$, $\Delta t=3$ GeV$^2$, $\Delta\phi = \pi/3$, $\Delta p_{st}=0.04$ GeV/c,
the cross section is about $5$ fb.
The reason for such a small cross section is the QC rule, according to which the elastic cross section $d\sigma_{pp}/dt$ for a fixed $t/s$ drops
$\propto s^{-10}$ for $s \to \infty$, see Eq.(\ref{dsigma_pp^QC/dt_par}).
Hence, for the luminosity of $2\cdot10^{31}$ cm$^{-2}$ s$^{-1}$ (see Ref. \cite{Abramov:2021vtu}), the event rate
of about $3$ events/year will be reached which is certainly too low.
The event rate can, however, be increased for $\Theta_{c.m.} < 90\degree$.
As shown in Fig.~\ref{fig:sig_180deg}, for $\Theta_{c.m.}=53\degree$ the cross section is more than two orders of magnitude larger than
for $90\degree$.
On the other hand, the effects of CT experience only little change
as can be seen by comparing the beam momentum dependence of the transparency for $\Theta_{c.m.}=90\degree$ and $53\degree$
in Fig.~\ref{fig:T_vs_plab}.
As we checked, the same is true also for the $p_{st}$- and $\phi$-dependence of $T$ and
of the tensor analyzing power.
Thus, at $\Theta_{c.m.} \simeq 50\degree$ the event rate in the above kinematic range
will reach values of several events/day which seems to be sufficient for the studies of CT.

\section{Summary and conclusions}
\label{summary}

The effects of color transparency in the process $p d \to p p n$ with production of fast forward protons
and a slow spectator neutron in the transverse direction have been considered.
This process is caused by hard elastic $pp$ scattering.
CT reduces the ISI of the incoming proton and the FSI of the outgoing protons with the neutron.
Thus, the gross effect of CT is to bring the results closer to the impulse approximation limit.

Beyond the IA, the present calculations include the amplitudes with single rescattering of the incoming proton
and either of the outgoing protons, as well as the amplitudes with rescattering of both outgoing protons.
The partial amplitudes with rescattering were obtained using the method of generalized eikonal approximation
developed in Refs. \cite{Frankfurt:1996uz,Larionov:2019xdn}.
CT effects were taken into account within the quantum diffusion model \cite{Farrar:1988me}
by performing the calculations in the coordinate space and including the position dependence
of the elementary rescattering amplitudes.
The amplitude of hard elastic $pp$ scattering was described by the sum of the quark counting and Landshoff components.
Accordingly, in calculations with QDM, the interference of quark configurations of small and large sizes
was taken into account.
The interference appears because the ISI/FSI of small-size configurations are reduced by CT
while the ISI/FSI of large-size ones remain unchanged and effectively lead to their absorption by a nucleus
(nuclear filtering effect predicted in Ref. \cite{Ralston:1988rb}).

The dependence of the four-differential cross section and of the transparency $T$ on the
transverse momentum $p_{st}$ of the spectator neutron and on the relative azimuthal angle between
the scattered proton and the neutron was systematically studied in the beam momentum range from 6 to 75 GeV/c. 
The GEA results (i.e. without CT) for $T$ at $p_{\rm lab}=6$ and 15 GeV/c agree with those of Ref. \cite{Frankfurt:1996uz}
reasonably well.
It is shown that the main trend of CT to enhance $T$ at $p_{st} \ltsim 0.2$ GeV/c and to reduce it at $p_{st} \gtsim 0.3$ GeV/c
predicted in Ref. \cite{Frankfurt:1996uz} remains valid also at higher beam momenta.
In light of the recent JLab results \cite{HallC:2020ijh} the uncertainty range of the proton coherence length became very large \cite{Li:2022uvf}  
which hinders definite predictions on the beam energy when CT effects become sizable.
Nevertheless, it is possible to conclude that at $p_{\rm lab} \gtsim 30$ GeV/c the CT effects on $T$ should be already
measurable on the level of at least $30\%$.

Due to the nuclear filtering effect the transparency at small transverse momentum of the spectator and the scaled $pp$ cross section at $\Theta_{c.m.}=90\degree$
oscillate out-of-phase as functions of $p_{\rm lab}$, in agreement with predictions of Ref. \cite{Ralston:1988rb}
for the process $A(p,pp)$ with heavy nuclear targets.
As a result, $T$ at $p_{st}=0.2$ GeV/c has maxima at $p_{\rm lab} \simeq 9, 25,$ and 75 GeV/c. The maximum at 9 GeV/c
is in-line with BNL data \cite{Aclander:2004zm}. Thus, the nuclear filtering effect should be observable at the NICA energy
range.

At $p_{\rm lab} \geq 65$ GeV/c, $T$ grows above unity even for small $p_{st}$ values in the calculation with CT.
Such an "antiabsorptive" behavior is due to out-of-phase QC and Landshoff components of the hard amplitude,
so that the Landshoff components of the single-rescattering amplitudes become of the same sign with the IA amplitude.

The calculation of the tensor analyzing power $A_{zz}$ for the longitudinal deuteron polarization was performed.
Significant effects of CT on $A_{zz}$ at $p_{st} \simeq 0.3-0.4$ GeV/c become measurable already at $p_{\rm lab} = 15$ GeV/c.

Finally, the event rate at NICA-SPD was estimated. It is shown that the reduction of the c.m. scattering angle (in the $pp$ hard
elastic scattering) from $90\degree$ to $53\degree$ increases the differential cross section by about two orders of magnitude
and makes the observation of CT feasible.

\begin{acknowledgments}
The author is grateful to Prof. Mark Strikman and Prof. Yuri Uzikov for stimulating discussions.
\end{acknowledgments}

\bibliography{pd2ppn}

\newpage

\appendix

\section{The four-differential cross section of $pd \to ppn$}
\label{d^4sigma}

Let us now outline the derivation of Eq.(\ref{dsig/dalpha}) for the four-differential cross section.
The fully differential cross section is defined by the standard expression (see Ref. \cite{BLP}):
\begin{equation}
  d\sigma =  
  \frac{(2\pi)^4\overline{|M|^2}}{4p_{\rm lab}m_d} d\Phi_3~,          \label{dsigma}
\end{equation}
where 
\begin{equation}
  d\Phi_3 = \delta^{(4)}(p_1+p_d-p_3-p_4-p_s)
  \frac{d^3p_3}{(2\pi)^32E_3}   \frac{d^3p_4}{(2\pi)^32E_4}   \frac{d^3p_s}{(2\pi)^32E_s}   \label{dPhi_3}
\end{equation}
is the invariant three-body phase space volume element. It is convenient to perform calculation in the IMF where
the four-momentum of the $p + d$ system is ${\cal P} = p_1 + p_d = ({\cal P}^0,\bvec{0},P)$ with $P \to +\infty$.
The total energy can be rewritten as follows:
\begin{equation}
  {\cal P}^0 = P + \frac{{\cal P}^2}{2P} + O(P^{-3})~.     \label{calP^0}
\end{equation}
Introducing the momentum fractions
\begin{equation}
  \beta_i = \frac{2p_i^z}{P}~,~~~i=3,4,s,       \label{beta_i}
\end{equation}
one can rewrite the energies of outgoing particles in a similar form:
\begin{equation}
  E_i = \frac{\beta_i P}{2} + \frac{m_{it}^2}{\beta_i P} + O(P^{-3})~,   \label{E_i}
\end{equation}
where $m_{it} = \sqrt{m_i^2+p_{it}^2}$. Using the relation
\begin{equation}
  \frac{d^3p_i}{E_i} = \frac{d\beta_i d^2p_{it}}{\beta_i}~,     \label{d^3p/E}
\end{equation}
it is now easy to obtain the following expression for the invariant phase space volume element:
\begin{equation}
  d\Phi_3 = 4\delta\left(\sum_i\beta_i -2\right) \delta\left(\sum_i \frac{2m_{it}^2}{\beta_i} - {\cal P}^2\right)
            \delta^{(2)}\left(\sum_i \bvec{p}_{it}\right) \prod_i \frac{d\beta_i d^2p_{it}}{(2\pi)^3 2\beta_i}~.     \label{dPhi_3_LC}
\end{equation}
Integration over $d\beta_4 d^2p_{4t} dp_{3t}$ removes all $\delta$-functions giving
\begin{equation}
  d\Phi_3 = \frac{d\beta_3 p_{3t} d\phi d\beta_s d^2p_{st}}{2(2\pi)^9|\partial{\cal F}/\partial p_{3t}|\beta_4\beta_3\beta_s}~,
               \label{dPhi_3_LC_int}
\end{equation}
where
\begin{eqnarray}
   && {\cal F} \equiv \sum_i \frac{2m_{it}^2}{\beta_i} - {\cal P}^2~,     \label{calF}\\
   && \partial{\cal F}/\partial p_{3t} = \frac{4}{\beta_4} \left[(\beta_4/\beta_3+1)p_{3t} + p_{st} \cos\phi\right]~.   \label{dercalF}
\end{eqnarray}
Note that variables $\beta_i$ are defined in the IMF but can be calculated in any frame related with the IMF by a longitudinal boost.
In particular, in the deuteron rest frame  they are given by the following expression:
\begin{equation}
  \beta_i = \frac{2(E_i+p_i^z)}{E_1 + m_d + p_{\rm lab}}~.          \label{beta_i_lab}
\end{equation}
The azimuthal angle of the spectator can be integrated out, since the cross section does not depend on it.
Thus, in variables $\beta_s, \beta_3, \phi, p_{st}$, the four-differential cross section can be expressed as follows:
\begin{equation}
  \beta_s \beta_3 \frac{d^4\sigma}{d\beta_s d\beta_3 d\phi p_{st} dp_{st}}
  = \frac{\overline{|M|^2} p_{3t}}{8(2\pi)^4p_{\rm lab}m_d|\partial{\cal F}/\partial p_{3t}| \beta_4}~.     \label{d^4sigma_1}
\end{equation}
Instead of variables $\beta_s,\beta_3$ one can equivalently use variables $\alpha_s$ and $\beta$ defined by Eqs.(\ref{alpha_s}),(\ref{beta}),
since
\begin{eqnarray}
  && \left|\frac{d\alpha_s}{\alpha_s}\right| = \left|\frac{d\beta_s}{\beta_s}\right| = \left|\frac{dp_s^z}{E_s}\right|~,    \label{alpha_s_beta_s}\\
  && \left|\frac{d\beta}{\beta}\right| = \left|\frac{d\beta_3}{\beta_3}\right| = \left|\frac{dp_3^z}{E_3}\right|~.      \label{beta_beta_3}
\end{eqnarray}
This gives
\begin{equation}
  \alpha_s \beta \frac{d^4\sigma}{d\alpha_s d\beta d\phi p_{st} dp_{st}}
  = \frac{\overline{|M|^2} p_{3t}}{16(2\pi)^4p_{\rm lab}m_d \kappa_t}~,     \label{d^4sigma_2}
\end{equation}
where $\kappa_t \equiv |\partial{\cal F}/\partial p_{3t}| \beta_4/2$
is expressed by Eq.(\ref{kappa_t}).

At large $s=(p_3+p_4)^2$, variable $\beta$ is simply related to the c.m. scattering angle,
$\beta \simeq 1 + \cos\Theta_{c.m.}$. However, in the energy region studied in this work the accuracy of this relation
is not enough to keep $\Theta_{c.m.}$ and thus also $t$ fixed, in particular, at large transverse momenta of the spectator.
Therefore, one needs to perform one more variable transformation $\beta \to t$. To this end,
keeping $\alpha_s, \phi$ and $p_{st}$ to be constant we can write the following differential relations:
\begin{eqnarray}
  && dt = -2E_1dE_3 + 2p_{\rm lab} dp_3^z~,     \label{rel1}\\
  && d\beta = \frac{2d(E_3+p_3^z)}{\tilde{\cal P}^+}~,  \label{rel2}
\end{eqnarray}
where $\tilde{\cal P}^+ = E_1+m_d-E_s+p_{\rm lab}-p_s^z$ is the ``+'' momentum
of the system of colliding protons.
Using (\ref{rel1}),(\ref{rel2}) one can write:
\begin{equation}
   dt = -E_1\tilde{\cal P}^+ d\beta + 2(p_{\rm lab}+E_1) dp_3^z~.     \label{rel3}
\end{equation}
The relation between $d\beta$ and $dp_3^z$ can be obtained from the requirement that the total energy of the
two outgoing protons is constant, i.e.
\begin{equation}
   dE_3+dE_4=0~,      \label{rel4}
\end{equation}
where
\begin{eqnarray}
  && dE_3 = \frac{p_{3t}dp_{3t} + p_3^zdp_3^z}{E_3}~,   \label{rel5}\\
  && dE_4 = \frac{p_{4t}dp_{4t} + p_4^zdp_4^z}{E_4}~.   \label{rel6}
\end{eqnarray}
Substituting Eqs.(\ref{rel5}),(\ref{rel6}) in Eq.(\ref{rel4}) and using the relation
\begin{equation}
  p_{4t} dp_{4t} = p_{3t} dp_{3t} + p_{st} \cos\phi\, dp_{3t}~,    \label{rel7}
\end{equation}
that is obtained by differentiating the equation $p_{4t}^2 = p_{st}^2 + p_{3t}^2 + 2p_{st}p_{3t}\cos\phi$,
together with the condition $dp_3^z+dp_4^z=0$ leads to the relation
\begin{equation}
    p_{3t} dp_{3t} = \lambda dp_3^z~,     \label{rel8}
\end{equation}
where $\lambda$ is defined by Eq.(\ref{lambda}). Substituting Eq.(\ref{rel5}) in Eq.(\ref{rel2})
and using Eq.(\ref{rel8}) leads to the following formula:
\begin{equation}
   d\beta = \frac{2}{\tilde{\cal P}^+}\left(\frac{\lambda+p_3^z}{E_3}+1\right)dp_3^z~.  \label{rel9}
\end{equation}
From Eqs.(\ref{rel3}),(\ref{rel9}) we finally obtain the expression
\begin{equation}
   dt = \kappa_t^\prime d\beta/\beta~,    \label{rel10}
\end{equation}
with $\kappa_t^\prime$ defined by Eq.(\ref{kappa_t^prime}) which allows us to rewrite the
four-differential cross section, Eq.(\ref{d^4sigma_2}), in the form of Eq.(\ref{dsig/dalpha}).

\end{document}